\documentclass[prb,onecolumn,showpacs,preprintnumbers,amsmath,aps,floatfix]{revtex4}
\usepackage{graphicx,hyperref}
\usepackage{xcolor}
\usepackage{amssymb}
\usepackage{dsfont}
\usepackage{bm}
\usepackage{subfigure}
\def\nbC{{\mathchoice {\setbox0=\hbox{$\displaystyle\rm C$}%
\hbox{\hbox to0pt{\kern0.4\wd0\vrule height0.9\ht0\hss}\box0}}
{\setbox0=\hbox{$\textstyle\rm C$}\hbox{\hbox
to0pt{\kern0.4\wd0\vrule height0.9\ht0\hss}\box0}}
{\setbox0=\hbox{$\scriptstyle\rm C$}\hbox{\hbox
to0pt{\kern0.4\wd0\vrule height0.9\ht0\hss}\box0}}
{\setbox0=\hbox{$\scriptscriptstyle\rm C$}\hbox{\hbox
to0pt{\kern0.4\wd0\vrule height0.9\ht0\hss}\box0}}}}
\def\nbQ{{\mathchoice {\setbox0=\hbox{$\displaystyle\rm
Q$}\hbox{\raise
0.15\ht0\hbox to0pt{\kern0.4\wd0\vrule height0.8\ht0\hss}\box0}}
{\setbox0=\hbox{$\textstyle\rm Q$}\hbox{\raise
0.15\ht0\hbox to0pt{\kern0.4\wd0\vrule height0.8\ht0\hss}\box0}}
{\setbox0=\hbox{$\scriptstyle\rm Q$}\hbox{\raise
0.15\ht0\hbox to0pt{\kern0.4\wd0\vrule height0.7\ht0\hss}\box0}}
{\setbox0=\hbox{$\scriptscriptstyle\rm Q$}\hbox{\raise
0.15\ht0\hbox to0pt{\kern0.4\wd0\vrule height0.7\ht0\hss}\box0}}}}
\def\nbT{{\mathchoice {\setbox0=\hbox{$\displaystyle\rm
T$}\hbox{\hbox to0pt{\kern0.3\wd0\vrule height0.9\ht0\hss}\box0}}
{\setbox0=\hbox{$\textstyle\rm T$}\hbox{\hbox
to0pt{\kern0.3\wd0\vrule height0.9\ht0\hss}\box0}}
{\setbox0=\hbox{$\scriptstyle\rm T$}\hbox{\hbox
to0pt{\kern0.3\wd0\vrule height0.9\ht0\hss}\box0}}
{\setbox0=\hbox{$\scriptscriptstyle\rm T$}\hbox{\hbox
to0pt{\kern0.3\wd0\vrule height0.9\ht0\hss}\box0}}}}
\def\nbS{{\mathchoice
{\setbox0=\hbox{$\displaystyle     \rm S$}\hbox{\raise0.5\ht0%
\hbox to0pt{\kern0.35\wd0\vrule height0.45\ht0\hss}\hbox
to0pt{\kern0.55\wd0\vrule height0.5\ht0\hss}\box0}}
{\setbox0=\hbox{$\textstyle        \rm S$}\hbox{\raise0.5\ht0%
\hbox to0pt{\kern0.35\wd0\vrule height0.45\ht0\hss}\hbox
to0pt{\kern0.55\wd0\vrule height0.5\ht0\hss}\box0}}
{\setbox0=\hbox{$\scriptstyle      \rm S$}\hbox{\raise0.5\ht0%
\hboxto0pt{\kern0.35\wd0\vrule height0.45\ht0\hss}\raise0.05\ht0%
\hbox to0pt{\kern0.5\wd0\vrule height0.45\ht0\hss}\box0}}
{\setbox0=\hbox{$\scriptscriptstyle\rm S$}\hbox{\raise0.5\ht0%
\hboxto0pt{\kern0.4\wd0\vrule height0.45\ht0\hss}\raise0.05\ht0%
\hbox to0pt{\kern0.55\wd0\vrule height0.45\ht0\hss}\box0}}}}
\def\nbZ{{\mathchoice {\hbox{$\sf\textstyle Z\kern-0.4em Z$}}
{\hbox{$\sf\textstyle Z\kern-0.4em Z$}}
{\hbox{$\sf\scriptstyle Z\kern-0.3em Z$}}
{\hbox{$\sf\scriptscriptstyle Z\kern-0.2em Z$}}}}

\begin{document}
\date{\today}

\title{Emergence of a random field at the yielding transition of a mean-field Elasto-Plastic model}

\author{ Saverio Rossi} \email{saverio.rossi@sorbonne-universite.fr}
\affiliation{LPTMC, CNRS-UMR 7600, Sorbonne Universit\'e, 4 Place Jussieu, 75252 Paris cedex 05, France}
\author{ Gilles Tarjus} \email{tarjus@lptmc.jussieu.fr}
\affiliation{LPTMC, CNRS-UMR 7600, Sorbonne Universit\'e, 4 Place Jussieu, 75252 Paris cedex 05, France}

\begin{abstract}
We study the mean-field limit of an elasto-plastic model introduced to describe the yielding transition of athermally and quasi-statically sheared 
amorphous solids. We focus on the sample-to-sample fluctuations, which we characterize analytically, and investigate in detail the analogy with the 
athermally driven random-field Ising model. We stress that the random field at the yielding transition is an emerging disorder and we investigate the 
various factors that determine its strength.
\end{abstract}

\maketitle

\section{Introduction}
\label{sec:introduction}

Many phenomena in physics involve a system which is slowly driven in the presence of some frozen disorder. When temperature is low enough, thermal 
relaxation is virtually impossible and the system only moves under the influence of the applied force. Situations of this kind are encountered for instance
in the deformation and yielding of an amorphous solid under shear, in the depinning of an interface in a random environment, or in the hysteresis and 
Barkhausen noise of a disordered magnetic material driven by an applied field. Idealized models have been proposed that capture the main features of the phenomenology: elasto-plastic models for the deformation and flow of amorphous solids~\cite{baret02,picard05,lin14,barrat_review}, random elastic manifold models for depinning~\cite{larkin,fisher85, fisher86,nattermann,fisher_review,chauve_creep,wiese_review}, and the athermally driven random-field Ising model (RFIM) for hysteresis and avalanches~\cite{sethna93,sethna01,sethna05}.

In this paper we focus on the yielding transition that takes place in sheared amorphous solids and separates a pseudo-elastic regime at low deformation from a flowing plastic regime at larger deformation. Analogies with the depinning transition of an  elastic manifold have already been put forward, especially for the description of the flowing stationary state~\cite{lin14,tyukodi16,jagla18,aguirre18,ferrero19}. 
We are more specifically interested in the yielding transition as it manifests itself in the transient, nonstationary athermal quasi-static (AQS) evolution, 
where its characteristics then depend on the initial state of the material, e.g., its preparation or degree of annealing. It has been suggested that the universal aspects of the transition are analogous to those of the transition found in an AQS driven RFIM~\cite{ozawaPNAS,ozawa_2D}. This conclusion is directly supported by the study of mean-field elasto-plastic models~\cite{ozawaPNAS,popovic-wyart18}. Here, we further examine this analogy at the mean-field level by investigating the sample-to-sample fluctuations in an elasto-plastic model and in an RFIM, both models for which we are able to derive a full analytic characterization. For an elasto-plastic model at a given state point in the close vicinity of the yielding transition, there exists an RFIM with essentially the same correlation functions as long as the variance of the sample-to-sample fluctuations is matched to a suitable effective random-field strength in the RFIM. We emphasize that the effective random field that is at play at the yielding transition and acts 
as a random source linearly coupled to the local stress drops is an emergent property. ``Emergent" refers to the fact that the random field is absent in the system prior to deformation and arises at a later stage of the process from a combination of the initial conditions and the history of 
deformation involving a whole sequence of local plastic events. We then study how the strength of the effective random field depends on the preparation of the amorphous solids which is encoded in distributions of local initial stresses, local yield stresses, local stress jumps, etc., 
in elasto-plastic models. (For a different approach using unsupervised machine learning, see Ref.~[\onlinecite{papanikolaou}].)

The rest of the paper is organized as follows. In Sec.~\ref{sec:models} we introduce the models in their mean-field limit and describe their average 
behavior (stress-strain curve for the elasto-plastic model and magnetization curve for the RFIM). We contrast the characteristics of the two models and 
we also discuss the relevance of mean-field elasto-plastic models for describing the yielding of strained amorphous solids and/or the depinning of an interface in finite dimensions. In Sec.~\ref{sec_RFstrength} we present the formalism allowing one to assess the strength of an effective random field from susceptibilities characterizing the fluctuations of the local order parameter and we illustrate the outcome in the case of the RFIM. Sec.~\ref{sec_suscept_EPM} is then devoted to the analytic calculation of the disconnected susceptibility associated with sample-to-sample fluctuations in the mean-field elasto-plastic model. We discuss the influence of the various types of disorder (initial local stresses, local stress jumps, ...) which are present in the elasto-plastic model on the loading curves (average stress versus applied strain), the connected and disconnected susceptibilities, and the variance of the effective random field near the yielding 
transition in Sec.~\ref{sec_results_EPM}. We find that, as in the RFIM, decreasing the strength of the effective random field makes yielding change from a continuous crossover to a discontinuous transition via a critical point. Finally, we give some concluding remarks in Sec.~\ref{sec_conclusion}. 
The details concerning the analytic calculation of the averaged quantities and those associated with sample-to-sample fluctuations for all considered 
models are given in two appendices. We also devote an appendix to a discussion of  a direct mapping between correlation functions in the  mean-field elasto-plastic model and a mean-field RFIM in a restricted region of deformation near yielding.

\section{The models and their mean-field limit}
\label{sec:models}

\subsection{Elasto-plastic models}

\subsubsection{Definition}

An elasto-plastic model (EPM) is a mesoscopic model in which the amorphous solid is decomposed into smaller blocks evolving with a simple local dynamics 
and in interaction with each other~\cite{barrat_review}. Each such block, associated with a vertex $i$ in a $d$-dimensional lattice with $i=1,\cdots, N$, is described by a local stress $\sigma_i$. (We consider for simplicity the case of a simple shear deformation for which it is sufficient to describe the local shear by a scalar variable.) In the athermal quasi-static (AQS) protocol, a strain $\gamma$ is imposed on each block and the system responds both elastically, i.e., 
with a change of the local stresses proportional to the change of strain, and plastically.  A local plastic event takes place at site $i$ whenever the 
stress $\sigma_i$ exceeds some threshold value $\sigma^{th}_i$ at which it loses mechanical stability. The local stress then abruptly decreases to a lower value. This local stress drop $\Delta\sigma_i$  influences the stress at other sites through an elastic interaction 
described by the Eshelby propagator which is long-ranged, with a decay with distance as $1/r_{ij}^d$, and anisotropic, with a quadrupolar symmetry (e.g., with a dependence on the relative angle $\theta_{ij}$ in $\cos(4\theta_{ij})$ in $d=2$). This interaction may then trigger plastic events at other sites and lead to collective avalanches of such events.

In these EPMs, randomness may appear in several places: the distribution of initial configurations of local stresses at zero strain, $\gamma=0$, the distribution of local thresholds $\sigma^{th}$, and the distribution of local stress drops $\Delta\sigma$. They describe at a phenomenological level the intrinsic heterogeneity and the degree of stability of the amorphous solid at the beginning of the deformation process. This depends on the preparation of the material such as the degree of annealing. For simplicity, we will consider in most of this work that all local thresholds are identical and set $\sigma^{th}=1$. The distribution of initial configurations $\mathcal P_{\gamma=0}(\{\sigma_i\})$ characterizes the randomness in the initial local stability. (By definition of a mechanically stable solid, one  requires that there are no values of the local stress larger than the threshold.) Finally, the distribution of local stress drops $\mathcal G(\{\Delta\sigma_i\})$ could be chosen as dependent on the deformation and the resulting material's history in order to include or reinforce phenomena such as stress hardening and weakening. However, a simpler and more generic picture, which is sufficient to investigate the properties of the yielding transition, is provided by assuming that it is independent of the degree of deformation. One generally further 
assumes that the distributions are independent and factorize as products of identical distributions on every site, e.g.,  
$\mathcal P_{\gamma=0}(\{\sigma_i\})=\prod_i P_{\gamma=0}(\sigma_i)$ and $\mathcal G(\{\Delta\sigma_i\})=\prod_i g(\Delta\sigma_i)$. This is what we will consider in the following.

\subsubsection{Mean-field limit and AQS evolution}
\label{sec:MF-EPMaverage}

A common way to define the mean-field limit of a system is to replace the $d$-dimensional Euclidean lattice by a fully connected lattice (akin to the limit $d\to \infty$). A site variable then interacts with all other site variables. The difficulty in the present EPM context stems from the anisotropic, quadrupolar-like, nature of the Eshelby interaction kernel. This property is a priori lost in a fully connected lattice. There are then two ways to handle the situation. The simplest 
mean-field description~\cite{ozawaPNAS} is to assume that after a local plastic event stress is redistributed uniformly over all sites with a very small amplitude proportional to $1/N$. A refined description consists in introducing some additional quenched disorder: a site receives small random kicks due to all the plastic events that take place elsewhere in the sample; these kicks, which are drawn from a given distribution, can be either positive or negative, therefore 
pushing the site closer or farther from its instability~\cite{lin-wyart16}. As a result the anisotropic character of the stress redistribution persists. The refined model is necessary to describe features of deformed amorphous solids such that their so-called marginal stability characterized by a nonzero pseudo-gap exponent~\cite{karmakar10,lin14,lin-wyart15,lin-wyart16,aguirre18}. However, the yielding transition itself and its dependence on the sample preparation are very similar in the two mean-field models~\cite{popovic-wyart18}. Since our purpose here is to describe the sample-to-sample fluctuations near yielding, we will consider the simpler model with a uniform positive redistribution. Studying these fluctuations in the random-kick model is much more involved and we defer it to future work.

Consider now the AQS evolution of the mean-field EPM under strain. It is convenient to switch variable from the local stress $\sigma_i$ to the local stability, or distance to threshold, $x_i=1-\sigma_i\geq 0$ (where we have used that $\sigma^{th}=1$). When the applied strain changes from 
$\gamma$ to $\gamma+d\gamma$, the local stress at each site increases linearly with $d\gamma$ with a proportionality factor equal to twice the shear modulus $\mu_2$, and accordingly $x_i \to x_i - 2 \mu_2 d\gamma$. Plastic events then take place at sites that reach the threshold; on these sites the stress drops (or, equivalently, the distance to instability $x_i$ jumps) by a random amounts $\hat x_i$ drawn from a distribution $g(\hat x)$ and is then kept constant for the rest of the process up to $\gamma+d\gamma$. The amount $\hat x_i$ is further redistributed on all other sites which, as a result, get closer to instability by an amount $[\mu_2/(\mu_1+\mu_2)]\hat x_i/N$ where $\mu_2/(\mu_1+\mu_2)$ (with $\mu_1>0$) is the amplitude of the interaction kernel. Some of these sites may now become unstable and yield. This process is repeated until no more sites become unstable. A key quantity is then the total stress increment per site that did not yield during the change from 
$\gamma$ to $\gamma+d\gamma$, which is denoted by $dy$. Due to the fully connected nature of the lattice, all sites can be considered as statistically equivalent. However, the increment $dy$ depends on both the sample, i.e., the initial configuration which is denoted by $\alpha$, and the sequence of stress drops, or equivalently, of random variables $\{\hat x_i^1,\cdots, \hat x_i^M\}$ with $\gamma=M d\gamma$. Note that,  
since the stress drop on each site $i$ is independently drawn for each plastic event from the same distribution, it is convenient to define a sequence of random 
drops for site $i$ that assigns a drop for each infinitesimal step $d\gamma$ even if the site does not actually yield at this step: many random variables in 
the sequence $\{\hat x_i^1,\cdots, \hat x_i^M\}$ then do not appear in the evolution equations and will be averaged away in the final quantities.
The total stress increment (for sites that did not yield) $dy^{\alpha,[\hat{x}]_{M+1} }$ then reads
\begin{equation}
\label{eqn:microscopic_dy_evolution_EPM_RJ}
dy^{\alpha,[\hat{x}]_{M+1} } = 2 \mu_{2} d\gamma + 
\frac{\mu_{2}}{N(\mu_{1}+\mu_{2}) } \sum_{i=1}^{N} \theta( dy^{\alpha,[\hat{x}]_{M+1} }-x_{i}^{\alpha,[\hat{x}]_{M}}(\gamma)) \hat{x}_i^{M+1},
\end{equation}
where $[\hat{x}]_{M}\equiv  \{\hat x_i^1,\cdots, \hat x_i^M\}$ for all sites (but keep in mind that the random jumps on each site are independent) and $\theta(x)$ is the Heaviside step function. The first term of the right-hand side corresponds to the linear elastic change and the second is the contribution of the plastic events. (Note that $dy^{\alpha,[\hat{x}]_{M+1} }$ is infinitesimal for an infinitesimal $d\gamma$ except when there is a macroscopic avalanche of unstable sites, which only occurs at the spinodal point associated with brittle yielding; in the mean-field setting one can nonetheless constrain the cumulative parameter $y$ to be continuous, provided one carefully accounts for the presence of a spinodal as a function of $\gamma$.)

Due to the positivity of the interaction, one may invert the above expression and relate a now sample- and history-dependent strain 
increase $d\gamma$ to a fixed $dy$:
\begin{equation}
\label{eqn:dgamma_disorder}
d \gamma^{\alpha,[\hat{x}]_{M+1}} 
= \frac{1}{2 \mu_2} \left[  dy - \frac{\mu_2}{(\mu_2+\mu_1)N} \sum_{i=1}^{N} \theta(d y - x_{i}^{\alpha,[\hat{x}]_{M}}(y))\hat{x}_i^{M+1} \right].
\end{equation}
With either control parameter, the initial condition is $\gamma=y=0$. Considering the evolution of the system as a function of the control parameter $y$ will prove much more convenient than as a function of $\gamma$ because the evolution at each site is then independent of what happens at the others and follows the simple rule 
\begin{equation}
    x_i^{\alpha,[\hat{x}]_{M+1}}(y+dy)=
    \begin{cases}
            x_i^{\alpha,[\hat{x}]_{M}}(y) - dy & \text{if $x_i^{\alpha,[\hat{x}]_{M}}(y) > dy$}\\
            \hat{x}^{M+1}_i & \text{otherwise.}
    \end{cases}
\end{equation}
In the end, one will of course have to switch back to the physical control parameter $\gamma$ by using the above expressions.

The dynamics of the system when changing from $y$ to $y+dy$ (where as an intermediate stage we discretize $y$ in $M$ steps $dy$) can be summarized 
by the following equation that describes the evolution of the stability at site $i$:
\begin{equation}
\theta(x - x_i^{\alpha,[\hat{x}]_{M+1}}(y+dy))
= \theta(x - (x_i^{\alpha,[\hat{x}]_{M}}(y) - dy))\theta(x_i^{\alpha,[\hat{x}]_{M}}(y) -dy) +  \theta(x - \hat{x}_i^{M+1})\theta(dy-x_i^{\alpha,[\hat{x}]_{M}}(y)).
\end{equation}
It expresses that the block $i$ can either move elastically if it is far enough from its instability threshold or jump by a random amount $\hat{x}_i^{M+1}$  if 
it reaches its threshold. After rearranging the Heaviside functions by using that $x\geq 0$, one can recast the equation as
\begin{equation}
\label{eq_evolution_theta_y}
\theta(x - x_i^{\alpha,[\hat{x}]_{M+1}}(y+dy))
= \theta(x+dy - x_i^{\alpha,[\hat{x}]_{M}}(y)) -  \theta(\hat{x}_i^{M+1}-x)\theta(dy-x_i^{\alpha,[\hat{x}]_{M}}(y)).
\end{equation}
From the above equation, one can for instance derive the evolution with $y$ of the fraction of sites that have a stability less than $x$, $F^{\alpha,[\hat{x}]_M}_y(x) = 
(1/N) \sum_{i=1}^N \theta(x-x^{\alpha,[\hat{x}]_M}_i(y))$. It reads
 \begin{equation}
    \label{eqn_f_ev}
    F_{y+dy}^{\alpha,[\hat{x}]_{M+1}}(x) = F_{y}^{\alpha,[\hat{x}]_M}(x + dy) - \frac{1}{N} \sum_{i=1}^N \theta(dy - x^{\alpha,[\hat{x}]_M}_i(y)) \theta(\hat{x}_i^{M+1} - x).
\end{equation}

We first consider the average evolution of the system. Fluctuations will be studied and discussed later on. Since there are two (independent) sources of randomness 
in the present model, one should distinguish two types of averages. For a generic quantity $A^{\alpha,[\hat{x}]_M}$, one defines an average over samples $\alpha$ 
and an average over sequences of random stress drops $[\hat{x}]_M$,
\begin{equation}
\overline{A^{\alpha,[\hat{x}]_M}} = \int_0^{\infty} \left(\prod_{i=1}^N {\rm d} x_i \, P_{0}(x_i) \right) A^{\alpha,[\hat{x}]_M}, 
\end{equation}
\begin{equation}
\label{eq_av_random-jumps}
\left< A^{\alpha,[\hat{x}]_M} \right>_{[\hat{x}]_M} = \int_0^{\infty} \left(\prod_{n=1}^M \prod_{i=1}^N {\rm d} \hat{x}^n_i \, g(\hat{x}^n_i) \right) A^{\alpha,[\hat{x}]_M},
\end{equation}
as well as a full average,
\begin{equation}
A = \overline{\left< A^{\alpha,[\hat{x}]_M} \right>}_{[\hat{x}]_M}. 
\end{equation}
In the above expressions, $A^{\alpha,[\hat{x}]_M}$ may be either taken as a function of $\gamma$ (then, $\gamma=M d\gamma$) or of $y$ (and 
$y=M dy$). Note that $P_0(x)=P_{\gamma=0}(x)=P_{y=0}(x)$.

Fully averaging Eq.~(\ref{eqn_f_ev}) and using the fact that the random jumps are independent from site to site and from event to event 
[see Eq.~(\ref{eq_av_random-jumps})] provide an equation for the cumulative probability $F_y(x)$ for a site to have a stability less than $x$. Deriving it with 
respect to $x$ gives an equation for the probability $P_y(x)=(1/N)\overline{<\sum_{i=1}^N \delta(x-x^{\alpha,[\hat{x}]_M}_i(y))>}$,
\begin{equation}
\label{eq_Py(x)diff_equation}
\partial_y P_y(x)=\partial_x P_y(x)+g(x)P_y(0),
\end{equation}
which can be formally solved in the form
\begin{equation}
\label{eqn:Py(x)evolution}
P_y(x) = P_0(x+y) +\int_x^{x+y} d\hat x\, g(\hat x) P_{x+y-\hat x}(0).
\end{equation}
This result has already been given in Ref.~[\onlinecite{ozawaPNAS}], together with its solution for an exponential distribution of random jumps $g(\hat x)$. 
For a general distribution $g(\hat x)$ it can be recast as
\begin{equation}
\label{eqn:generalP}
P_y(x) = P_0(x+y)+g(x)F_0(y) + \int_0^y dy' F_0(y') R_{y-y'}(x),
\end{equation}
with $F_0(x)=\int_0^x dx' P_0(x')$ and 
\begin{equation}
\label{eqn:R_y}
R_y(x) = g'(x+y) + g(y)g(x)+ \int_0^{y} dy' g(y')R_{y-y'}(x),
\end{equation}
where a prime on a function indicates a derivative with respect to its explicit argument. For a simple exponential distribution, $R_y(x)=0$ for every $y$ and $x$, 
which gives back the solution of  Ref.~[\onlinecite{ozawaPNAS}], and for a more general  distribution $g(\hat x)$ the above equations can be solved by a Laplace transform.

We define the volume-averaged local stability as
\begin{equation}
\label{eq_def_volume-averaged}
m^{\alpha,[\hat{x}]_M}=\frac 1N \sum_i x_i^{\alpha,[\hat{x}]_M}=\int_0^\infty dx\, xP_y^{\alpha,[\hat{x}]_M}(x),
\end{equation}
where $P_y^{\alpha,[\hat{x}]_M}(x)=\partial F^{\alpha,[\hat{x}]_M}_y(x)/\partial x = 
\frac{1}{N} \sum_{i=1}^N \delta(x-x^{\alpha,[\hat{x}]_M}_i(y))$. Averaging over the samples and the random jumps then leads to
\begin{equation}
\label{eq:m_y_fullaverage}
m(y)=\int_0^\infty dx\, xP_y(x),
\end{equation}
where $P_y(x)$ is given by Eqs.~(\ref{eqn:generalP},\ref{eqn:R_y}). (The notation $m$ anticipates the analogy with the magnetization in the RFIM.)

\begin{figure}
\centering
\includegraphics[width=0.6\linewidth]{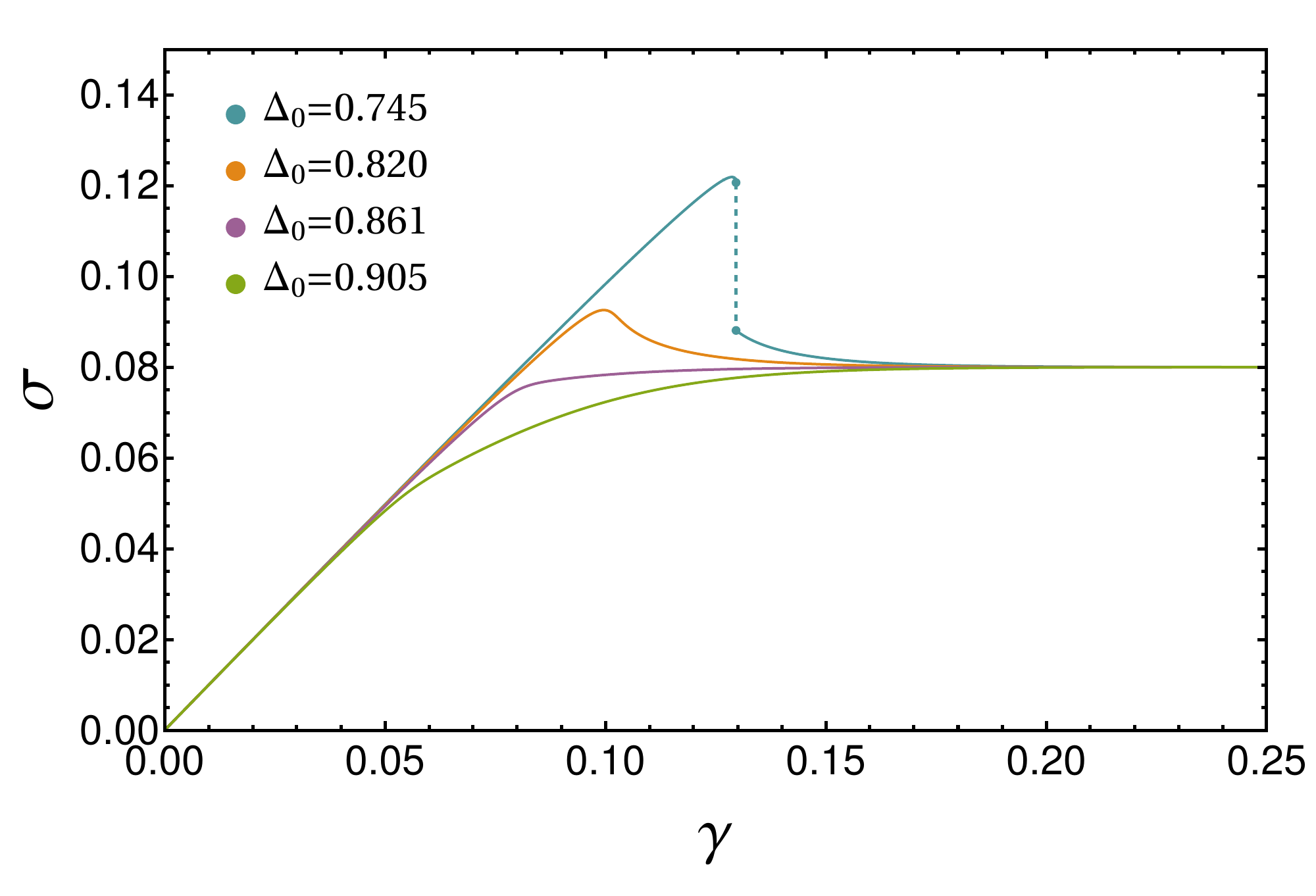}
\caption{Average stress $\sigma$ versus applied strain $\gamma$ for the mean-field EPM with a  distribution of random jumps chosen as an 
exponential and that of the initial local stress (or stability) as a linear combination of two exponentials~\cite{ozawaPNAS}. As the strength $\Delta_0$ 
of the disorder increases (for fixed parameters of the random jump distribution), one passes from 
a regime with a discontinuous stress jump to a continuous behavior with a stress overshoot to finally a monotonically 
increasing regime. The discontinuous and continuous regimes are separated by a critical point.}
\label{fig:stress_EPM}
\end{figure}

To go back to the strain $\gamma$ as the control parameter, one has to take the average of Eq.~(\ref{eqn:dgamma_disorder}). This is done by first averaging 
$d \gamma^{\alpha,[\hat{x}]_{M+1}}$ over $\hat x_i^{M+1}$, which, by using the independence of the variables at each step, simply brings a factor $<\hat x>$ 
$\forall i$ in the second term. Full averaging then further gives
\begin{equation}
\label{eqn:dgamma_fullaverage}
d \gamma
= \frac{1}{2 \mu_2} [  dy - \frac{\mu_2}{(\mu_2+\mu_1)} F_{y}(dy)<\hat{x}> ]=\frac{dy}{2\mu_2}[1-\frac{\mu_2}{(\mu_2+\mu_1)} P_{y}(0)<\hat{x}>],
\end{equation}
from which one directly obtains an expression for the derivative $\gamma'(y)$ and, through an integration starting from $\gamma=0$ at $y=0$,
\begin{equation}
\label{eqn:gamma_fullaverage}
\gamma(y)=\frac{1}{2\mu_2}[y-x_c\int_0^y dy' P_{y'}(0)]
\end{equation}
where we have defined
\begin{equation}
x_c=<\hat{x}>\frac{\mu_2}{(\mu_2+\mu_1)}.
\end{equation}

The loading curve, i.e., the average stress versus applied strain $\sigma(\gamma)$, is then obtained by combining Eqs.~(\ref{eq:m_y_fullaverage}), (\ref{eqn:gamma_fullaverage}), (\ref{eqn:generalP}) 
and (\ref{eqn:R_y}) with the relation
\begin{equation}
\label{eq_sigma_y_general}
\sigma(y)=1-m(y)=\int_0^\infty dx(1-x)P_y(x).
\end{equation}
It is illustrated in Fig.~\ref{fig:stress_EPM} when the distribution of random jumps $g(\hat x)$ is an exponential and the initial distribution $P_0(x)$ is the same 
linear combination of two exponentials as in Ref.~[\onlinecite{ozawaPNAS}]. More details are given in Sec.~\ref{subsec_stress-strain}, where other distributions 
are considered as well. The generic behavior is as follows: As the initial stability of the amorphous solid, which goes inversely with the variance of the initial stress distribution $\Delta_0=\int_0^{\infty}dx P_0(x)x^2-(\int_0^{\infty}dx P_0(x)x)^2$, decreases, yielding of the material changes from brittle-like with a discontinuous jump of the average stress to ductile-like with a continuous evolution, the 
two regimes being separated by a critical point for a specific value of the initial disorder strength. 
\\

\subsubsection{Relevance of the mean-field EPM for the yielding of amorphous solids and the depinning of an interface}

As already stressed in the preceding subsection, the mean-field version that we study here overlooks the anisotropic character of the interactions associated 
with stress redistribution after plastic events. Nonetheless, this shortcoming is not central for describing the yielding transition at the mean-field level since 
the improved model in which sites receive small random kicks of both signs from plastic events taking place in the whole sample yields the same qualitative 
description of the yielding transition as a function of the material's initial degree of stability~\cite{popovic-wyart18}. What may be more troublesome for 
applications to finite-dimensional systems is the fact that the mean-field EPMs are predicted to encounter a linear instability leading to a discontinuous stress 
drop whenever an overshoot (i.e., a local maximum) appears in the average stress-strain curve~\cite{fieldingPRL,fielding}. However, the outcome of this 
instability in real materials is still debated\cite{ozawa_seeds} as it is not clear whether or not this instability can be pinned (and the average stress still be a continuous function of the strain) by the disorder associated with the nonuniform structure of amorphous solids.

The mean-field EPM with a uniform (ferromagnetic-like) redistribution of stress has a more direct connection to the problem of the depinning of an elastic interface 
in a random environment. At the mean-field (fully connected lattice) level, the evolution of the height $h_i$ of the interface at base site $i$ is given 
by~\cite{jagla18,aguirre18,landes14}
\begin{equation}
\partial_t h_i(t)=k(\frac 1N \sum_{j}h_j-h_i)+m^2(w-h_i)-\frac{\partial \mathcal V_i(h_i)}{\partial h_i}
\end{equation}
where the first term of the right-hand side corresponds to the elastic interaction, the second to the drive and the last one to the pinning potential which is 
statistically translation-invariant along each coordinate $h_i$. When the potential $\mathcal V_i(h_i)$ is modeled as a collection of narrow pinning wells 
separated by random intervals $z$ taken from a given distribution $g(z)$, the height at site $i$ is pinned until the driving force exceeds a local threshold 
$f_i^{th}$ associated with the well depth. The evolution equation can then be reformulated in a way similar to that of the mean-field EPM by introducing the 
stability of site $i$ as $x_i=f_i^{th}-m^2(w-h_i)-k(\frac 1N \sum_{j}h_j-h_i)$~\cite{landes14}. A local depinning event is like a plastic event,  
the jump $z_i$ to a new well controlled by the distribution $g(z)$ being analogous to the local stress jump $\hat x_i$ controlled by $g(\hat x)$ and each 
site $\neq i$ getting a (negative) kick in its stability of $kz_i/N$ equivalent to the kick of $[\mu_2/(\mu_1+\mu_2)]\hat x_i/N$ in the EPM. 

The equivalence between depinning-like models and EPMs has been stressed for instance by Jagla and coworkers~\cite{jagla18,aguirre18,ferrero19,landes14} 
and exploited mostly for studying the properties of the stationary state in both models. However, one has to be more careful when considering the transient behavior and the yielding transition that may then take place. Indeed, one knows from Middleton's theorem~\cite{middleton} 
that when starting from $w\to-\infty$ with a flat interface the approach to the steady state must be monotonic (in the thermodynamic limit) and no overshoot 
in the average ``stress'', $\sigma(w)=m^2(w-\overline h)$, is possible when the disorder is (statistically) translationally invariant. To observe an overshoot and 
the equivalent of a discontinuous or a critical yielding transition, one must therefore start, say, at $w=0$, with a very special initial condition $\{h_i(0)\}$: 
the $h_i(0)$ must be such that $(1/N)\sum_i h_i(0)=0$ (implying that each sample starts with a volume-averaged stress equal to $0$) and such that 
$\sigma_i(0)=-(k+ m^2)h_i(0)<f_i^{th}$ (stability requirement on each site when $\overline h=w=0$). In the case of the EPMs these conditions are easily 
satisfied with an appropriate distribution $P_0(x)$. However, for an interface the $h_i(0)$'s must correspond to locations of pinning wells, which for each base 
site are randomly distributed from $-\infty$ to $+\infty$, leading to a statistical translational invariance. This puts strong constraints on the allowed initial 
configurations which then correspond to rare and not typical samples. (Note that, again because of Middleton's theorem, the stress of an interface starting from $w=-\infty$ is always larger than that of an interface starting at $w=w_0$ finite. Since the former is already in its steady state, the presence of an overshoot when starting from $w_0$ just corresponds to a particularly large avalanche in the stationary evolution, and, being untypical, it would vanish after taking the average over samples~\cite{footnote_depinning}.)

Finally, we also note that in the case of pinned interfaces, a critical depinning transition characterized by scale-free avalanches is only obtained in 
the limit $m^2\to 0$. Otherwise, the distribution of avalanche sizes has a cutoff set by $m^2$.

\begin{figure}
    \centering
    \includegraphics[width=0.6\linewidth]{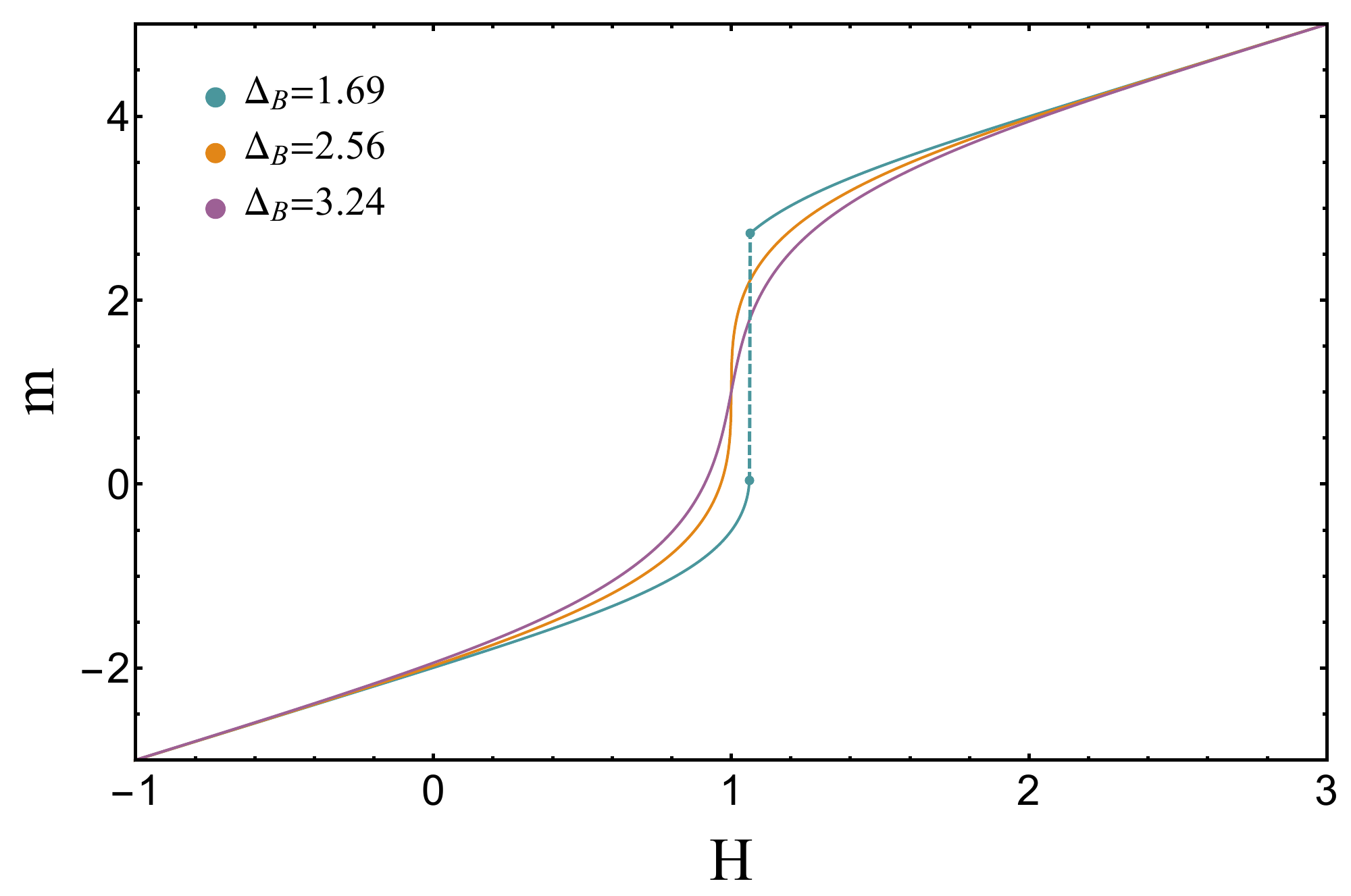}
    \caption{Quasi-statically driven mean-field RFIM at zero temperature: average magnetization $m=\overline{m}$ versus applied field $H$ for the ascending 
    branch of the hysteresis loop and for different values of the disorder strength $\Delta_B$. For these curves, $k=2$ and $J=1$.}
    \label{fig:magn_RFIM}
\end{figure}

\subsection{Driven Random-field Ising model}
\label{sec_driven-RFIM}

\subsubsection{Definition}

The athermal evolution of the RFIM when quasi-statically driven by an applied magnetic field was introduced by Sethna and coworkers~\cite{sethna93,sethna01,sethna05,dahmen96,perkovic96} to describe 
hysteresis, avalanches, and disorder-controlled criticality. The Hamiltonian of the RFIM in its soft-spin version reads
\begin{equation}
\label{eq:RFIM_hamiltonian}
 \mathcal{H}[\left\{ s_{i} \right\}] = - J \sum_{\langle i,j \rangle} s_{i} s_{j} + \sum_{i=1}^N V(s_i) - \sum_{i=1}^{N} h_{i} s_{i}.    
\end{equation}
where the $N$ spins $s_i \in  \mathbb{R}$ are placed on the vertices of a $d$-dimensional lattice, $<ij>$ indicates nearest-neighbor sites in the lattice, 
$V(s)$ is a generic symmetric double-well potential with minima in $s=\pm 1$, and the $h_i$'s are local random fields that are independent from site to site 
and are drawn from the same distribution $\rho(h)$ such that $\overline{h}=0$ and $\overline{h^2}=\Delta_B$. (Here and below the overline denotes an average over the quenched disorder.) In the standard version that we will consider below, the interactions are ferromagnetic and $J>0$.

When quasi-statically driven at zero temperature by an applied magnetic field $H$, the system evolves out of equilibrium by following a sequence 
of dynamically accessible metastable states which is described by the following equation of evolution:
\begin{equation}
\label{eq:RFIM_evolution}
\partial_t s_{i}(t)= J\sum_{j/i}s_j-V'(s_i) +h_i +H(t),
\end{equation}
where the magnetic field $H(t)$ is infinitely slowly ramped up or down (so that the configuration of spins has time to settle in a metastable state before the 
field is changed again) and the sum denoted $j/i$ is over the nearest neighbors of site $i$ in the lattice.

(Note the key differences of the RFIM with the depinning model: the nonrandom potential $V(s_i)$ is not purely Gaussian and the random potential   
$\mathcal V_i(s_i)$, which is simply given by $h_is_i$, breaks the statistical translational invariance.)

\subsubsection{Mean-field limit}

An easily tractable mean-field limit is obtained by considering a fully connected lattice and a double-well 
potential in the form of two joining pieces of parabolas,
\begin{align}
\label{eqn:potential}
    V(s_{i}) = \begin{cases}
                \frac{k}{2}(s_{i}+1)^{2} & s_{i} <0, \\
                \frac{k}{2}(s_{i}-1)^{2} & s_{i} >0,
               \end{cases}
\end{align}
with $k>J>0$. This model was introduced by Dahmen and Sethna~\cite{dahmen96} and further investigated in Refs.~[\onlinecite{mlr1,mlr2,truski}]. It displays a 
rich phenomenology with history-dependent hysteresis effects, avalanches in the evolution of the magnetization, and a sequence of qualitatively different 
behavior separated by an out-of-equilibrium critical point as a function of the disorder strength. In what follows we consider the protocol in which the 
magnetic field is ramped up, i.e., changes according to $H(t)=H+\Omega t$ with $\Omega\to 0^+$; this corresponds to the ascending branch of the 
hysteresis loop in the magnetization versus applied field diagram.

This mean-field model can be solved analytically~\cite{dahmen96}. On the ascending branch of the hysteresis loop, starting for instance from 
a negatively polarized configuration at large negative field, one finds that the solution of Eq.~(\ref{eq:RFIM_evolution}) when $\Omega\to 0^+$ 
is given by
\begin{align}
\label{eqn:spin}
s_{i}^\alpha(H) = \begin{cases}
                 \frac{H+h_{i}+Jm^\alpha(H)}{k} + 1, & h_{i} > -J m^\alpha(H) - H + k, \\
                 \frac{H+h_{i}+Jm^\alpha(H)}{k} - 1, & h_{i} < -J m^\alpha(H) - H + k,
                 \end{cases}
\end{align}
where the superscript $\alpha$ indicates that the quantities depend on the sample $\alpha$, i.e., on the realization of the random fields $\{h_i^\alpha\}$,  and 
$m^\alpha(H)=(1/N)\sum_i s_i^\alpha(H)$ is the volume-averaged magnetization which is given by the self-consistent equation
\begin{equation}
\label{eqn:magnetiz_RFIM}
 m^\alpha(H)  = -1 + \frac{H + J m^\alpha(H)}{k} + \frac{1}{N} \sum_{i=1}^{N} \left[  \frac{h_{i}}{k} + 2 \theta \big(h_{i} +H+Jm^\alpha(H) - k \big) \right].
\end{equation}
Details on the calculations are given in Appendix~\ref{app:MF-RFIM}. For illustration we display in Fig.~\ref{fig:magn_RFIM} the disorder-averaged 
magnetization, which is obtained from
\begin{equation}
\label{eqn:mean_magnetiz_RFIM}
m(H)=\overline{m^\alpha}(H) = \big (\frac k{k-J}\big )\big [-1 + \frac Hk+ 2 \int_{-(H+Jm(H) - k)}^{\infty}dh \rho(h)\big],
\end{equation}
as a function of the applied field $H$ on the ascending branch of the hysteresis curve for several values of the bare variance of the random field $\Delta_B$. 
We have chosen a Gaussian for the random-field distribution. As first shown in [\onlinecite{dahmen96}], the curve has a discontinuity, i.e., a magnetization jump, 
at low disorder, is continuous at large disorder and for a specific value $\Delta_{B,c}=\sqrt{2/\pi}\, kJ/(k-J)$ goes through a critical point at which the slope is 
infinite.

\begin{figure}
    \centering
    \includegraphics[width= 0.4 \linewidth]{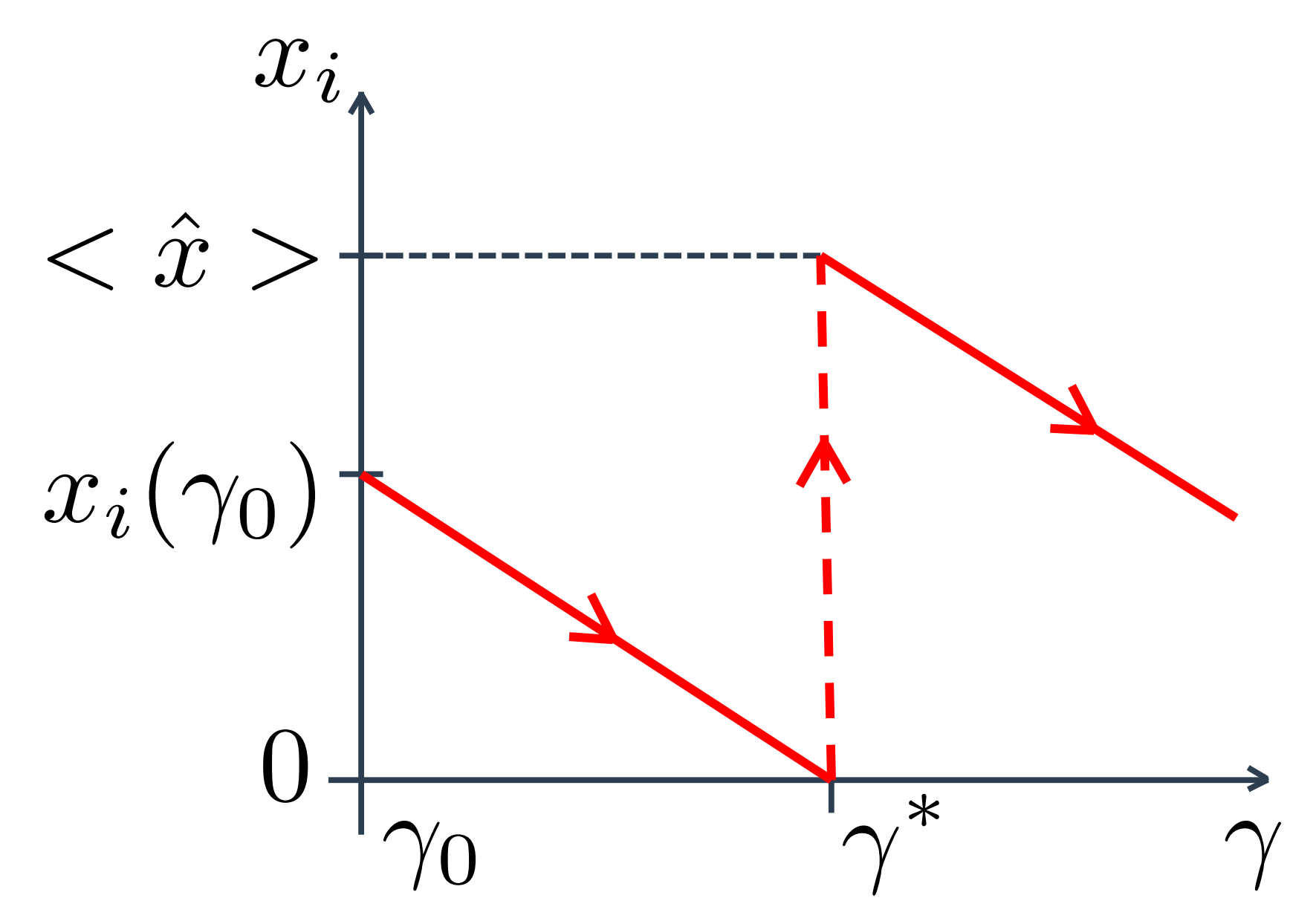} \includegraphics[width= 0.4 \linewidth]{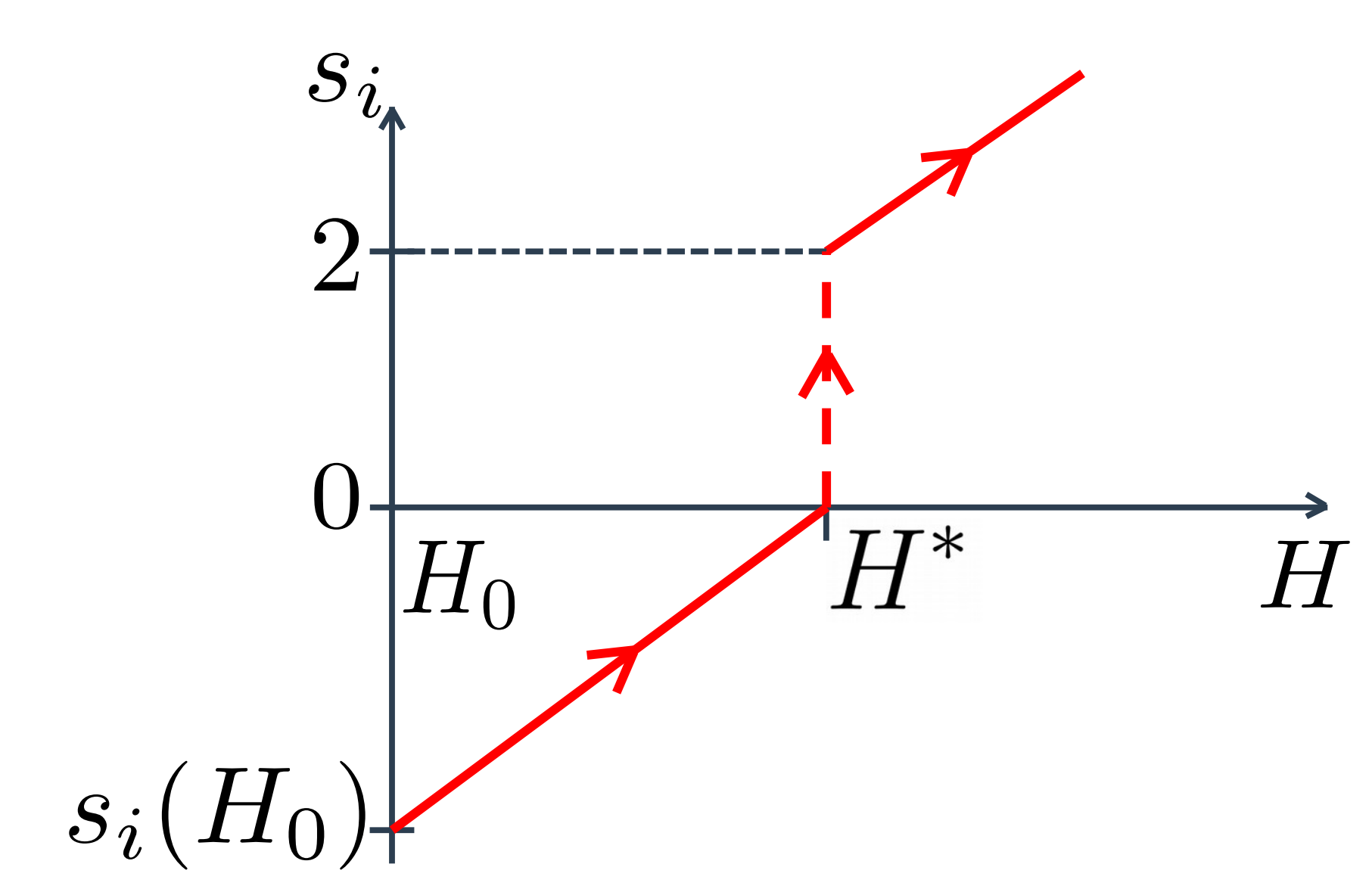}
    \caption{Sketch of the difference between the evolution of the local stability $x_i=1-\sigma_i$ versus applied strain $\gamma$ in 
    an EPM (left) and that of the local magnetization $s_i$ versus applied magnetic field $H$ in an RFIM (right). In both cases the evolution starts from some initial 
    value of the drive, $\gamma_0$ or $H_0$.}
    \label{fig_EPM-RFIM_sketch}
\end{figure}

\subsection{Contrasting EPM and RFIM}
\label{subsec_contrasting}

As we will elaborate more, the comparison between the EPM and the RFIM is meaningful, at least for the mean-field models,  
in the region of the average stress-strain curve between the overshoot, if present, and the steady state. What has been shown~\cite{ozawaPNAS} is that the yielding transition which takes place in this region is in the same universality class as the transition of the driven athermal RFIM.

It is instructive to make a direct comparison between the equations for the AQS evolution of the EPM and of the RFIM. They are indeed rather similar when starting from an initial condition at some intermediate strain $\gamma_0$ or magnetic field $H_0$. Consider the mean-field models introduced above. For simplicity we choose the EPM without randomness in the thresholds and the jumps, so that disorder only comes from the initial condition, but this does not affect the 
conclusions. The equations for the local stability in a given sample $\alpha$ starting from $x_i^\alpha(\gamma_0)$ can then be written
\begin{equation}
\begin{aligned}
\label{eq_contrast_EPM}
x_i^\alpha(\gamma)&=x_i^\alpha(\gamma_0)-2\mu_2(\gamma-\gamma_0)-x_c \,n^\alpha(\gamma_0,\gamma),
\;\;{\rm if}\;\; \gamma< \gamma_*^\alpha,\\&
=x_i^\alpha(\gamma_0) -2\mu_2(\gamma-\gamma_0) -x_c \,n^\alpha(\gamma_0,\gamma)+ <\hat x>,\;\;{\rm if}\;\; \gamma> \gamma_*^\alpha,
\end{aligned}
\end{equation}
where $n^\alpha(\gamma_0,\gamma)$ is the fraction of sites that have yielded between $\gamma_0$ and $\gamma$, and $\gamma_*^\alpha$ is defined by 
$x_i^\alpha(\gamma_0)-2\mu_2(\gamma_*^\alpha-\gamma_0)-x_c \,n^\alpha(\gamma_0,\gamma_*^\alpha)=0$. We implicitly assume here that within the chosen interval of strain the sites yield only once; this will be further discussed and checked below. 

On the other hand, the equations for the driven RFIM starting from an initial spin value $s_i(0)<0$ can be cast as
\begin{equation}
\begin{aligned}
\label{eq_contrast_RFIM}
s_i^\alpha(H)&=s_i^\alpha(H_0)+ (H-H_0) + J[m^\alpha(H)-m^\alpha(H_0)],\;\;{\rm if}\;\; H< H_*^\alpha,\\&
=s_i^\alpha(H_0)+ (H-H_0) + J[m^\alpha(H)-m^\alpha(H_0)]+2,\;\;{\rm if}\;\; H>H_*^\alpha,
\end{aligned}
\end{equation}
where $m^\alpha$ is as before the volume-averaged magnetization in sample $\alpha$ and $H_*^\alpha$ is defined by $s_i^\alpha(H_0)+ 
(H_*^\alpha-H_0) + J[m^\alpha(H_*^\alpha)-m^\alpha(H_0)]=0$. Eqs.~(\ref{eq_contrast_EPM}) and (\ref{eq_contrast_RFIM}) have a 
similar form, except for the sign of the linear drive and for the fact that $n^\alpha(\gamma_0,\gamma)$ is not quite the difference $[m^\alpha(\gamma)-m^\alpha(\gamma_0)]$.

The evolutions corresponding to the above equations are sketched in Fig.~\ref{fig_EPM-RFIM_sketch}. They both correspond to continuous (linear) segments as the control parameter, either $\gamma$ or $H$, is increased which are interrupted by a discontinuous jump associated with a local plastic event (jump of size $<\hat x>$) or a local spin flip (of size $2$). One however notes that the relative sign of the jumps and the linear evolutions are {\it opposite} in the EPM and the RFIM. In a sense, the RFIM has a negative elasticity and, as a result, the spins can only flip once, whereas a site in an EPM will yield many times if one continues to increase the strain (not shown in the sketch), allowing the system to reach a stationary state controlled by the competition between linear elastic increases and plastic stress drops.

The difference just outlined precludes a direct mapping between the EPM and the RFIM at the level of individual dynamical realizations. Nonetheless,  provided one stays in a restricted region of strain where most sites yield only once, the mapping between the mean-field EPM and the mean-field RFIM can be further pushed at the level of correlation functions: see Appendix C. As we show, a direct mapping is possible at the averaged level, e.g., the average stress-versus-strain curve, but is more demanding for sample-to-sample fluctuations. We also show that the assumption of a single plastic event per site is very good 
up to yielding but deteriorates beyond, although this somewhat depends on the random-jump distribution: see Fig.~\ref{fig:mapping} in Appendix~\ref{app_direct-mapping}. 

In the following we rather focus on the sample-to-sample fluctuations and investigate the emergence and the properties of an effective random field in the vicinity of the yielding transition.

\section{Sample-to-sample fluctuations and strength of the effective random field}
\label{sec_RFstrength}

\subsection{Framework}

Our purpose is to derive a workable expression for estimating the strength of the effective random field  (if present) in an athermally driven disordered system 
in terms of quantities that characterize the spatial fluctuations present in the system, i.e., susceptibilities.

Consider a model in the presence of an explicit random field,
\begin{equation}
\label{eq_Hamiltonian_genericRF}
\mathcal H_{{\rm dis}}(\bm s)=\mathcal H_{{\rm pure}}(\bm s)-\sum_i h_i s_i\,,
\end{equation}
where $\bm s\equiv\{s_i\}_{i=1\cdots N}$ are scalar variables,  $\mathcal H_{{\rm pure}}(\bm s)$ is the Hamiltonian of the pure system as for instance 
described in the preceding subsection, and the random fields $h_i$'s are chosen for simplicity i.i.d. Gaussian variables with $h_i=0$ and 
$\overline{h_i h_j}=\delta_{ij} \Delta_B$. Consider also an athermal quasi-static protocol according to which the system is driven by an infinitely  
slowly applied magnetic field,
\begin{equation}
\partial_t s_{i}(t)=-\frac{\partial \mathcal H_{{\rm pure}}(\bm s(t))}{\partial s_{i}(t)} +h_i +H(t)
\end{equation}
with $H(t)=H+\Omega t$ with, e.g., for a ramped up field, $\Omega\to 0^+$ [see Eq.~(\ref{eq:RFIM_evolution})].

This problem can be cast in the framework of generating functionals thanks to the Martin-Siggia-Rose-Janssen-deDominicis formalism~\cite{MSR,janssen-dedom}. 
Through standard manipulations and the introduction of auxiliary variables $\widehat s_i$, one can describe the process by means of a dynamical partition function 
and a dynamical action~\cite{balog_activated,balog_eqnoneq},
\begin{equation}
\begin{aligned}
\mathcal Z_h[\bm H,\bm{\widehat H}]\equiv e^{\mathcal W_h[\bm H,\bm{\widehat H}]}=
\int \prod_i \mathcal D s_i\prod_i \mathcal D\widehat s_i e^{-\mathcal S_h[\bm s,\bm{\widehat s}]+
\sum_i \int_t[\widehat H_i(t) s_i(t)+H_i(t)\widehat s_i(t)]},
\end{aligned}
\end{equation}
where we have used the Ito prescription~\cite{zinn-justin}, with
\begin{equation}
\mathcal S_h[\bm s,\bm{\widehat s}]=\sum_i\int_t \widehat s_i(t)[\partial_t s_i(t)+\frac{\partial \mathcal H_{{\rm pure}}(\bm s(t))}{\partial s_{i}(t)} 
-h_i -H(t)],
\end{equation}
where $\int_t\equiv \int_{-\infty}^{+\infty}dt$. We have added site-dependent sources $H_i(t)$ and $\widehat H_i(t)$ to generate all the correlation 
functions. (In the end, to describe the physical process, one should take $\widehat H_i(t)=0$ and $H_i(t)=H+\Omega t$ with $\Omega\to 0^+$). Brackets 
indicate that we are now dealing with functionals of the variables considered at all times.

The action $\mathcal S_h$ and the free energy $\mathcal W_h$ both depend on the random field (i.e., on the sample $\alpha\equiv \{h_i^\alpha\}$) and can be 
characterized by their cumulants. With the choice of Gaussian random fields, the action is also a Gaussian random functional which is characterized by its first two cumulants given by
\begin{equation}
\begin{aligned}
&S_1[\bm s,\bm{\widehat s}]=\overline{\mathcal S_h[\bm s,\bm{\widehat s}]}=\sum_i\int_t \widehat s_i(t)[\partial_t s_i(t)+
\frac{\partial \mathcal H_{{\rm pure}}(\bm s(t))}{\partial s_{i}(t)}-H(t)] ,\\&
S_2[\bm s,\bm{\widehat s}]=\overline{\mathcal S_h[\bm s,\bm{\widehat s}]^2}^{{\rm cum}}= \Delta_B \sum_i\int_t\int_{t'}\widehat s_i(t)\widehat s_i(t'),
\end{aligned}
\end{equation}
where the superscript ``cum'' indicates a cumulant average. For further use, we note that the variance of the bare random field is then obtained by 
differentiating  the second cumulant twice,
\begin{equation}
\label{eq:variance_bare}
\Delta_B \delta_{ij}=\frac{\partial^2 S_2[\bm s,\bm{\widehat s}]}{\partial \widehat s_i(t)\partial \widehat s_j(t')},
\end{equation}
or equivalently, since the second derivative is purely local, 
\begin{equation}
\label{eq:variance_bare}
\Delta_B=\frac 1N\sum_{ij}\frac{\partial^2 S_2[\bm s,\bm{\widehat s}]}{\partial \widehat s_i(t)\partial \widehat s_j(t')},
\end{equation}
where the second cumulant is here independent of time: this expresses the property that the disorder is purely static (quenched).

The cumulants of the random free energy $\mathcal W_h$ can also be introduced, with $W_1[\bm H,\bm{\widehat H}]$ the first cumulant, 
$W_2[\bm H,\bm{\widehat H}]$ the second cumulant, etc., and one then has
\begin{equation}
\label{eq:expansion_W}
W[\bm H,\bm{\widehat H}]=\ln\overline{\mathcal Z_h[\bm H,\bm{\widehat H}]} =
W_1[\bm H,\bm{\widehat H}]+\frac 12 W_2[\bm H,\bm{\widehat H}]+\frac 1{3!}W_3[\bm H,\bm{\widehat H}]+\cdots
\end{equation}
For a complete description of the functional dependence of the cumulants on their arguments and a more transparent introduction of the expansion 
in cumulants one could introduce copies or replicas of the system with the same random field but coupled to different sources: see 
Refs.~[\onlinecite{balog_activated,balog_eqnoneq,tarjus_review}]. However, this is not really needed here and we proceed without replicas to alleviate the notations.

From the functional $W[\bm H,\bm{\widehat H}]$ one obtains all the correlation functions by functional differentiation. For our purpose we only 
need the susceptibilities, which due to the presence of the random field come as the so-called connected susceptibility, which is also the response of the average magnetization to the applied field, $\chi_{{\rm con}}=\partial\overline{m}/\partial H$, and can be expressed as
\begin{equation}
\label{eq:con_suscept}
\chi_{{\rm con}}(H)=\frac 1N\sum_{i,j}FT_{\omega=0}\, \overline{\widehat s_i(t)s_j(t')}\big\vert_{{\rm unif}}^{{\rm cum}}
=\frac 1N\sum_{i,j}FT_{\omega=0}\,\frac{\partial^2 W_1[\bm H,\bm{\widehat H}]}{\partial H_i(t)\partial \widehat H_j(t')}\Big\vert_{{\rm unif}}\,,
\end{equation}
and the so-called disconnected susceptibility,
\begin{equation}
\label{eq:dis_suscept}
\chi_{{\rm dis}}(H)=\frac 1N\sum_{i,j}\lim_{\vert t-t'\vert\to +\infty}\,\overline{s_i(t)s_j(t')}\big\vert_{{\rm unif}}^{{\rm cum}}
=\frac 1N\sum_{i,j}\lim_{\vert t-t'\vert\to +\infty}\,\frac{\partial^2 W_2[\bm H,\bm{\widehat H}]}{\partial \widehat H_i(t)\partial \widehat H_j(t')}\Big\vert_{{\rm unif}}\,,
\end{equation}
where the subscript ``unif'' indicates uniform sources with, for all sites $i$, $\widehat H_i=0$ and $H_i(t)=H$ (in the limit $\Omega\to 0^+$), and 
$FT_\omega$ denotes a Fourier transform over the time difference $t-t'$.

If one wishes to evaluate the strength of the renormalized random field once fluctuations on all scales have been taken into account, one must substitute the bare action with the effective action or Gibbs free energy $\Gamma[\bm{\widehat m},\bm m]$. The latter is obtained from the free energy 
$W[\bm H,\bm{\widehat H}]$ through a Legendre transform, 
\begin{equation}
\Gamma[\bm{\widehat m},\bm m]=-W[\bm H,\bm{\widehat H}]+\sum_{i}\int_t [H_i(t)\widehat m_i(t)+\widehat H_i(t) m_i(t)]
\end{equation}
where $\widehat m_i(t)=\partial W[\bm H,\bm{\widehat H}]/\partial H_i(t)=\overline{\widehat s_i(t)}$ and $m_i(t)=
\partial W[\bm H,\bm{\widehat H}]/\partial \widehat H_i(t)=\overline{s_i(t)}$. 
Just like the free energy $W[\bm H,\bm{\widehat H}]$ in Eq. (\ref{eq:expansion_W}), the effective action can be expanded in cumulants 
(again, a more precise connection is obtained by introducing replicas: see Refs.~[\onlinecite{balog_activated,balog_eqnoneq,tarjus_review}]),
\begin{equation}
\label{eq:expansion_Gamma}
\Gamma[\bm{\widehat m},\bm m]=\Gamma_1[\bm{\widehat m},\bm m]-\frac 12 \Gamma_2[\bm{\widehat m},\bm m]+
\frac 1{3!}\Gamma_3[\bm{\widehat m},\bm m]-\cdots
\end{equation}
with
\begin{equation}
\begin{aligned}
\label{eq:cumulants_Gamma}
&\Gamma_1[\bm{\widehat m},\bm m]=-W_1[\bm H,\bm{\widehat H}]+\sum_{i}\int_t [H_i(t)\widehat m_i(t)+\widehat H_i(t) m_i(t)],\\&
\Gamma_2[\bm{\widehat m},\bm m]=W_2[\bm H[\bm{\widehat m},\bm m],\bm{\widehat H}[\bm{\widehat m},\bm m]],
\end{aligned}
\end{equation}
etc., where  $H_{i,t}[\bm{\widehat m},\bm m]=\partial \Gamma_1[\bm{\widehat m},\bm m]/\partial \widehat m_i(t)$ and 
$H_{i,t}[\bm{\widehat m},\bm m]=\partial \Gamma_1[\bm{\widehat m},\bm m]/\partial m_i(t)$ are nonrandom sources.

By analogy with Eq.~(\ref{eq:variance_bare}), one can define the component of the variance of the renormalized random field that is local in space and independent of time as
\begin{equation}
\label{eq:variance_eff}
\Delta_{{\rm eff}}(m)=\frac 1N \sum_{ij}\frac{\partial^2 \Gamma_2[\bm m,\bm{\widehat m}]}{\partial \widehat m_i(t)\partial \widehat m_j(t')}\Big\vert_{{\rm unif}},
\end{equation}
where the subscript  ``unif'' now indicates uniform variables with $\widehat m_i=0$ and $m_i(t)=m$. (Remember that we consider the quasi-static limit, 
for which it can be shown that the above defined quantity for such uniform variables is indeed purely static~\cite{balog_eqnoneq}.)

By using the properties of the Legendre transform and Eq.~(\ref{eq:cumulants_Gamma}), it is straightforward 
to relate $\Delta_{{\rm eff}}(m\,{\rm or}\, H)$ to the connected and disconnected susceptibilities introduced in  Eqs.~(\ref{eq:con_suscept}, \ref{eq:dis_suscept}):
\begin{equation}
\label{eq:strength_effectiveRF}
\Delta_{{\rm eff}}(H)=\frac{\chi_{{\rm dis}}(H)}{\chi_{{\rm con}}(H)^2}\,.
\end{equation}
This allows us to define the strength of the renormalized random field, i.e., the effective random field obtained after having included all fluctuations. This 
effective random field of course need not be Gaussian nor purely local, and the renormalized disorder may include other forms of randomness, but 
Eq.~(\ref{eq:strength_effectiveRF}) provides us with an estimate of the dominant contribution, which will be sufficient for mean-field models. Note also that 
right at the critical point of the RFIM in finite dimensions, below the upper critical dimension and above the lower critical dimension, $\Delta_{{\rm eff}}$ 
diverges. However, this  divergence is very slow in $d=3$~\cite{balog_eqnoneq,tarjus_review} and is absent at the mean-field level.

Eq.~(\ref{eq:strength_effectiveRF}) will be taken as a general means to assess the strength of an emerging random field in systems where the random field is not present or not easily detectable at the microscopic level, as in a strained amorphous solid. In the latter, even in the simple EPM description, the quenched disorder that is initially present as a result of the amorphous nature of the material does not a priori take the form of a random field coupled, say, to the local stress. Besides Eq.~(\ref{eq:strength_effectiveRF}) and the ensuing necessary condition that there exists a nonzero 
disconnected susceptibility on top of the usual connected one, there are additional requirements to be satisfied before concluding to the presence of a random field: 
\begin{itemize}
    \item First, once a local order parameter is identified, the associated connected susceptibility which quantifies the linear response of this order parameter to an applied field must be strictly positive (as it is for systems described by Eq.~(\ref{eq_Hamiltonian_genericRF})). The quantities appearing in the numerator and the denominator of Eq.~(\ref{eq:strength_effectiveRF}) should indeed be describable as the variance of the local order parameter and the response of the average order parameter to a uniform field.
    \item Second, as will be illustrated below in the case of the AQS driven RFIM, the variation of $\Delta_{{\rm eff}}(H)$ should be limited (at least away from the critical point) for the random-field description to be of any use.
\end{itemize}

\subsection{Illustration for the mean-field RFIM}

We first illustrate the outcome of Eq.~(\ref{eq:strength_effectiveRF}) in the case of the AQS driven mean-field RFIM introduced in 
Sec.~\ref{sec_driven-RFIM}. To do so we have to compute the connected and the disconnected (magnetic) susceptibilities.

The connected susceptibility [see Eq.~(\ref{eq:con_suscept})] can be simply expressed as the derivative of the average magnetization 
$m=\overline{m^\alpha}$ with respect to the applied field $H$. With the help of Eq.~(\ref{eqn:mean_magnetiz_RFIM}), this leads to
\begin{equation}
 \label{eqn:conn_MF-RFIM}
    \chi_{\rm con}(H) = \frac{d m(H)}{dH} = \frac{1 + 2k \rho(H+Jm(H)-k)}{ k-J-2kJ\rho(H+Jm(H)-k)},
\end{equation}
where we choose for the random-field distribution $\rho(h)$ a Gaussian (centered on $0$ and of variance $\Delta_B$). The susceptibility diverges both 
at the critical point when $\Delta_B=\Delta_{B,c}=\sqrt{2/\pi}\, kJ/(k-J)$ and $H=H_c$ and, when $\Delta_B<\Delta_{B,c}$, at the spinodal instability 
$H=H_s(\Delta_B)$ leading to the discontinuous jump of the magnetization. Both $H_c$ and $H_s$ are solutions of 
$\rho(H+Jm(H)-k)=(k-J)/(2kJ)$~\cite{dahmen96}.

We next compute the disconnected susceptibility, which requires a description of the sample-to-sample fluctuations and has not been calculated before. The 
disconnected susceptibility is defined by
\begin{equation}
\chi_{\rm dis}(H) = N \big[ \overline{m^\alpha(H)^2} - \overline{m^\alpha(H)}^{2}\big],
\end{equation}
which corresponds to Eq.~(\ref{eq:dis_suscept}) in the case of the mean-field RFIM. 

As standard for mean-field models, the calculation of all averaged quantities such as the mean magnetization and the connected susceptibility can be 
reduced to a saddle-point approximation in the thermodynamic limit where $N\to\infty$. To access the fluctuations,
\begin{equation}
\delta \widehat{m}^\alpha(H)=\sqrt N[m^\alpha(H)-m(H)],
\end{equation}
one needs to consider terms beyond the saddle-point approximation. As detailed in Appendix~\ref{app:MF-RFIM}, this is easily performed, and one ends 
up with the following result for the disconnected susceptibility, 
\begin{equation}
 \label{eqn:disc_MF-RFIM}  
\chi_{\rm dis}(H) = \overline{\delta \widehat{m}^\alpha(H)^2}= 
\frac{\left[\Delta_B + 
4 k^2\int_{-H-Jm(H)+k}^{+\infty} dh \rho(h) \int_{-\infty}^{-H-Jm(H)+k} dh \rho(h) + 4 k\int_{-H-Jm(H)+k}^{+\infty} dh h\rho(h) \right]}
{\left[  k-J-2kJ\rho(H+Jm(H)-k) \right]^{2}}.
\end{equation}
From Eqs.~(\ref{eq:strength_effectiveRF}), (\ref{eqn:conn_MF-RFIM}), and (\ref{eqn:disc_MF-RFIM}) one then obtains the variance of the renormalized 
random field as
\begin{equation}
\label{eqn:eff_RF_Ising}
\Delta_{\rm eff}(H) = \frac{\left[ \Delta_B+ 4k^2 \int_{-H-Jm(H)+k}^{+\infty} dh \rho(h) \int_{-\infty}^{-H-Jm(H)+k} 
dh \rho(h) + 4 k\int_{-H-Jm(H)+k}^{+\infty} dh h \rho(h) \right]}{\left[ 1 + 2 k\rho(H+ Jm(H) -k) \right]^{2}}.
\end{equation}

\begin{figure}
    \centering
    \includegraphics[scale = 0.45]{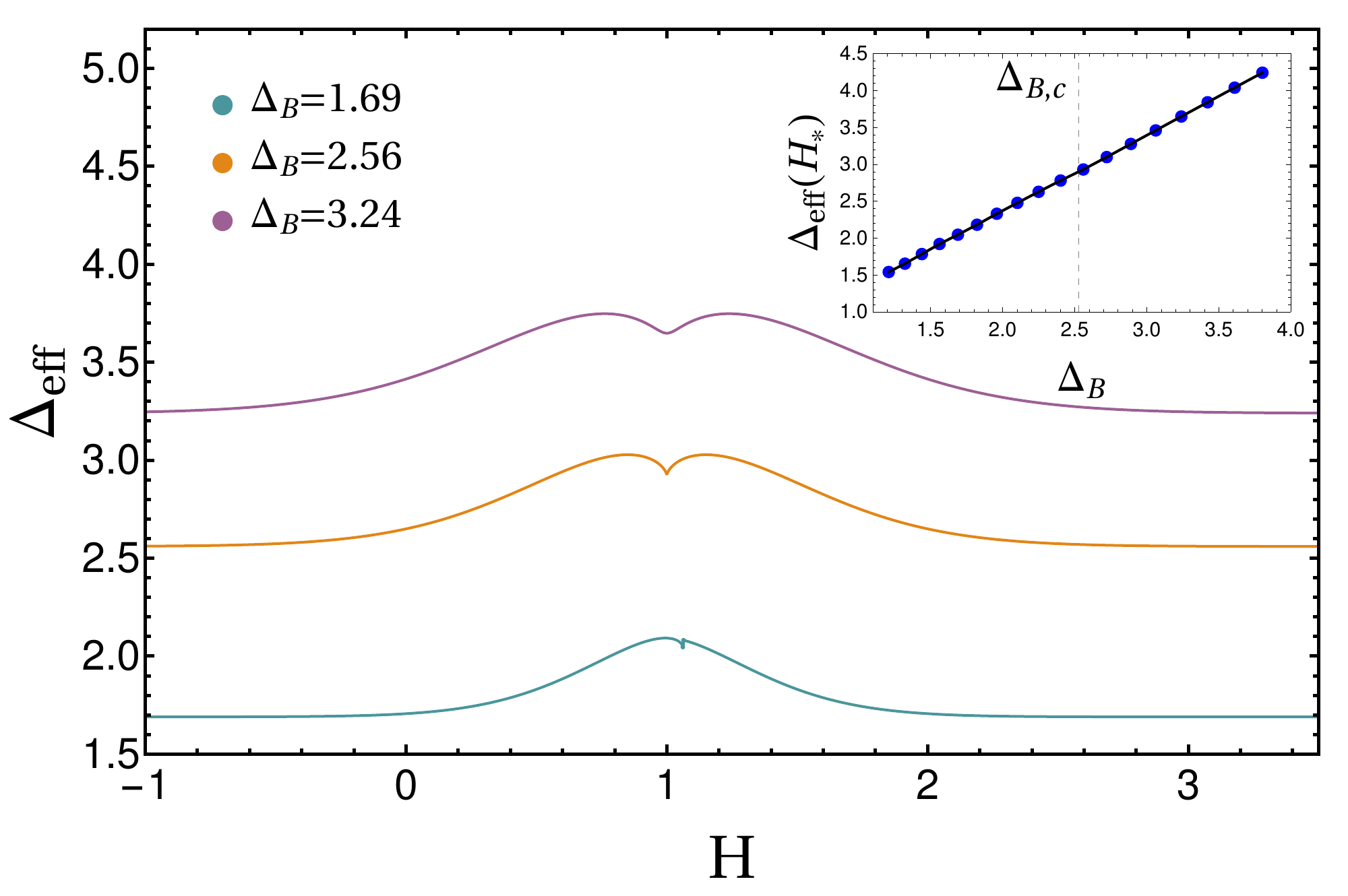}
    \caption{Variance of the renormalized random field $\Delta_{\rm eff}$ vs $H$ for the mean-field AQS RFIM for different 
    values of the bare variance $\Delta_B$. Inset: $\Delta_{\rm eff}(H_*)$ vs $\Delta_B$, where $H_*$ is the location of the transition (critical or spinodal point) or of the maximum of the susceptibilities.}
    \label{fig:eff_dis_RFIM}
\end{figure}

We plot the outcome of Eq.~(\ref{eqn:eff_RF_Ising}) 
in Fig.~\ref{fig:eff_dis_RFIM} for a range of values of $\Delta_B$. The deterministic AQS dynamics clearly renormalizes the random field, whose variance 
is equal to the bare one only in the limits where $H\to \pm \infty$ and is otherwise larger than $\Delta_B$. However, this renormalization is quantitatively 
very limited. The relative difference between $\Delta_{\rm eff}(H)$ and $\Delta_B$ is indeed always less than $30\%$.

If one looks at the effective random field variance at the transition (critical point when $\Delta_B=\Delta_{B,c}$ and spinodal point when 
$\Delta_B<\Delta_{B,c}$) or at the maximum of the susceptibilities when $\Delta_B>\Delta_{B,c}$ (the maxima take place at the same value of $H$ for 
the two susceptibilities), one clearly sees that it increases  with $\Delta_B$: see the inset in Fig.~\ref{fig:eff_dis_RFIM}. As a result, the sequence 
continuous curve/critical point/discontinuous curve for $m(H)$ corresponds to decreasing $\Delta_{\rm eff}$ as well as decreasing $\Delta_B$ (everything 
else being held constant).

\section{Susceptibilities in the mean-field EPM}
\label{sec_suscept_EPM}

The connected susceptibility, which is equal to the derivative of the averaged local stability (average distance to the local yield stress threshold) 
$m(\gamma)=\overline{\left< m^{\alpha,[\hat{x}]_M}(\gamma)\right>_{[\hat{x}]_M}}$ 
with respect to the applied strain $\gamma$, is easily obtained from the results of Sec.~\ref{sec:MF-EPMaverage} and Ref.~[\onlinecite{ozawaPNAS}]. It 
reads
\begin{equation}
\label{eq:conn_MF-EPM}
\chi_{{\rm con}}(\gamma)=m'(y(\gamma))y'(\gamma)=\frac{2\mu_2[-1+<\hat x> P_{y(\gamma)}(0)]}{[1-x_c P_{y(\gamma)}(0)]}
\end{equation}
where $y(\gamma)$ is obtained by inverting Eq.~(\ref{eqn:gamma_fullaverage}) and  we recall that a prime denotes a derivative with respect to the 
argument of the function. Note that we have defined the susceptibility such that it is negative when the derivative of the average stress-strain curve 
$\sigma'(\gamma)$ is positive (because $m=(1/N)\sum_i x_i=1-(1/N)\sum_i \sigma_i$).

We now need to  compute the disconnected susceptibility, which is associated with the sample-to-sample fluctuations,
\begin{equation}
\label{eqn:disc_def}
    \chi_{\text{dis}}(\gamma) = N \overline{\left< \bigg[ m^{\alpha,[\hat{x}]_M}(\gamma) - m(\gamma) \bigg]^2\right>}_{[\hat{x}]_M},
\end{equation}
and is the central focus of the present work.

As already discussed, it is more convenient to first consider the quantity $m^{\alpha,[\hat{x}]_M}$ as a function of the control parameter $y$. However, we are 
interested in the fluctuations at fixed applied strain $\gamma$, and we therefore need to properly switch from the situation at fixed $y$ to that at fixed 
$\gamma$. In the present mean-field model, we can introduce fluctuations as
\begin{equation}
   \label{eqs_flutuations}
    \begin{aligned}
    m^{\alpha,[\hat{x}]_M}(\gamma) &= m(\gamma) + \frac{1}{\sqrt{N}}\delta\widehat m^{\alpha,[\hat{x}]_M}(\gamma) \\
     \widetilde{m}^{\alpha,[\hat{x}]_M}(y) &= \widetilde{m}(y) + \frac{1}{\sqrt{N}} \delta\widehat{\widetilde{m}}^{\alpha,[\hat{x}]_M}(y) \\
      y^{\alpha,[\hat{x}]_M}(\gamma) &= y(\gamma) + \frac{1}{\sqrt{N}} \delta\widehat y^{\alpha,[\hat{x}]_M}(\gamma) \\
    \gamma^{\alpha,[\hat{x}]_M}(y) &= \gamma(y) + \frac{1}{\sqrt{N}} \delta\widehat\gamma^{\alpha,[\hat{x}]_M}(y),
    \end{aligned}
\end{equation}
where we have momentarily added a tilde on the volume-averaged local stability $m$ evaluated at fixed $y$.

From the identity $m^\alpha(\gamma)=\widetilde m^\alpha(y^\alpha(\gamma))$, where to alleviate the notation we have 
subsumed all of the disorder characterization $\alpha,[\hat{x}]_M$ in the single subscript $\alpha$, we obtain the following relation:
\begin{equation}
\begin{aligned}
\label{eqn:fluctuation_m_MF-EPM}
\delta\widehat m^{\alpha}(\gamma) &=  \sqrt N \big [ \widetilde{m}(y(\gamma) +\frac{1}{\sqrt{N}} \delta \widehat y^{\alpha}(\gamma))-\widetilde m(y(\gamma))\big ]+
\delta\widehat{\widetilde{m}}^{\alpha} (y(\gamma) +\frac{1}{\sqrt{N}} \delta \widehat y^{\alpha}(\gamma)) \\&
=  \widetilde{m}'(y(\gamma)) \delta \widehat y^{\alpha}(\gamma) +\delta\widehat{\widetilde{m}}^{\alpha} (y(\gamma))+ \epsilon(1/N),
\end{aligned}
\end{equation}
where $\epsilon(1/N)\to 0$ when $N\to \infty$.

By applying the same procedure to the identities $y = y^{\alpha}(\gamma^{\alpha}(y))$ and $\gamma=\gamma^\alpha(y^\alpha(\gamma))$, we also find 
relations between the fluctuations of the two control parameters,
\begin{equation}
    \begin{aligned}
        \label{eqn:fluct_MF-EPM}
    \gamma'(y(\gamma)) \delta\widehat y^{\alpha}(\gamma) &= -  \delta\widehat \gamma^{\alpha}(y(\gamma)) \\
    y'(\gamma(y)) \delta \widehat\gamma^\alpha(y) &= - \delta\widehat y^{\alpha}(\gamma(y)),
    \end{aligned}
\end{equation}
in the limit where $N\to\infty$.

We next consider the evolution with $y$ of the volume-averaged local stability $\widetilde{m}^{\alpha,[\hat{x}]_M}(y)$, where we reinstall the full explicit 
dependence on the disorder (sample and history of random jumps). The infinitesimal variation between $y$ and $y+dy$ is given by
\begin{equation}
d\widetilde{m}^{\alpha,[\hat{x}]_M}(y)=\frac 1N\sum_{i=1}^N\Big [-dy\,[1-\theta(dy-x_i^{\alpha,[\hat{x}]_M}(y))]+
 [\hat x_i^{M+1}-x_i^{\alpha,[\hat{x}]_M}(y)]\,\theta(dy-x_i^{\alpha,[\hat{x}]_M}(y)) \Big ],
\end{equation}
where the first term in the parentheses is the elastic contribution which decreases the local stabilities of the sites that do not yield by $-dy$ and the second is the plastic contribution that gives a jump $\hat x_i^{M+1}$ to each site $i$ that becomes unstable. At order $dy$, the above equation gives
\begin{equation}
\begin{aligned}
d\widetilde{m}^{\alpha,[\hat{x}]_M}(y)&=-dy + \frac 1N\sum_{i=1}^N\hat x_i^{M+1}\,\theta(dy-x_i^{\alpha,[\hat{x}]_M}(y)) \\&
= \frac{\mu_1}{\mu_2} dy -2 (\mu_1 + \mu_2) d\gamma^{\alpha,[\hat{x}]_{M+1}}(y) ,
\end{aligned}
\end{equation}
where the second line is obtained with the help of Eq.~(\ref{eqn:dgamma_disorder}). After integrating from $y=0$, this leads to
\begin{equation}
\label{eq_m_alpha_y}
     \widetilde{m}^{\alpha,[\hat{x}]_{M}}(y) =m^{\alpha} (0) + \frac{\mu_1}{\mu_2} y -2 (\mu_1 + \mu_2) \gamma^{\alpha,[\hat{x}]_{M}}(y).
\end{equation}
where the initial condition, $m^\alpha(0)=(1/N)\sum_i x_i^\alpha(0)$, only depends on the sample (and since $\sigma(0)=0$, $m(0)=1$). By combining 
Eqs.~(\ref{eqs_flutuations}), (\ref{eqn:fluctuation_m_MF-EPM}) and (\ref{eq_m_alpha_y}), one then obtains
\begin{equation}
\begin{aligned}
\delta\widehat m^{\alpha,[\hat{x}]_{M}}(\gamma)= \delta \widehat m^{\alpha}(0)  + \frac{\mu_1}{\mu_2} \delta\widehat y^{\alpha,[\hat{x}]_{M}}(\gamma),
\end{aligned}
\end{equation}
so that the disconnected susceptibility can be cast as the sum of three terms,
\begin{equation}
\label{eq_disc_suscept_3terms}
\begin{aligned}
\chi_{\text{dis}}(\gamma) &=  \overline{< \delta\widehat m^{\alpha,[\hat{x}]_{M}}(\gamma)^2>}_{[\hat{x}]_M} \\&
=\Delta_0 +  \left(\frac{\mu_1}{\mu_2} \right)^2 \overline{\left< \delta\widehat y^{\alpha,[\hat{x}]_{M}}(\gamma)^2\right>}_{[\hat{x}]_M} + 
2\left (\frac{\mu_1}{\mu_2}\right ) \overline{\left< \delta\widehat y^{\alpha,[\hat{x}]_{M}}(\gamma) \delta\widehat m^{\alpha}(0) \right>}_{[\hat{x}]_M},
\end{aligned}
\end{equation}
where
\begin{equation}
\Delta_0=\overline{\delta\widehat m^{\alpha}(0)^2}=\overline{[x_i^\alpha(0)-1]^2}=\int_0^\infty dx (x-1)^2P_0(x)=\overline{[\sigma_i^\alpha(0)]^2}
\end{equation}
is the variance of the initial sample-to-sample distribution of the local stability (and of the local stress as well) and we have used the fact that initially 
there are no correlations from site to site.

The calculations of the two nontrivial terms in Eq.~(\ref{eq_disc_suscept_3terms}) are rather unwieldy and are detailed in 
Appendix~\ref{app:solution_MF-EPM}. The final expressions read
\begin{equation}
    \label{eq_final_gamma_diff}
    \begin{aligned}
    & \left(\frac{\mu_1}{\mu_2} \right)^2\overline{\left< \delta\widehat y^{\alpha,[\hat{x}]_{M}}(\gamma)^2\right>}_{[\hat{x}]_M} =   
    \left(\frac{\mu_1}{2\mu_2(\mu_1+\mu_2)\gamma'(y(\gamma))} \right)^2 \Big ( <\hat{x}^2> \int_{0}^{y(\gamma)}d y' P_{y'}(0) 
    -<\hat x>^2\left(\int_{0}^{y(\gamma)} dy' P_{y'}(0) \right)^2 +\\& 
  2 <\hat x>\int_{0}^{y(\gamma)} dy' P_{y'}(0) \int_{y'}^{y(\gamma)} dy''\Big[g(0)[ <\hat x> - T(y''-y')] - T'(y''-y') +
    \int_0^{y''-y'} d\widehat{y}[<\hat x> - T(\widehat{y})]R_{y''-y'-\widehat{y}}(0) \Big ] \Big )
    \end{aligned}
\end{equation}
where $T(x)=\int_x^\infty d\hat x \hat x g(\hat x)$, and
\begin{equation}
\begin{aligned}
\label{eqn_mixed_EPM}
        &2 \left (\frac{\mu_1}{\mu_2}\right ) \overline{\left< \delta\widehat y^{\alpha,[\hat{x}]_{M}}(\gamma)\delta\widehat m^{\alpha}(0) \right>}_{[\hat{x}]_M} =\\&
        \frac{\mu_1<\hat x>}{\mu_2(\mu_1+\mu_2)\gamma'(y(\gamma))} \int_{0}^{y(\gamma)}dy'\Big[ (y'-1)P_0(y') + g(0)\int_0^{y'} dy'' (y''-1) P_0(y'') 
       + \int_0^{y'} dy''\, R_{y'-y''}(0) \int_0^{y''}d\widehat y (\widehat y-1) P_0(\widehat y)\Big],
\end{aligned}
\end{equation}
where $P_y(x)$ and $R_y(x)$ are solutions of Eqs.~(\ref{eqn:generalP},\ref{eqn:R_y}).

By putting all of the above equations together we obtain the expression for the disconnected susceptibility that we were looking for.

One can check that for $\gamma = 0$, the terms in Eqs.~(\ref{eq_final_gamma_diff}) and (\ref{eqn_mixed_EPM}) go to $0$ [as $y(\gamma=0)=0$], so that, 
as expected, the disconnected susceptibility at the beginning of the deformation reduces to the fluctuations associated with the initial distribution of local 
stresses, $\chi_{\rm dis}(0) = \Delta_0$. It is also instructive to study the limiting form of the disconnected susceptibility when 
$\gamma, y\to \infty$, which corresponds to the stationary state. Then, from Eqs.~(\ref{eqn:generalP},\ref{eqn:R_y}), one finds that $R_\infty(x)=0$ 
and $P_\infty(x)=1/<\hat{x}>\int_x^{\infty}g(x')dx'$. After some lengthy algebra, the disconnected susceptibility in the stationary state can be simply expressed as
\begin{equation}
\label{eq_disc_suscept_SS}
\begin{aligned}
\chi_{\text{dis}}(\gamma\to \infty) &= < x^2>_{\infty}-< x>_{\infty}^2\\
&= \frac{<\hat x^3>}{3<\hat x>} -\frac{<\hat x^2>^2}{4<\hat x>^2},
\end{aligned}
\end{equation}
where we have defined $<x^n>_{\infty}=\int_0^{\infty}dxP_{\infty}(x)x^n$. As expected, this expression does not depend on the initial condition. 

Both the connected and the disconnected susceptibilities  diverge at the yielding transition, corresponding either to a mean-field spinodal point or a critical point (see below). For a range of continuous, ductile, behavior the susceptibilities go through a local maximum at a finite, nonzero value of the strain $\gamma$. One can easily check that the locations of the maxima are not exactly the same for the two susceptibilities but asymptotically converge to the same value when the connected susceptibility becomes very large.
\\

\section{Results for the mean-field EPM}
\label{sec_results_EPM}

\subsection{Stress-strain curves and the role of the various types of disorder}
\label{subsec_stress-strain}

As we emphasized when introducing them, EPMs account for the disorder associated with the structure of the amorphous solid at the beginning of the  
deformation process through a distribution of initial local stresses (or stabilities), which we call $P_0(x)$, a distribution of local thresholds, which we take 
here as a delta-function $\delta(\sigma^{th}-1)$, and a distribution of stress jumps after local yielding, which we have introduced here as $g(\hat x)$. (More 
complicated schemes could be envisaged where the distributions of local thresholds and jumps evolve after the sites yield.) What is then the role of 
these various types of disorder and how do they conspire to generate an emergent random field? 

We first address the role of the various types of disorder (mainly, random initial stresses and random stress jumps) on the averaged stress-strain curves 
for the mean-field EPM that we have analyzed in the previous sections.  We consider three different distributions of the local random jumps,
\begin{equation}
\begin{aligned}
\label{eq_several_gRJ}
&g(\hat x)=\frac{e^{-\frac x{<\hat x>}}}{<\hat x>},\\&
g(\hat x)=\frac{e^{-\frac x t}-e^{-\frac x{<\hat x>-t}}}{2t- <\hat x>},\; {\rm with}\; t \in ]\frac {<\hat x>}2,<\hat x>[,\\&
g(\hat x)=\delta(x-<\hat x>),
\end{aligned}
\end{equation}
all with the same average value $<\hat x>$. In addition we consider two different distributions of the initial local stresses or stabilities,
\begin{equation}
\begin{aligned}
\label{eq_distrib_initial}
&P_0(x)=\frac{e^{-\frac xA}-e^{-\frac x{1-A}}}{2A-1},\; {\rm with}\; A \in ]\frac 12,1[,\\&
P_0(x)=\frac{1}{\sqrt{2 \pi \Delta_0}}e^{-\frac{(x-1)^2}{2\Delta_0}}.
\end{aligned}
\end{equation}
They both satisfy $<x>_0=1$, implying $<\sigma>_0=0$, as well as the requirement of plastic stability,  $P_0(0)=0$. (In the case of the Gaussian the latter requirement is only approximately satisfied but, since we use small values of $\Delta_0$, $P_0(0)$ is then negligible; we also tried a truncated Gaussian which exactly enforces the condition in $x=0$ but we found virtually no difference with the results obtained with the full Gaussian.) The initial variance is given by  
$<x^2>_0-<x>_0^2=\Delta_0$. For the combination of two exponentials, $\Delta_0=1-2A(1-A)$ with $1/2<\Delta_0<1$. 

For all of the above cases deriving the analytical formulas for $\sigma(\gamma)$ is quite involved and the details are given in Appendix~\ref{app:diff_mod}. In 
Fig.~\ref{fig_compare_stress-strain} we display the averaged stress-strain curve for several combinations of the above cases and for several values of 
the initial variance $\Delta_0$.

\begin{figure}
\centering
\begin{tabular}{cc}
 \includegraphics[width=.48\textwidth]{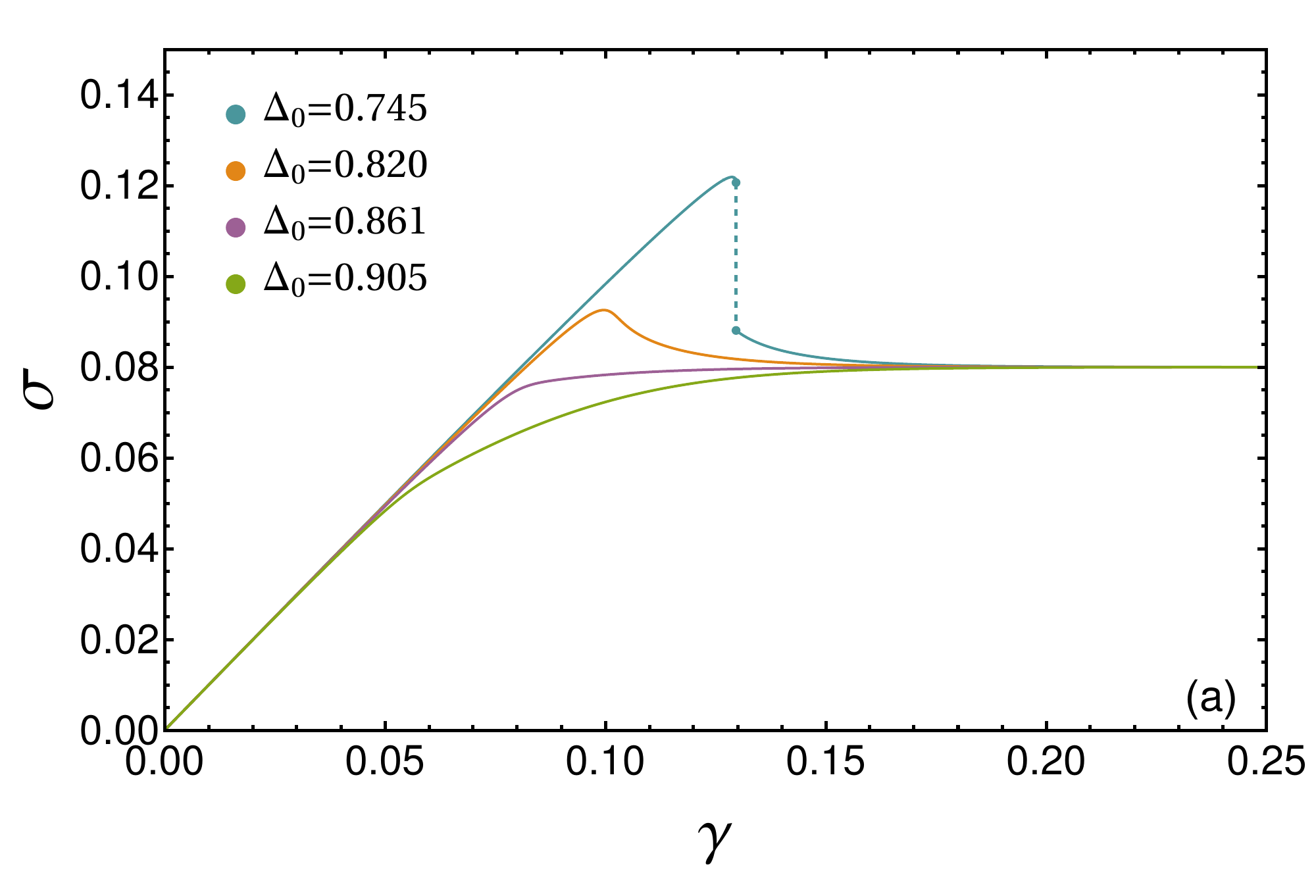}  &   \includegraphics[width=.48\textwidth]{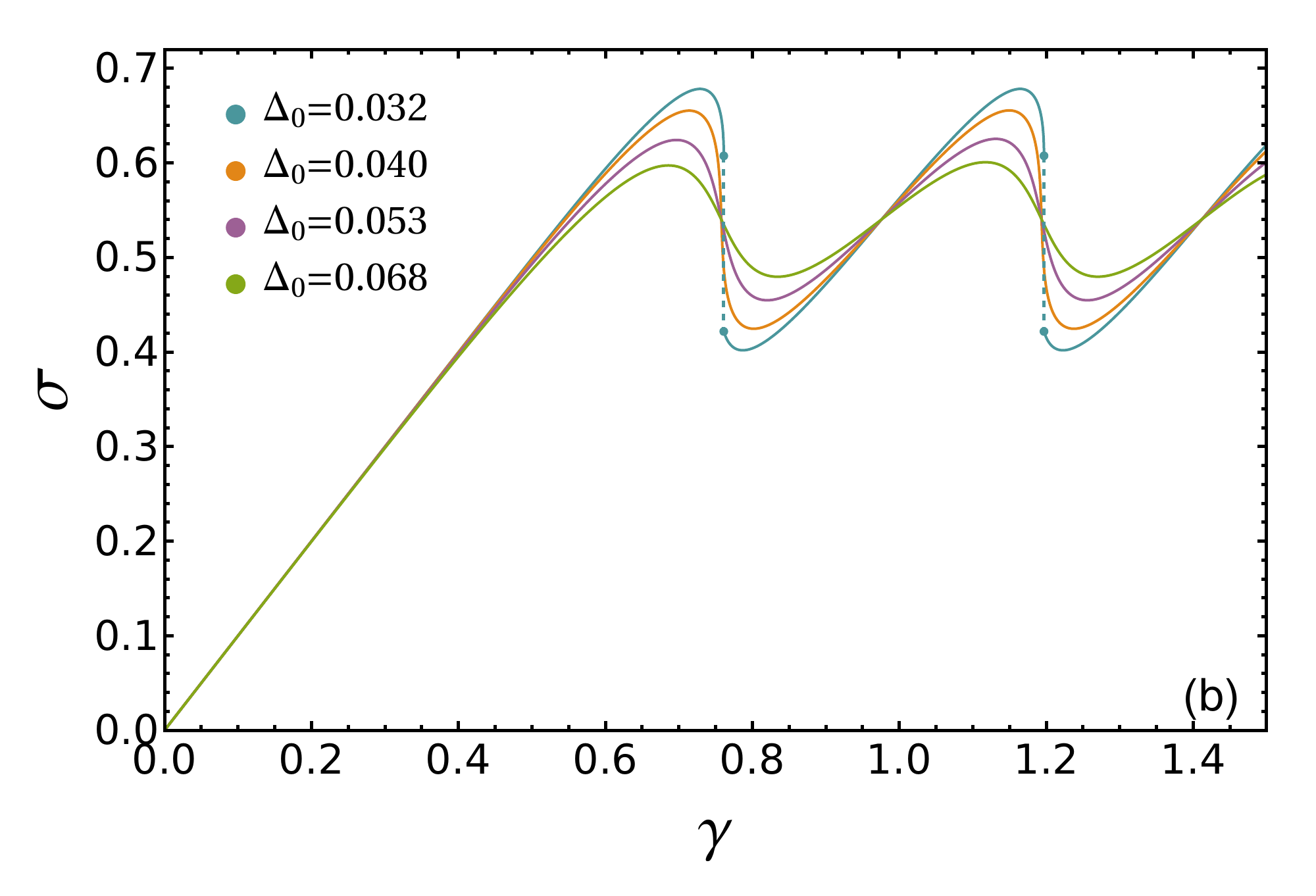}
 \\
   \includegraphics[width=.48\textwidth]{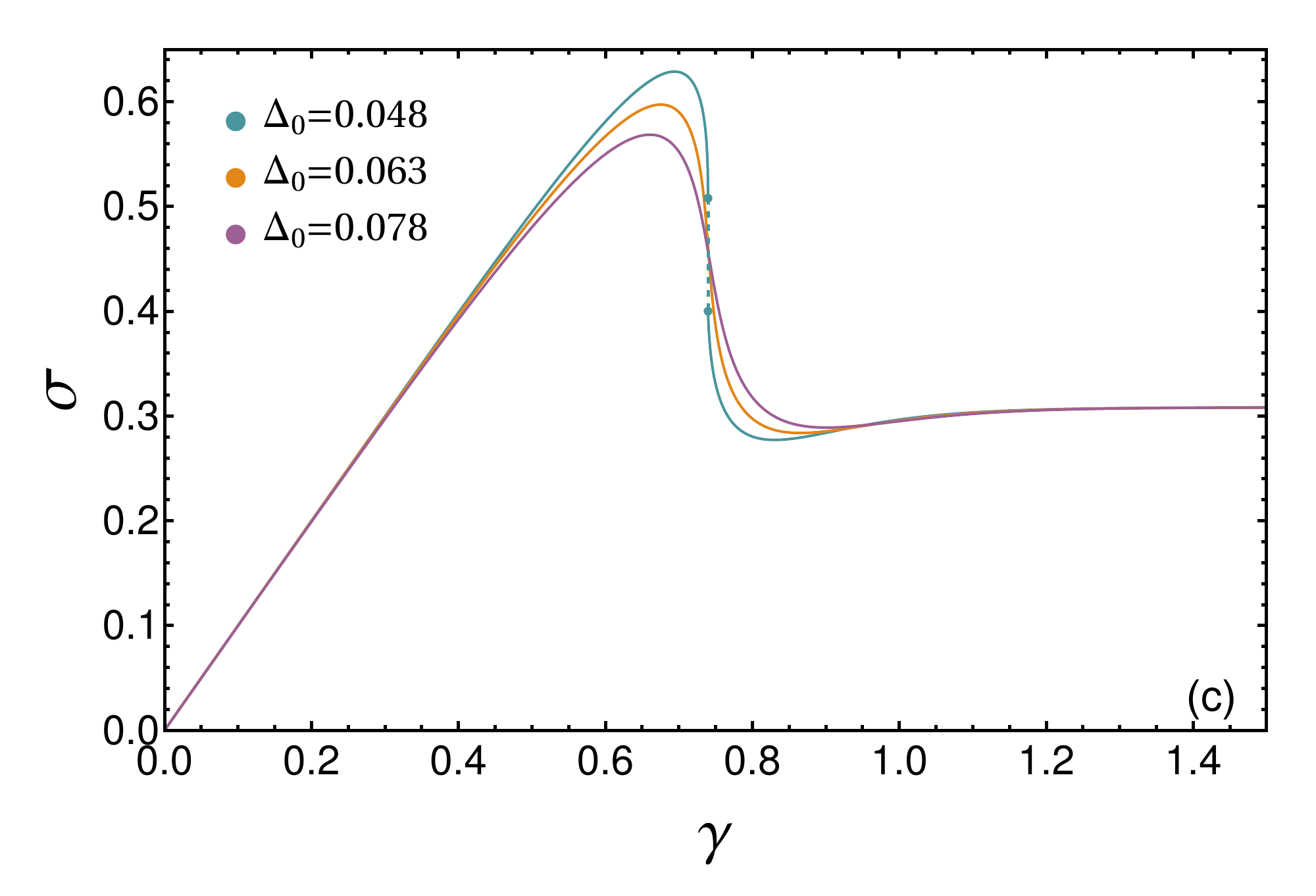}  & \includegraphics[width=.48\textwidth]{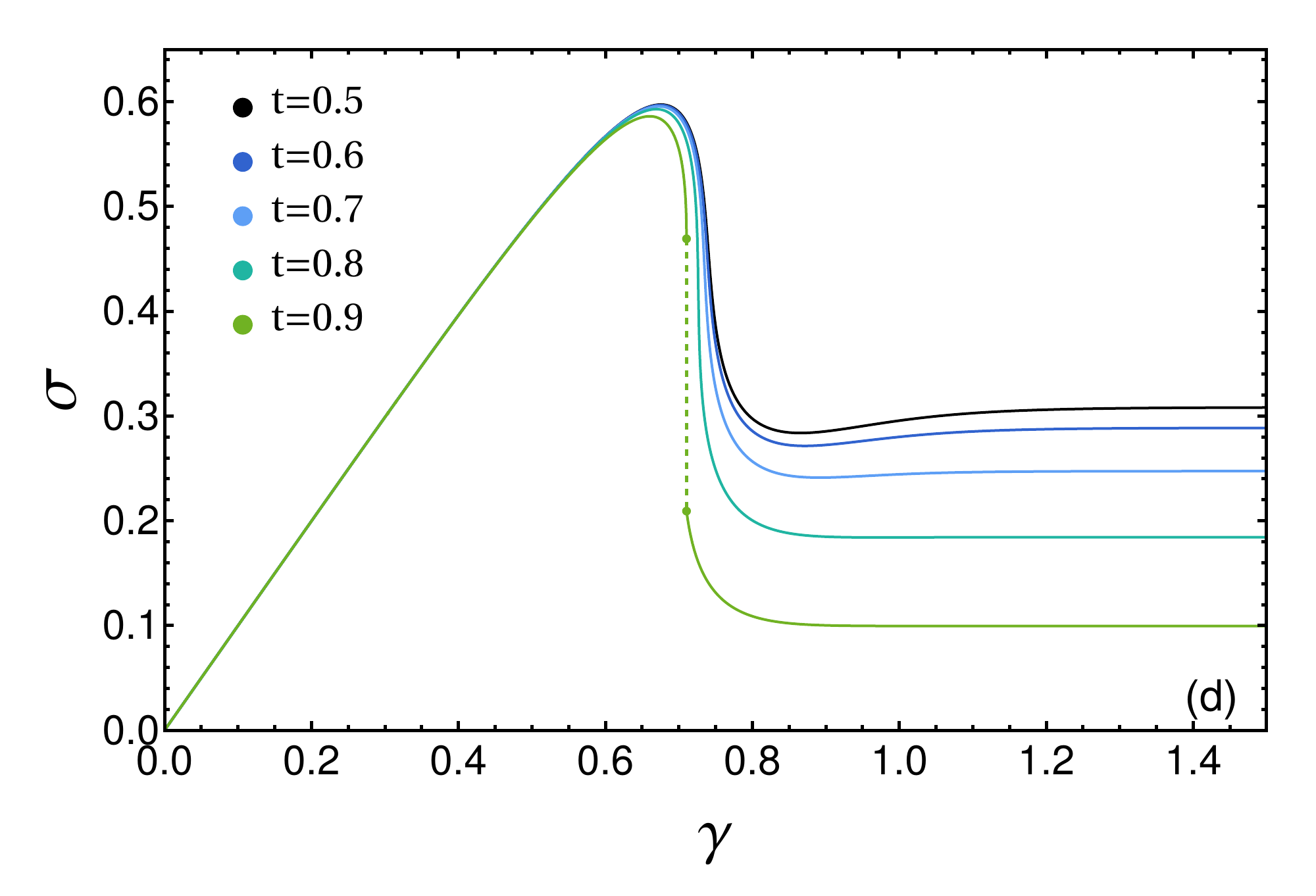}
  \\  
\end{tabular}
\caption{
Average stress $\sigma$ versus strain $\gamma$ for the MF-EPM with different distributions of the local 
random jumps $g(\hat x)$ [see Eq.~(\ref{eq_several_gRJ})] and  different distributions of the initial local stresses $P_0(x)$ [see Eq.~(\ref{eq_distrib_initial})]. 
(a): $P_0(x)$ is a 2-exponential combination and $g(\hat x)$ is a single exponential; curves are shown for several values of the initial disorder variance $\Delta_0$. (b-d): $P_0(x)$ is a Gaussian and $g(\hat x)$ is either a delta function (no randomness) in (b), a 2-exponential combination at fixed $t=0.5$ and several values of $\Delta_0$ in (c), or a 2-exponential combination at fixed $\Delta_0=0.25$ and several values of $t$ in (d).
In all cases, $2\mu_2=1$ and $<\hat x>=0.92$. When $P_0(x)$ is the combination of 2 exponentials, $\mu_1=0.0222\mu_2$, while in the Gaussian case we chose $\mu_1=0.9\mu_2$ in order to see both ductile and brittle behavior.}
\label{fig_compare_stress-strain}
\end{figure}

It is instructive to consider three different domains of deformation: (i) small deformation, $\gamma\ll1$, (ii) stationary state, $\gamma\gg 1$, which should 
physically correspond to the ``flowing state" of the sheared material when the average stress stays constant and equal to the macroscopic yield stress, 
and (iii) the region of the overshoot and of the yielding transition, when present.

In region (i) at very small deformation, the response is purely elastic and $\sigma(\gamma)\approx 2\mu_2 \gamma$, irrespective of the disorder(s). 

Region (ii) is obviously very different when there is randomness in the local jumps and where there is none [as in panel (b)]. In the former case, a 
{\it bona fide} steady state with an essentially constant stress is reached, with
\begin{equation}
\sigma(\gamma\to \infty)=1-\frac{<\hat x^2>}{2<\hat x>}.
\end{equation}
This expression is independent of the initial distribution $P_0(x)$, as it should. For a single exponential, the asymptotic value $\sigma(\infty)$ is simply equal 
to $1-<\hat x>$, but for a combination of two exponentials, $<\hat x^2>$ is in general different from $2<\hat x>^2$ and one can change the steady-state 
stress for a given $<\hat x>$ by varying the parameter $t$, as illustrated in Fig.~\ref{fig_compare_stress-strain} (d).  On the other hand, in the absence 
of randomness in the jumps, $\sigma(\gamma)$ is a periodic function at large deformation (of period $\hat x\mu_1/[(2\mu_2)(\mu_1+\mu_2)]$ when $y\gg 1$, as shown in Appendix~\ref{app:diff_mod}) and it does not reach a physically acceptable steady state [see Fig.~\ref{fig_compare_stress-strain} (b)]. The limiting state 
is however again independent of the initial distribution $P_0(x)$, as it should.

\begin{figure}
    \centering
     \includegraphics[width=.48\textwidth]{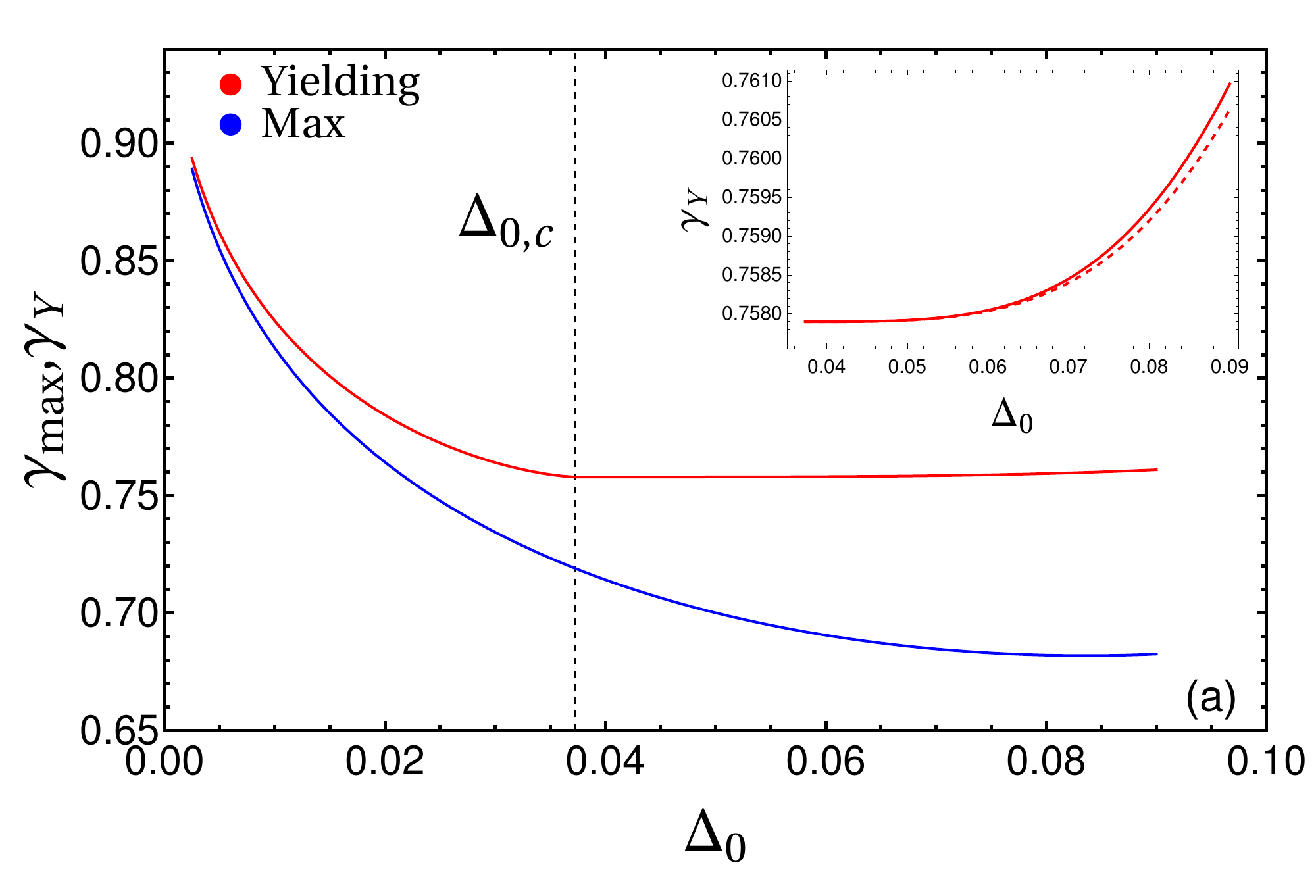}
     \includegraphics[width=.48\textwidth]{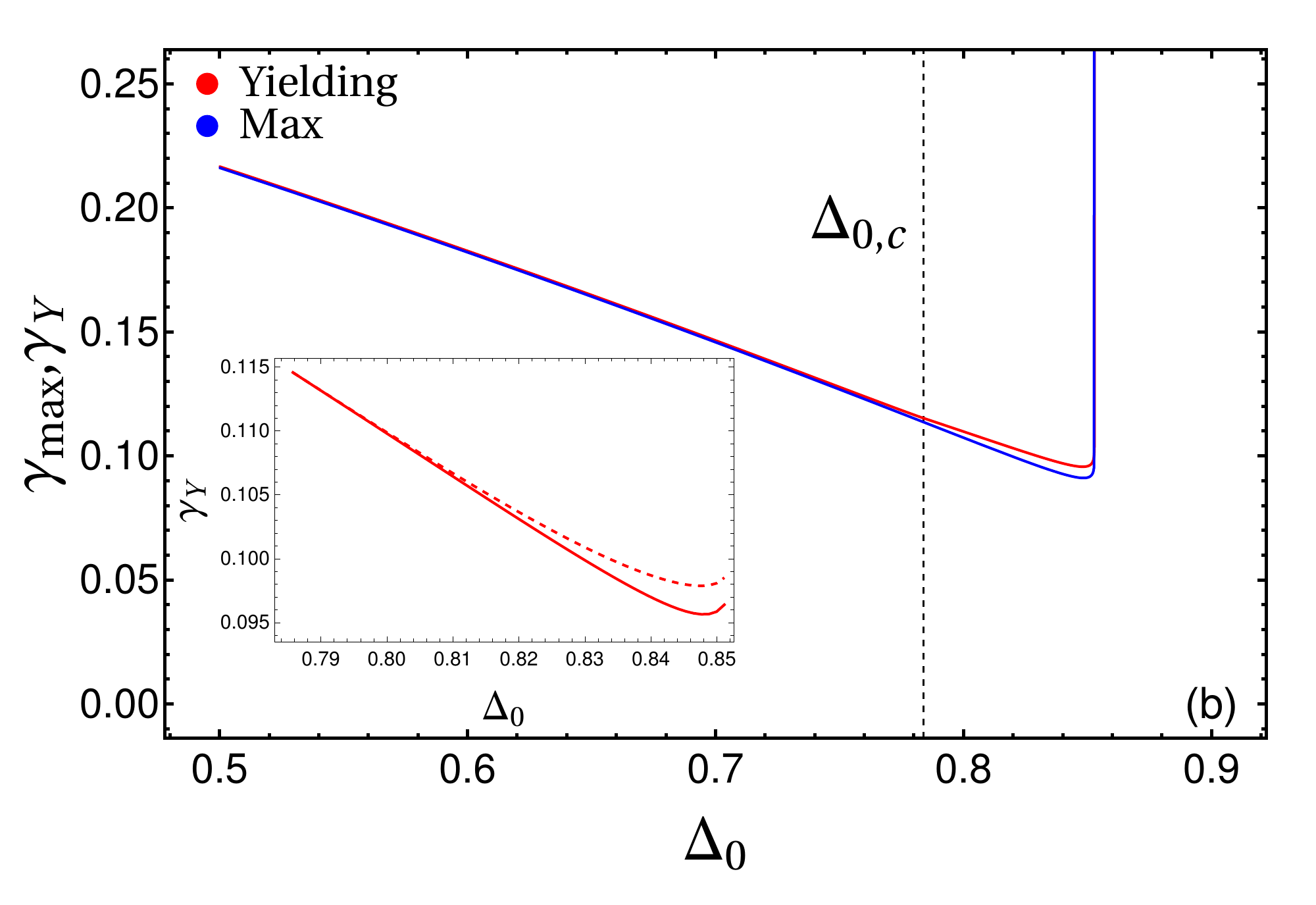} 
     \includegraphics[width=.48\textwidth]{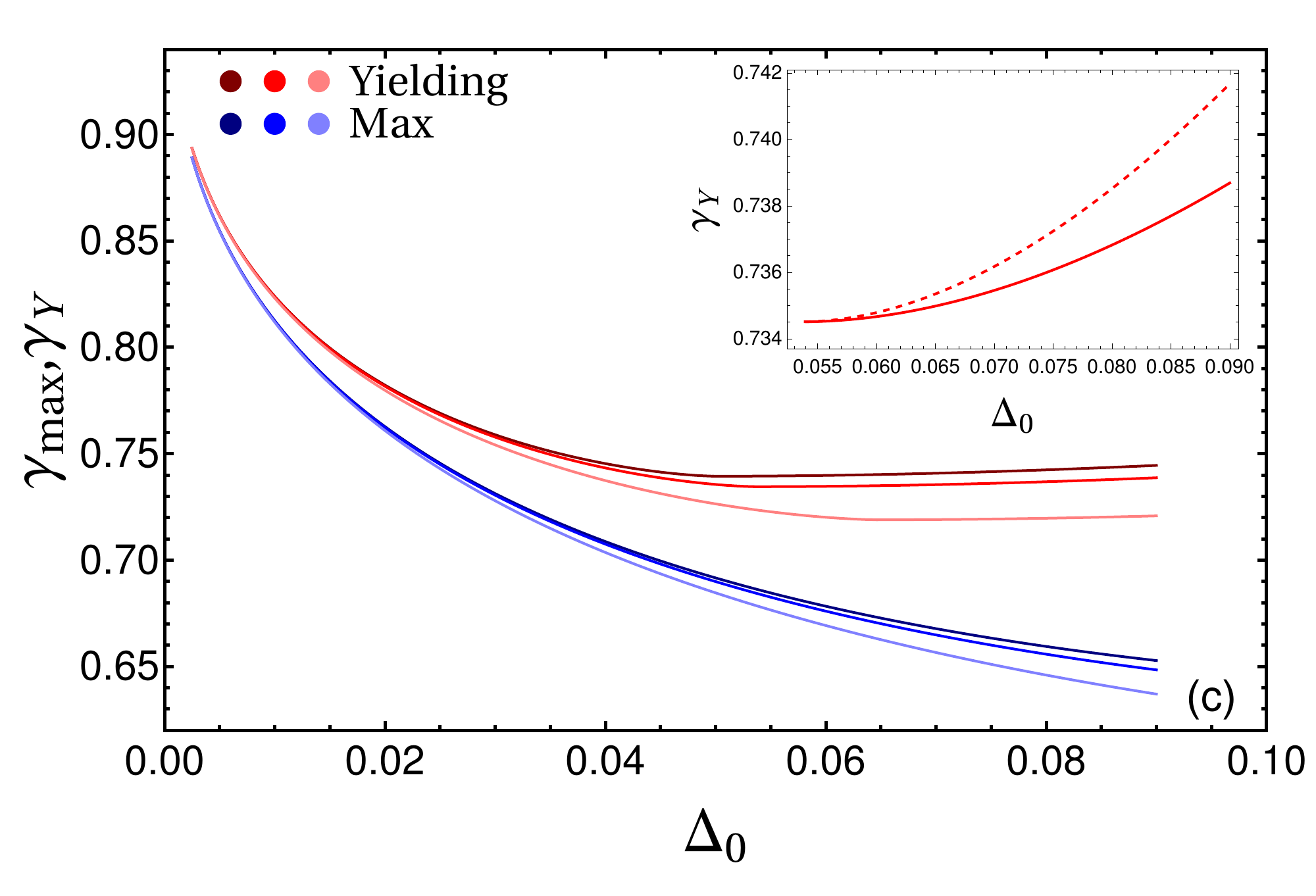}\\
\caption{(a): Location of the overshoot $\gamma_{\rm max}$ and of the yielding transition and/or the local maximum of the connected suceptibility $\gamma_Y$ versus the initial disorder variance $\Delta_0$ for the mean-field EPM. (a) Gaussian initial distribution of the initial local stresses and no random jumps. 
(b): Combination of two exponentials for the initial local stresses and single-exponential distribution of random jumps. In both cases the vertical dashed line marks the critical value of the bare disorder. 
(c): Dependence on the parameter $t$ when the distribution of random jumps is the 2-exponential combination and the initial distribution $P_0(x)$ is a Gaussian; from top to bottom: $t=0.5, 0.7,0.85$.
Insets: Comparison of the location of the maxima of the connected susceptibility $\gamma_Y$ (full line) and the disconnected susceptibility $\gamma_Y^{\rm dis}$ (dashed line). In (c), $t=0.7$. 
One can show that the divergence of $\gamma_{\rm max}$ in (b) goes as $\gamma_{\rm max} \sim \log(1/(<\hat x>-A))$. }
    \label{fig:gamma_y_gamma_max}
\end{figure}

Region (iii) is the one of most interest for our present purpose of investigating an emergent potential random field. One can see from 
Fig.~\ref{fig_compare_stress-strain} (and confirm by a direct analysis of the expressions in Appendix~\ref{app:diff_mod}) that the same sequence of 
behavior is observed for all distributions. As $\Delta_0$ decreases the behavior passes through a regime with an overshoot and a continuous evolution to finally 
reach a ``brittle" regime with a discontinuity in the stress. The brittle and the continuous regimes are separated by a critical point~\cite{ozawaPNAS}. Of 
course, the quantitative details vary from one distribution to another but the pattern is the same. We illustrate the quantitative differences in Fig.~\ref{fig:gamma_y_gamma_max} by plotting the location at which the stress is maximum, denoted by $\gamma_{{\rm max}}$, and that at which the 
slope of the stress-strain curve (i.e., the connected susceptibility) is negative and maximum,  $\gamma_Y$, for several models. The latter corresponds to either a spinodal point or the critical point when the slope is infinite. (When the slope is not infinite but a local maximum still exists in the susceptibilities, we keep the notation $\gamma_Y$ for the maximum of the connected susceptibility; the maximum of the disconnected susceptibility $\gamma_Y^{\rm dis}$ slightly deviates from $\gamma_Y$ as one moves away from the critical point in the ductile region, as seen from the insets in  Fig.~\ref{fig:gamma_y_gamma_max}.) Note that the maximum stress (overshoot) and the maximum negative slope disappear for a large initial disorder $\Delta_0$ when the stress-strain curve is monotonic. We also display in Fig.~\ref{fig:plot_Delta0,c} the variation of the critical value of the initial variance, $\Delta_{0,c}$, as a function of the parameter $t$ in the 2-exponential distribution of random jumps [see Eq.~(\ref{eq_several_gRJ})]. 

Region (iii) is where the analogy with an AQS driven RFIM may hold. It is clear however that the quantitative aspects and, as we will see below, the strength 
of the effective random field depend on the details of the disorder distributions. Before delving more into this issue, we consider the effect of the disorder 
distributions on the connected and disconnected susceptibilities computed in Sec.~\ref{sec_suscept_EPM}.

\begin{figure}
\centering
    \includegraphics[width=.48\textwidth]{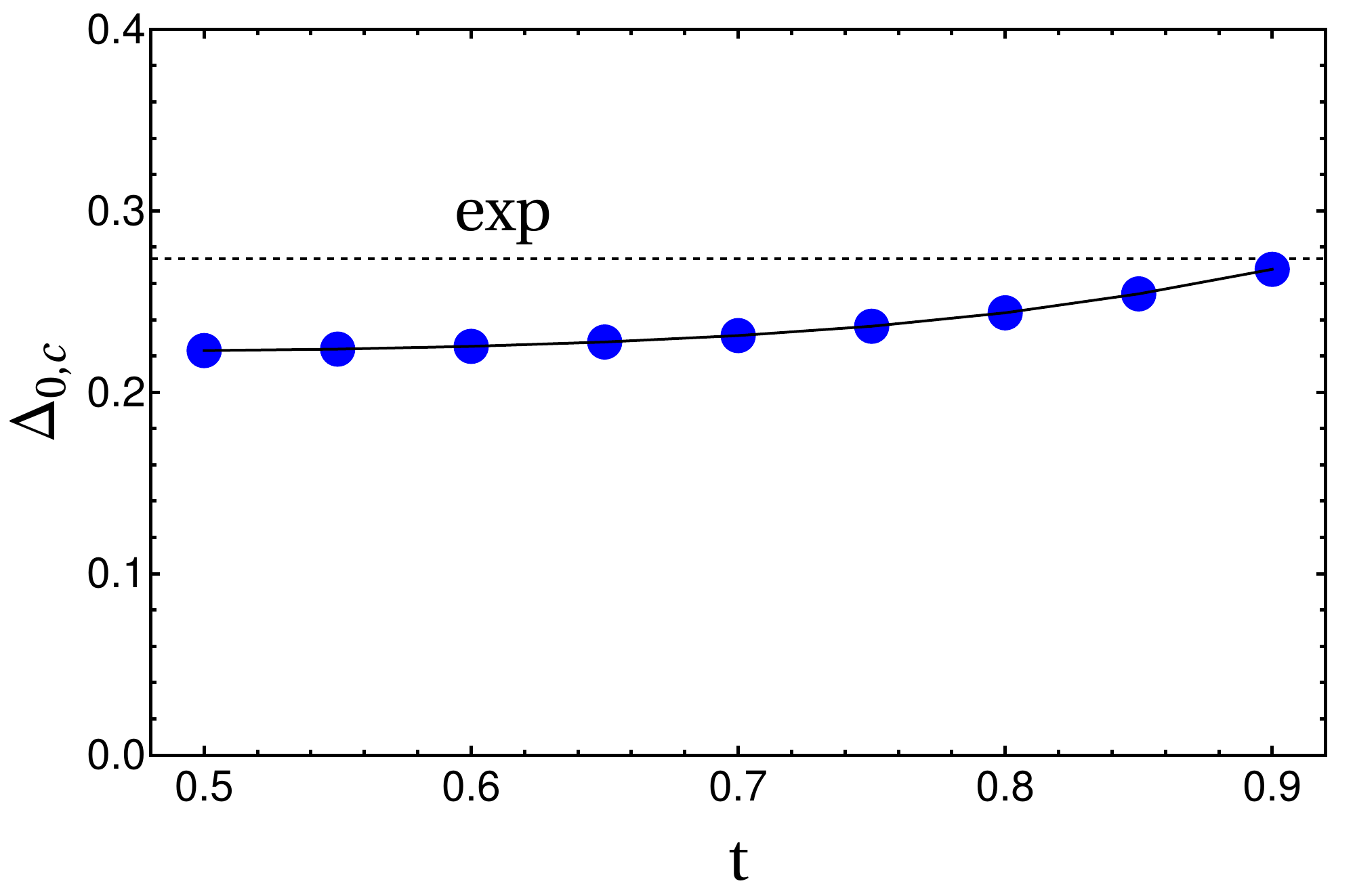}
\caption{Critical value of the initial variance of the local stresses, $\Delta_{0,c}$, as a function of the parameter $t$ that appears in the distribution of random 
jumps [see Eq.~(\ref{eq_several_gRJ})]. The initial distribution $P_0(x)$ is a Gaussian. The dashed line is the value $\Delta_{0,c}=0.273677$ obtained for the case of a single-exponential distribution of random 
jumps: it is reached when $t=<\hat x>=0.92$.}
\label{fig:plot_Delta0,c}
\end{figure}

\begin{figure}
\centering
\begin{tabular}{cc}
    \includegraphics[width=.48\textwidth]{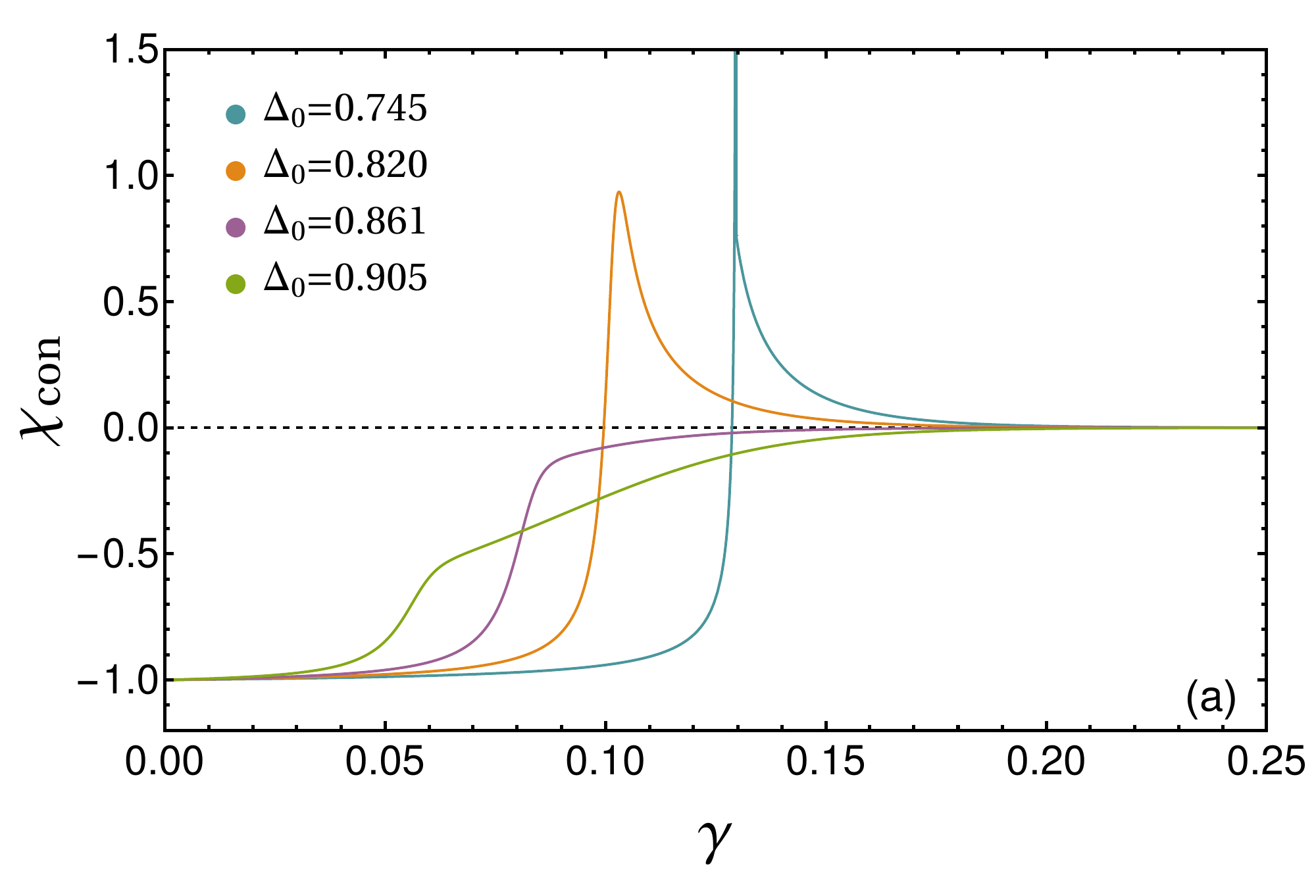}  &      \includegraphics[width=.48\textwidth]{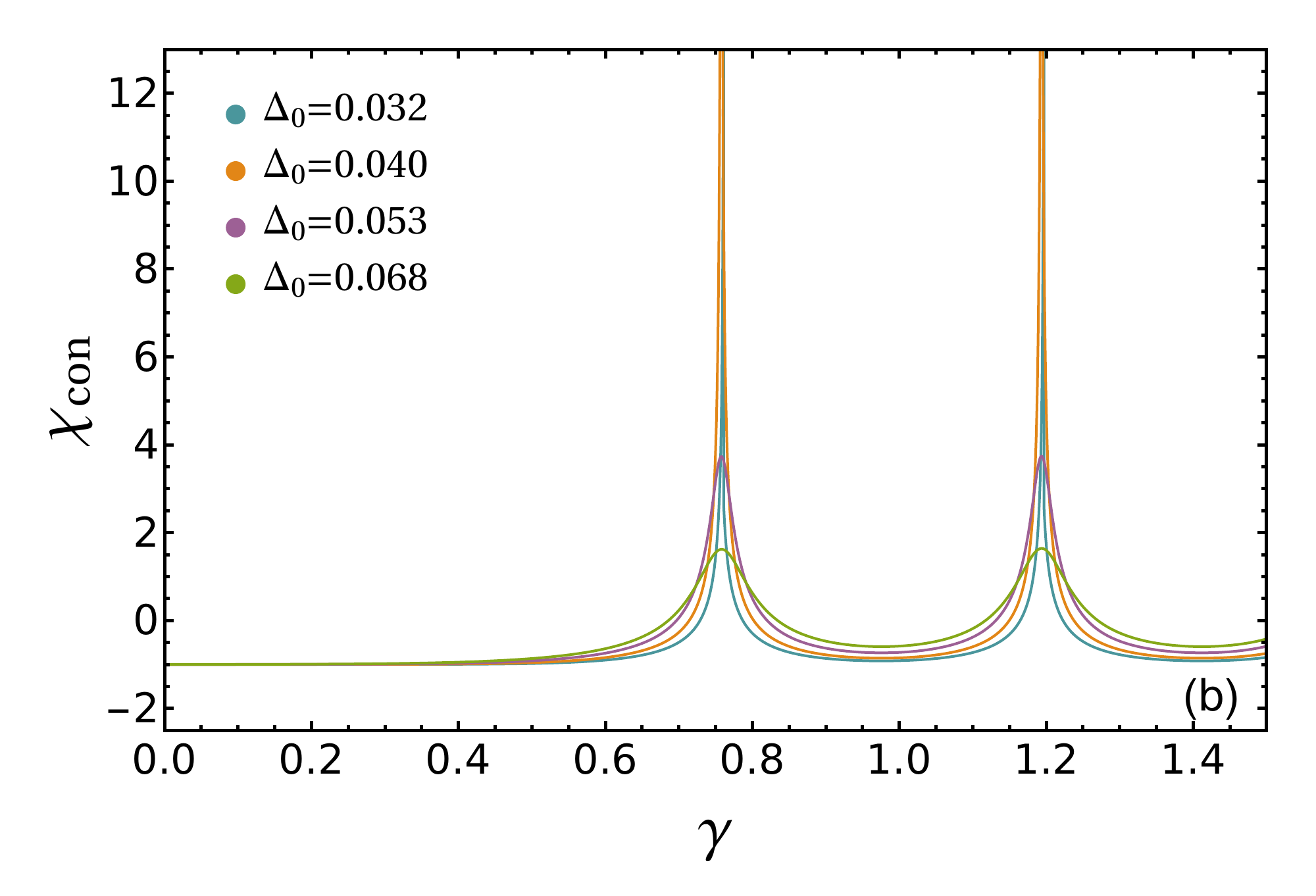}
    \\
    \includegraphics[width=.48\textwidth]{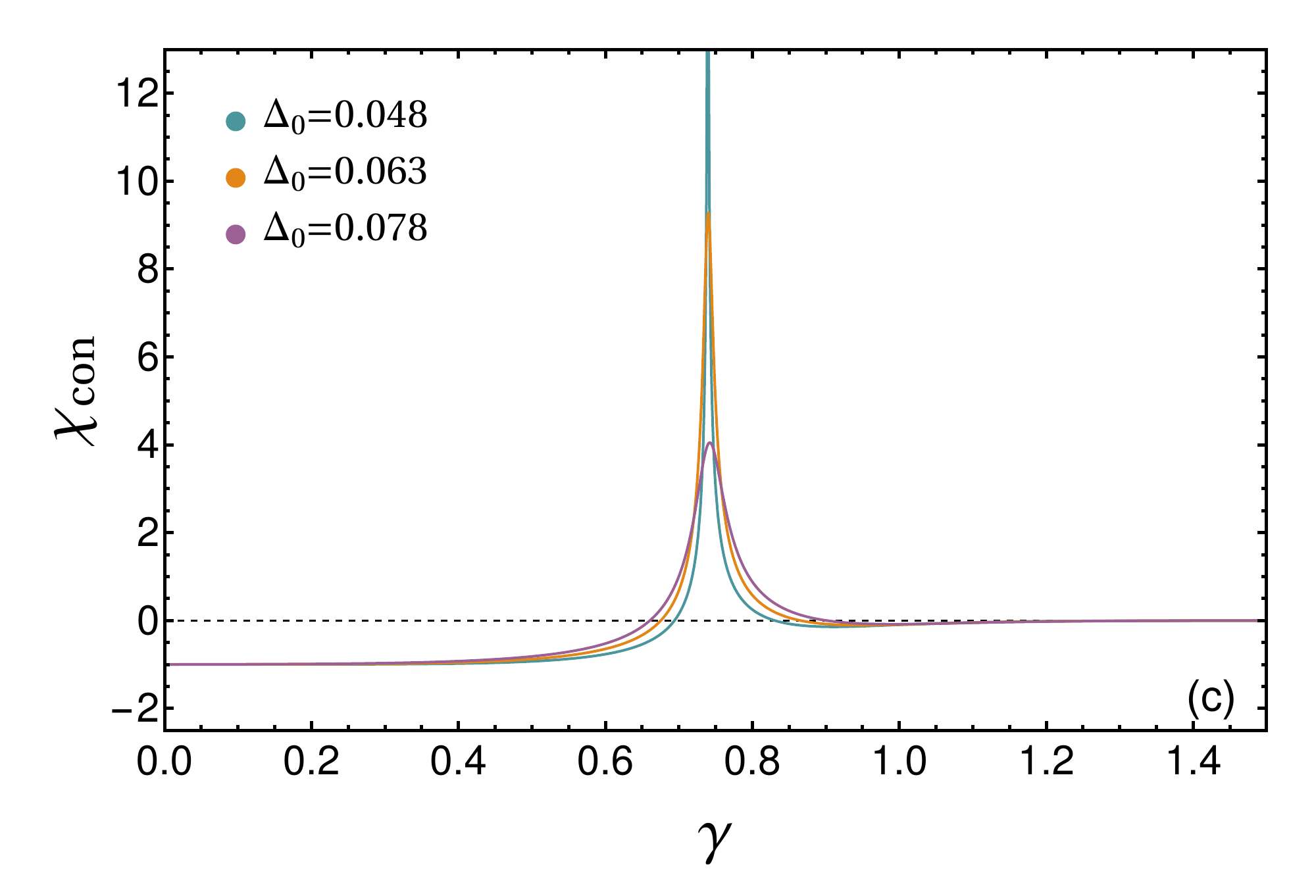}  & \includegraphics[width=.48\textwidth]{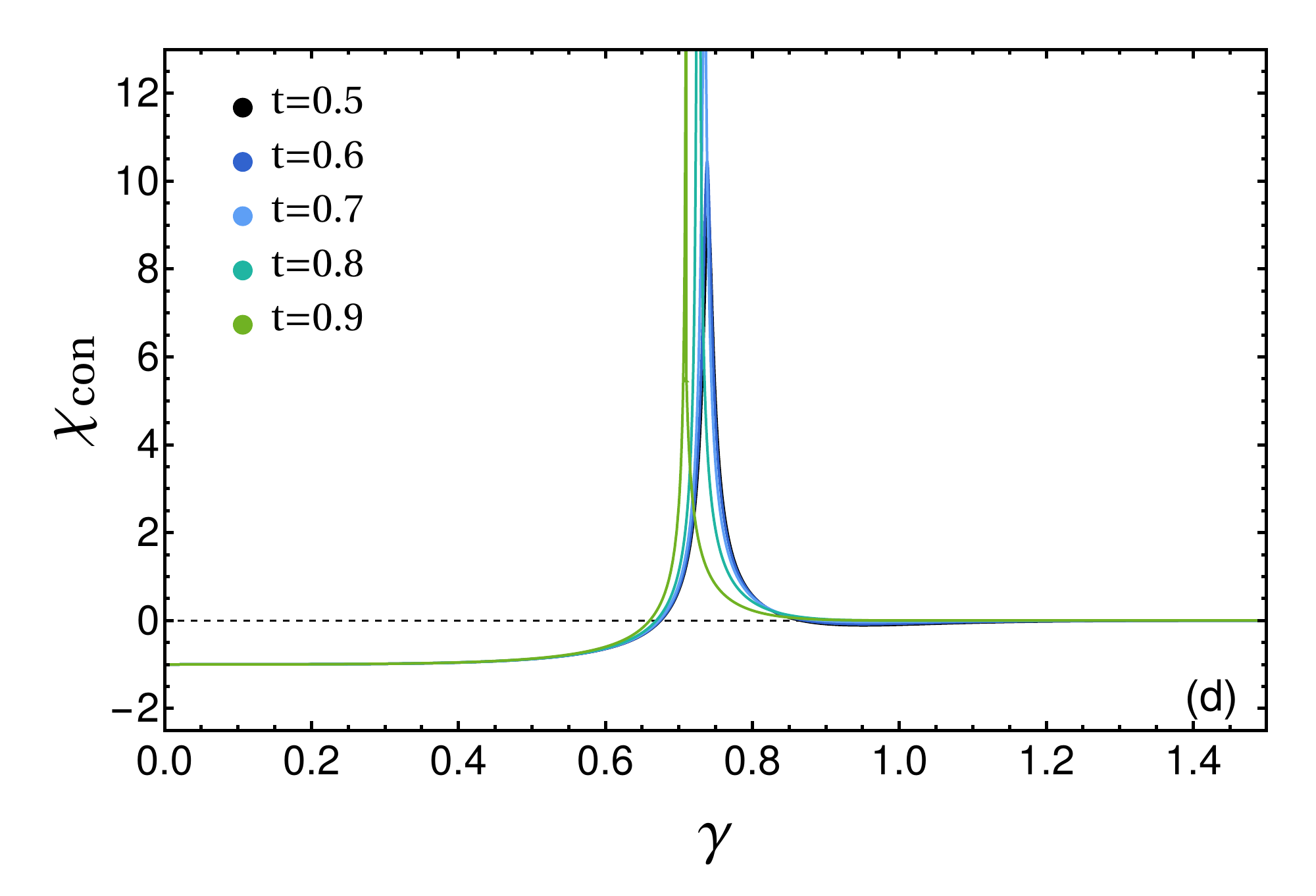}\\

\end{tabular}
\caption{
Connected susceptibility $\chi_{{\rm con}}(\gamma)$  for the MF-EPM with different distributions of the local random 
jumps $g(\hat x)$ [see Eq.~(\ref{eq_several_gRJ})] and  different distributions of the initial local stresses $P_0(x)$ [see Eq.~(\ref{eq_distrib_initial})]. 
(a): $P_0(x)$ is a 2-exponential combination and $g(\hat x)$ is a single exponential; curves are shown for several values of the initial disorder variance $\Delta_0$. (b-d): $P_0(x)$ is a Gaussian and $g(\hat x)$ is either a delta function (no randomness) in (b), a 2-exponential combination at fixed $t=0.5$ and several values of $\Delta_0$ in (c), or a 2-exponential combination at fixed $\Delta_0=0.25$ and several values of $t$ in (d).
In all cases, $2\mu_2=1$ and $<\hat x>=0.92$. When $P_0(x)$ is the combination of 2 exponentials, $\mu_1=0.0222\mu_2$, while in the Gaussian case we chose $\mu_1=0.9\mu_2$ in order to see both ductile and brittle behavior.}
\label{fig_compare_con_susc}
\end{figure}

\begin{figure}
\centering
\begin{tabular}{cc}
    \includegraphics[width=.48\textwidth]{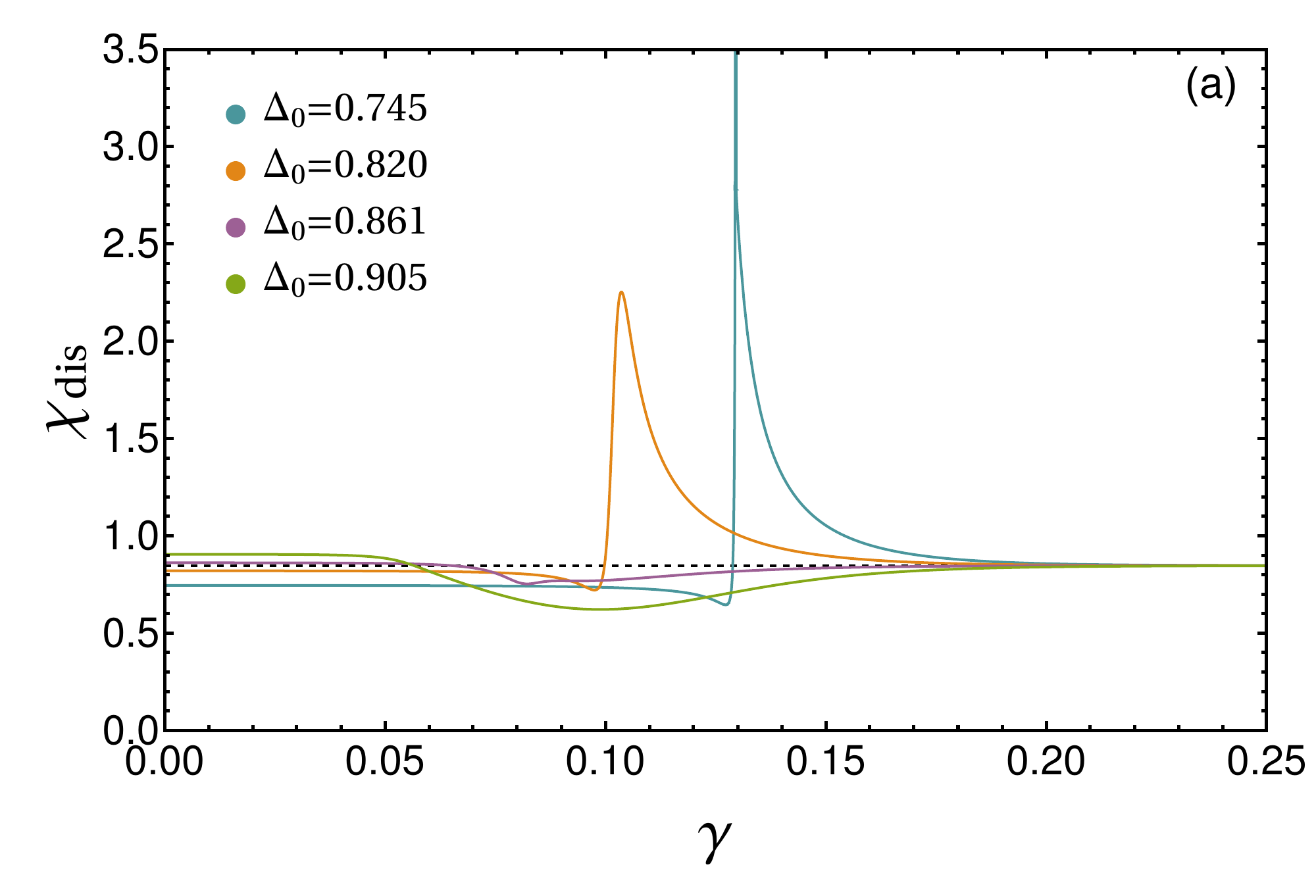}  &  \includegraphics[width=.48\textwidth]{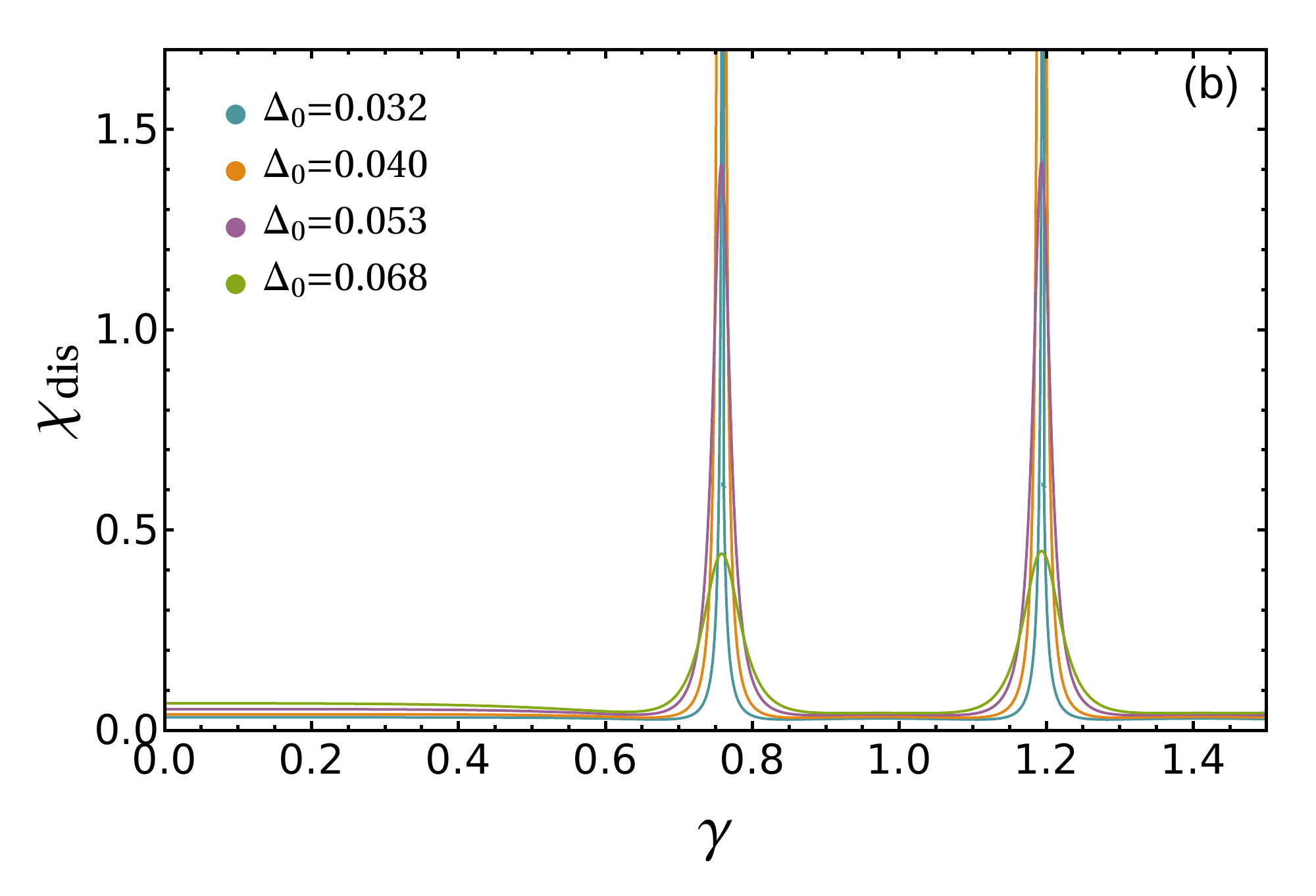}   
    \\
    \includegraphics[width=.48\textwidth]{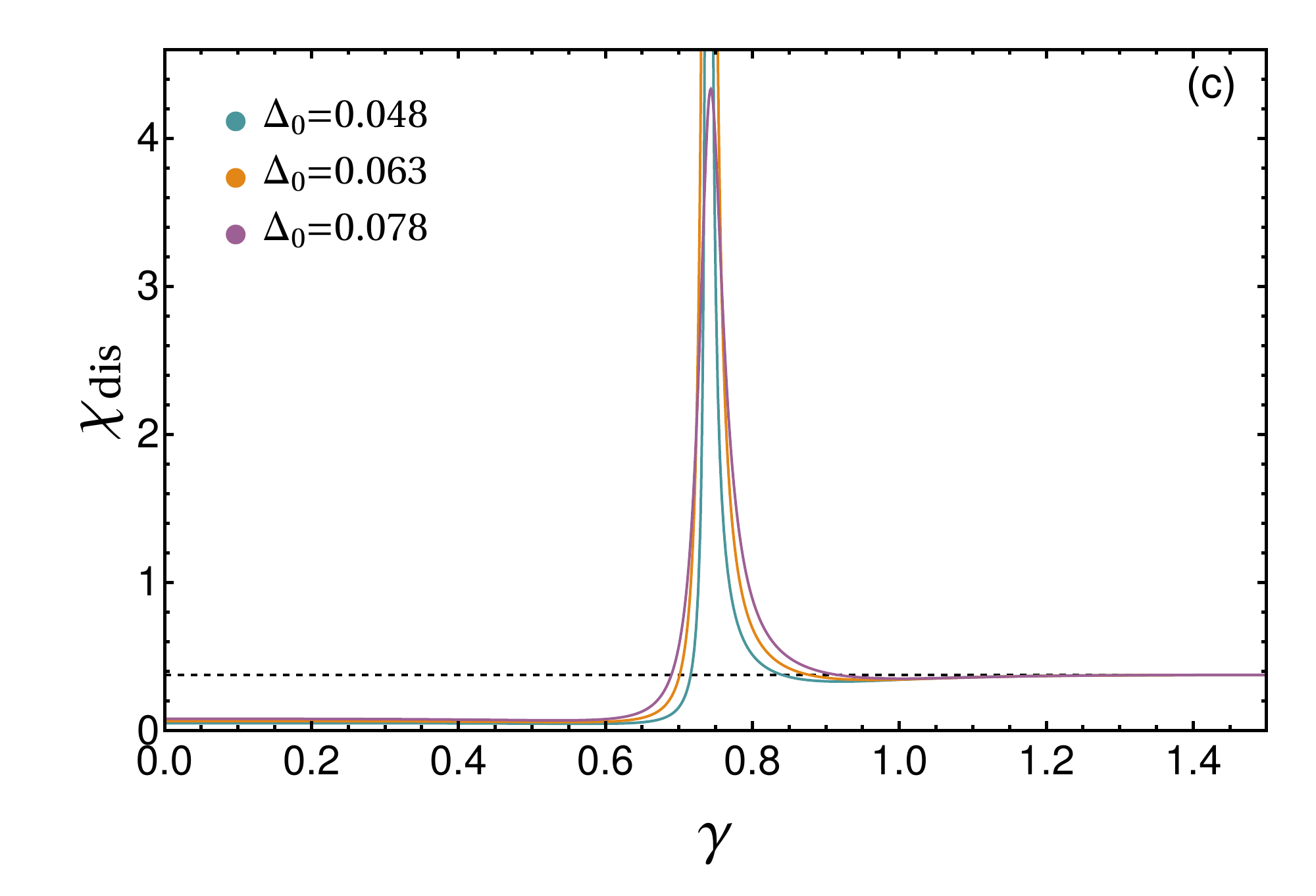}  & \includegraphics[width=.48\textwidth]{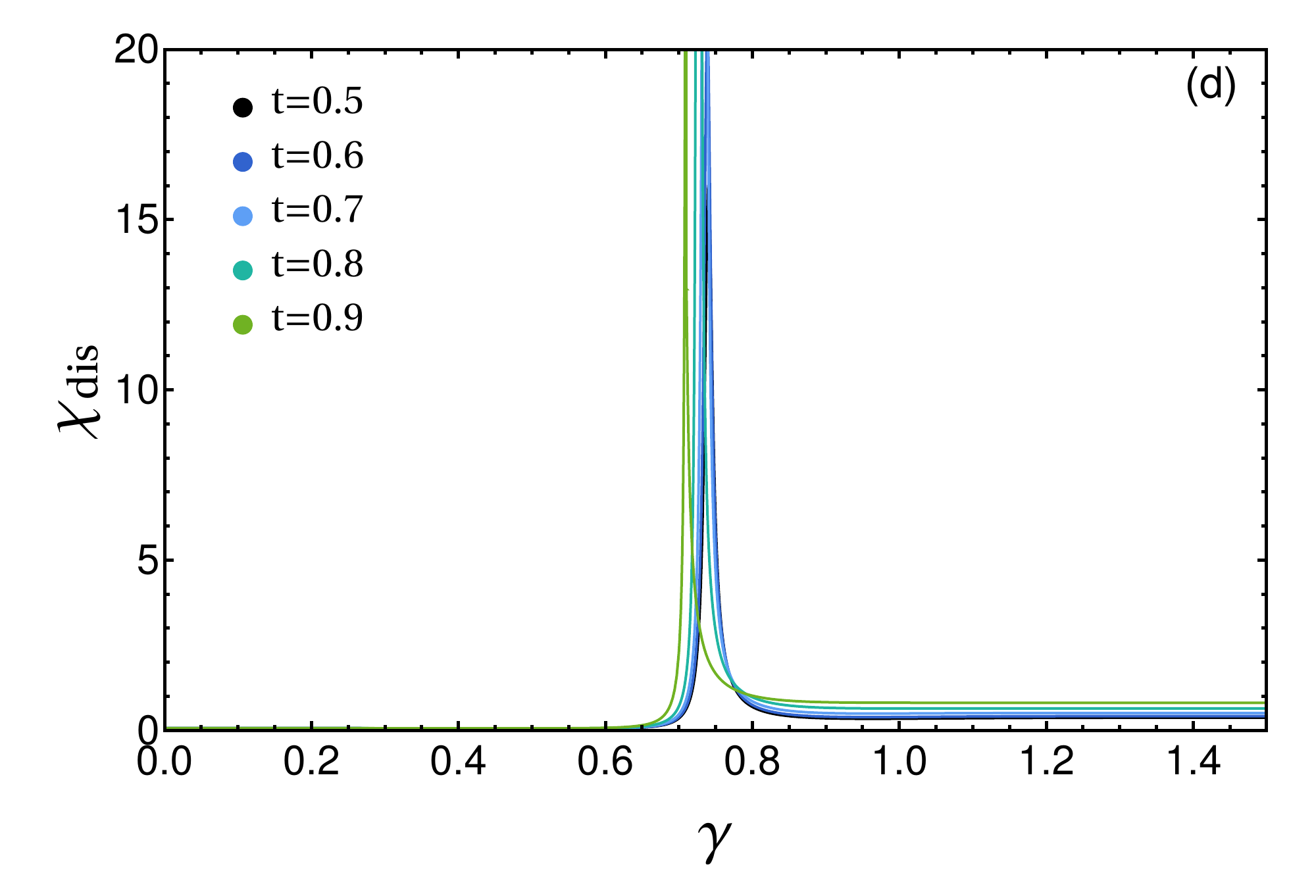} \\
 \end{tabular}
\caption{
Disconnected susceptibility $\chi_{{\rm dis}}(\gamma)$  for the MF-EPM with different distributions of the 
local random jumps $g(\hat x)$ [see Eq.~(\ref{eq_several_gRJ})] and  different distributions of the initial local stresses $P_0(x)$ [see 
Eq.~(\ref{eq_distrib_initial})]. 
(a): $P_0(x)$ is a 2-exponential combination and $g(\hat x)$ is a single exponential; curves are shown for several values of the initial disorder variance $\Delta_0$. (b-d): $P_0(x)$ is a Gaussian and $g(\hat x)$ is either a delta function (no randomness) in (b), a 2-exponential combination at fixed $t=0.5$ and several values of $\Delta_0$ in (c), or a 2-exponential combination at fixed $\Delta_0=0.25$ and several values of $t$ in (d).
In all cases, $2\mu_2=1$ and $<\hat x>=0.92$. When $P_0(x)$ is the combination of 2 exponentials, $\mu_1=0.0222\mu_2$, while in the Gaussian case we chose $\mu_1=0.9\mu_2$ in order to see both ductile and brittle behavior.}
\label{fig_compare_dis_susc}
\end{figure}

\begin{figure}
\centering
\begin{tabular}{cc}
    \includegraphics[width=.48\textwidth]{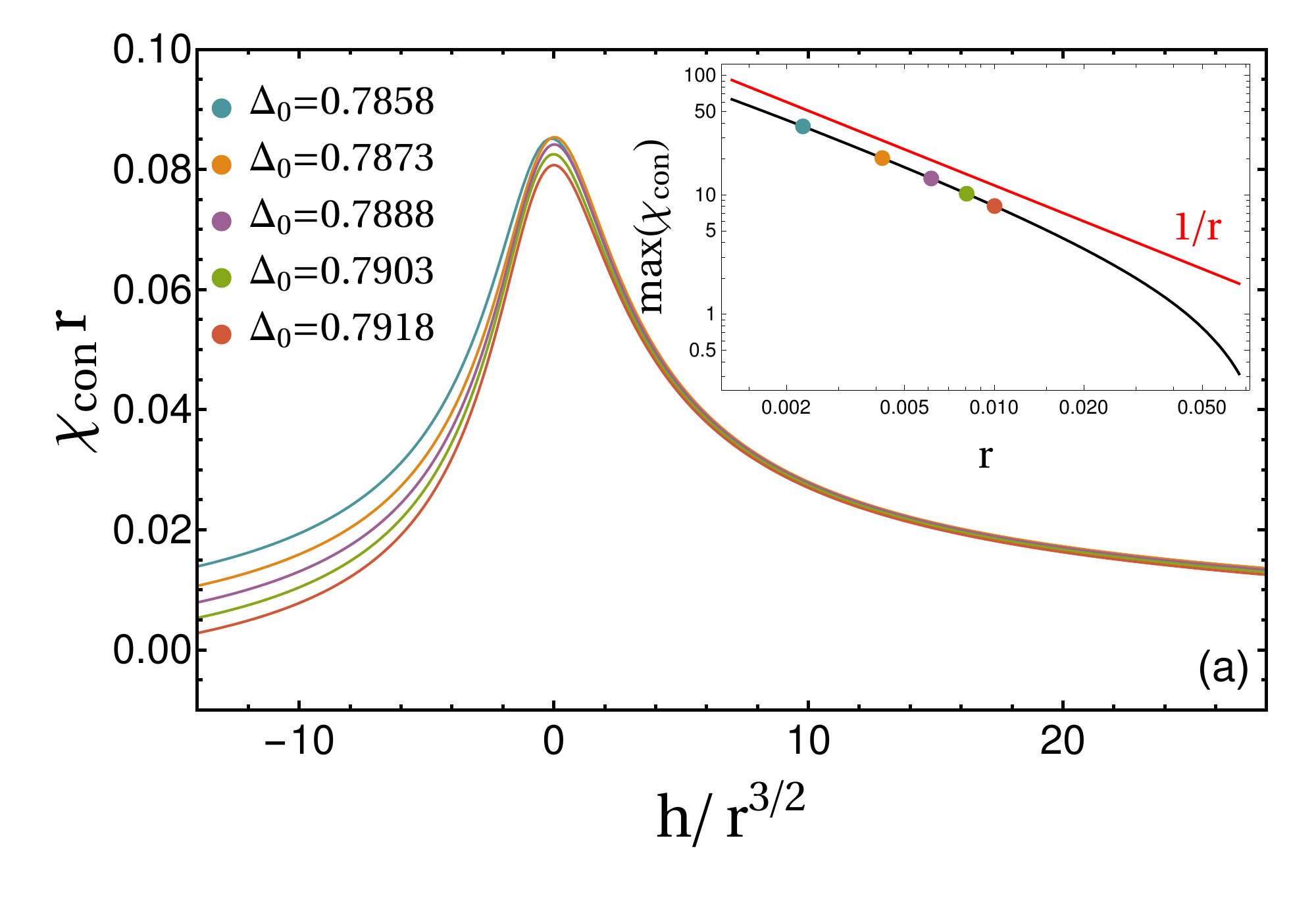}  & \includegraphics[width=.48\textwidth]{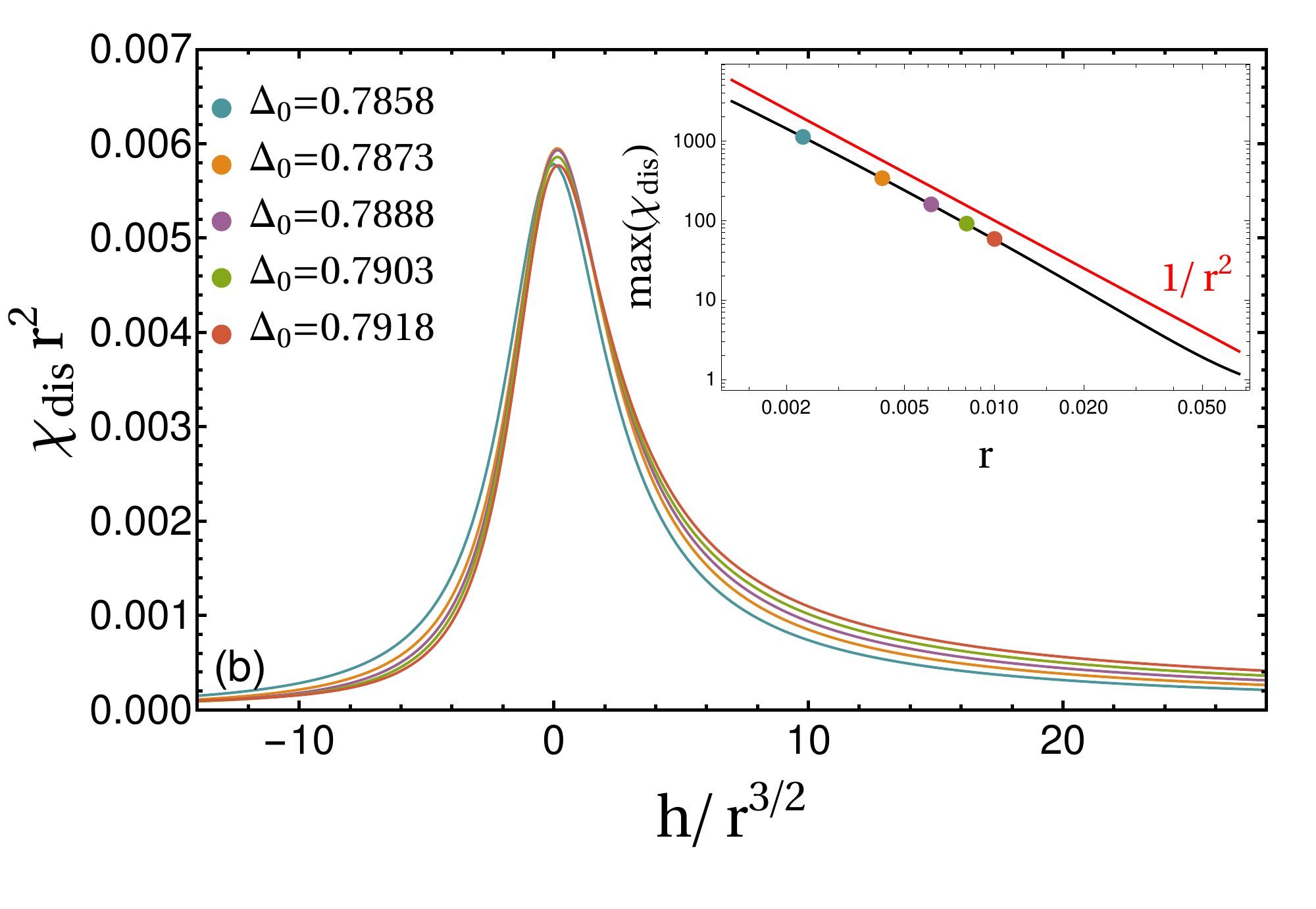}   \\
 \end{tabular}
\caption{Collapse of the rescaled connected susceptibility $\chi_{{\rm con}}(\gamma;\Delta_0)r$
(a) and rescaled disconnected susceptibility $\chi_{{\rm dis}}(\gamma;\Delta_0)r^2$ (b) versus 
$h/r^{3/2}$, with $r=[(\Delta_0-\Delta_{0,c})/\Delta_{0,c}]$ and $h=(\gamma-\gamma_Y)/\gamma_Y$, for the MF-EPM in the vicinity of the critical yielding transition at $\Delta_{0,c}$. (Here, $\gamma_Y$ is taken as the maximum of the connected (a) or disconnected (b) susceptibility, but the two maxima are extremely close to $\gamma_{Y,c}$in the vicinity of the critical point.). The insets show the scaling of the maxima of the susceptibilities with $r$ as $r\to 0$. The distribution of the local random stresses is the linear combination of two exponentials and that of the random jumps is the single-exponential one (then, $\Delta_{0,c}=0.78388$). Note that the collapse is not perfect but, as suggested in Ref.~[\onlinecite{perkovic96}] for the mean-field RFIM, could likely be improved by using a rotated scaling variable of the form $h'=h+b r$.}
\label{fig_scaling_susc}
\end{figure}

\subsection{Connected and disconnected susceptibilities}

We now discuss the results for the connected and disconnected susceptibilities in the mean-field EPM. Their expressions have been given above and are 
more explicitly derived in Appendix~\ref{app:diff_mod}). We again illustrate the results for the three different distributions of the local random jumps and 
the two distributions of initial local stresses considered above. The connected susceptibility $\chi_{{\rm con}}(\gamma)$ is shown in 
Fig.~\ref{fig_compare_con_susc} and the disconnected susceptibility in Fig.~\ref{fig_compare_dis_susc}, both for several values of the initial disorder 
variance $\Delta_0$.

Because it is defined as the opposite of the derivative of the stress-strain curve $\sigma(\gamma)$ (see above), the connected susceptibility is 
negative and equal to $-2\mu_2$ at small strain, goes through $0$ when there is an overshoot, is maximum in region (iii) previously defined, and goes to zero 
at large strain; it then approaches zero either from above or, when there is a local minimum in the stress-strain curve, from below after passing first through 
$0$ at a finite strain [see Fig.~\ref{fig_compare_con_susc} (c,d)]. In the case where there is no randomness in the jumps [Fig.~\ref{fig_compare_con_susc} (b)], 
the curve is periodic and the steady state is not physical.

The disconnected susceptibility is always positive. It starts from $\Delta_0$ at small strain, is maximum in region (iii), and goes to a positive value equal to 
the variance of the random jumps $<\hat x^2>-<\hat x>^2$, either from above or from below. Again, in the case where there is no randomness in the jumps [Fig.~\ref{fig_compare_dis_susc} (b)], the curve is periodic.

When there is a {\it bona fide} yielding transition, either discontinuous or critical, the two susceptibilities diverge. The divergence at the spinodal, which 
is the onset of the discontinuous jump in the average stress, is a consequence of the mean-field character of the model and can be expressed, e.g., as 
a function of the initial disorder variance $\Delta_0$, as 
\begin{equation}
\begin{aligned}
\label{eq_scaling_chi_spin}
&\chi_{\rm con}(\gamma;\Delta_0)\sim (\gamma_{Y,c}-\gamma)^{-\frac 12},\\&
\chi_{\rm dis}(\gamma;\Delta_0)\sim (\gamma_{Y,c}-\gamma)^{-1},
\end{aligned}
\end{equation}
when $\gamma\to\gamma_{Y,c}^-$ for $\Delta_{0}<\Delta_{0,c}$.

In the vicinity of the critical point at $\Delta_{0}=\Delta_{0,c}$, the susceptibilities can be described by scaling forms,
\begin{equation}
\begin{aligned}
\label{eq_scaling_chi_crit}
&\chi_{\rm con}(\gamma;\Delta_0)= \vert r\vert^{-(\beta\delta-\beta)} \mathcal F_{{\rm con},\pm}(\frac h{\vert r\vert^{\beta\delta}}),\\&
\chi_{\rm dis}(\gamma;\Delta_0)= \vert r\vert^{-2(\beta\delta-\beta)} \mathcal F_{{\rm dis},\pm}(\frac h{\vert r\vert^{\beta\delta}})
\end{aligned}
\end{equation}
where $r=(\Delta_0-\Delta_{0,c})/\Delta_{0,c}$ and $h=(\gamma-\gamma_{Y,c})/\gamma_{Y,c}$, and the exponents have their classical values, $\beta=1/2$ and 
$\delta=3$, hence $\beta\delta-\beta=1$. The scaling functions are different above and below the critical strain $\gamma_Y$ and are simply given by the smallest real root of cubic equations, just as in the mean-field RFIM~\cite{dahmen96}. These scaling forms are illustrated in Fig.~\ref{fig_scaling_susc}.

The fact that the exponent of the divergence of the disconnected susceptibility is twice that of the divergence of the connected susceptibility, either at the 
mean-field spinodal or at criticality, is a property that is characteristic of the mean-field RFIM and indicates the presence of an emerging random field at 
yielding. An additional signature of RFIM physics, also found in sample-to-sample fluctuations, is provided by the study of the size distribution of the 
avalanches (i.e., stress drops) present in the mean-field EPM. As shown in [\onlinecite{landes14,ozawaPNAS}], the distributions both at the critical 
and the spinodal points coincide with those of the mean-field RFIM~\cite{dahmen96}.

\begin{figure}
\centering
\begin{tabular}{cc}
    \includegraphics[width=.48\textwidth]{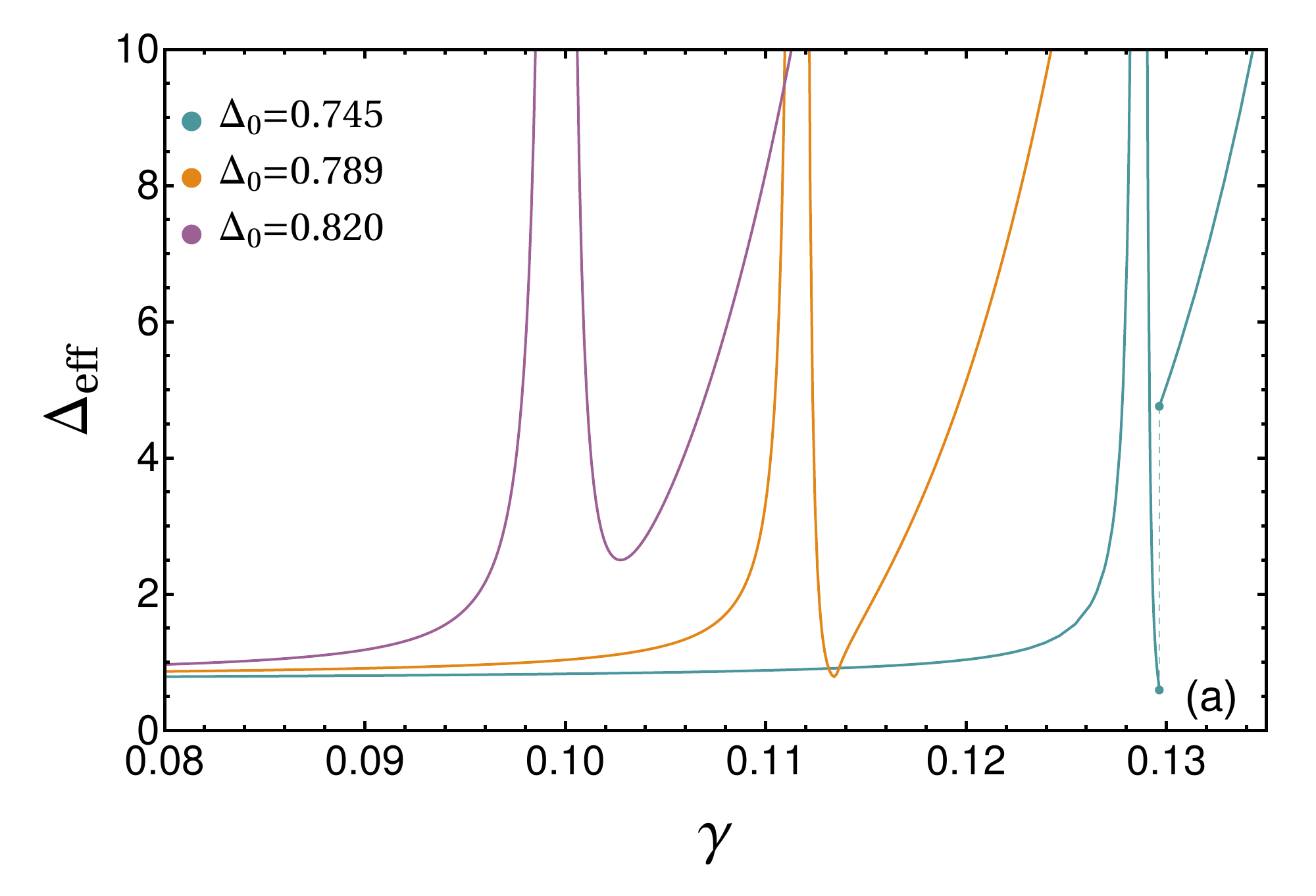}  &  \includegraphics[width=.48\textwidth]{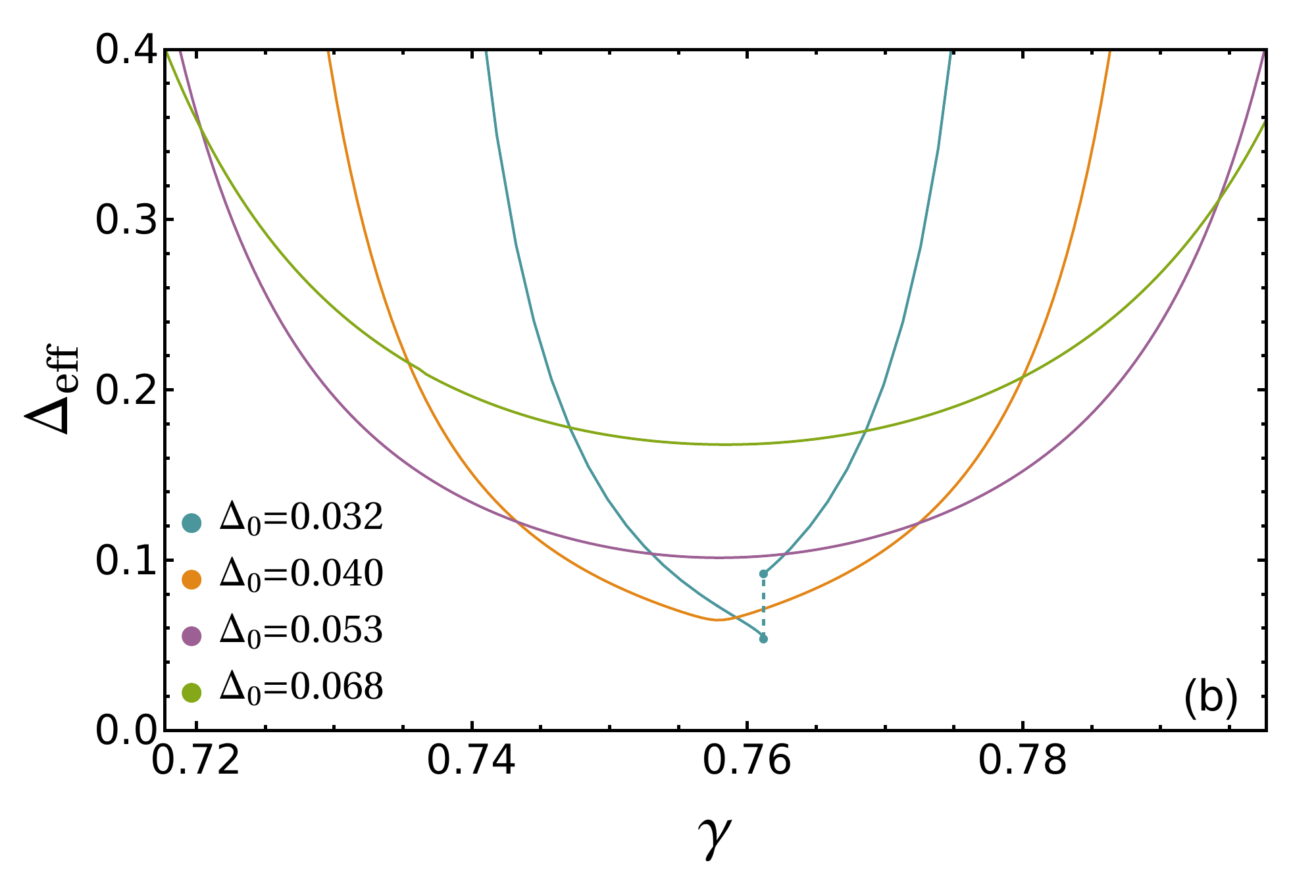}  
    \\
    \includegraphics[width=.48\textwidth]{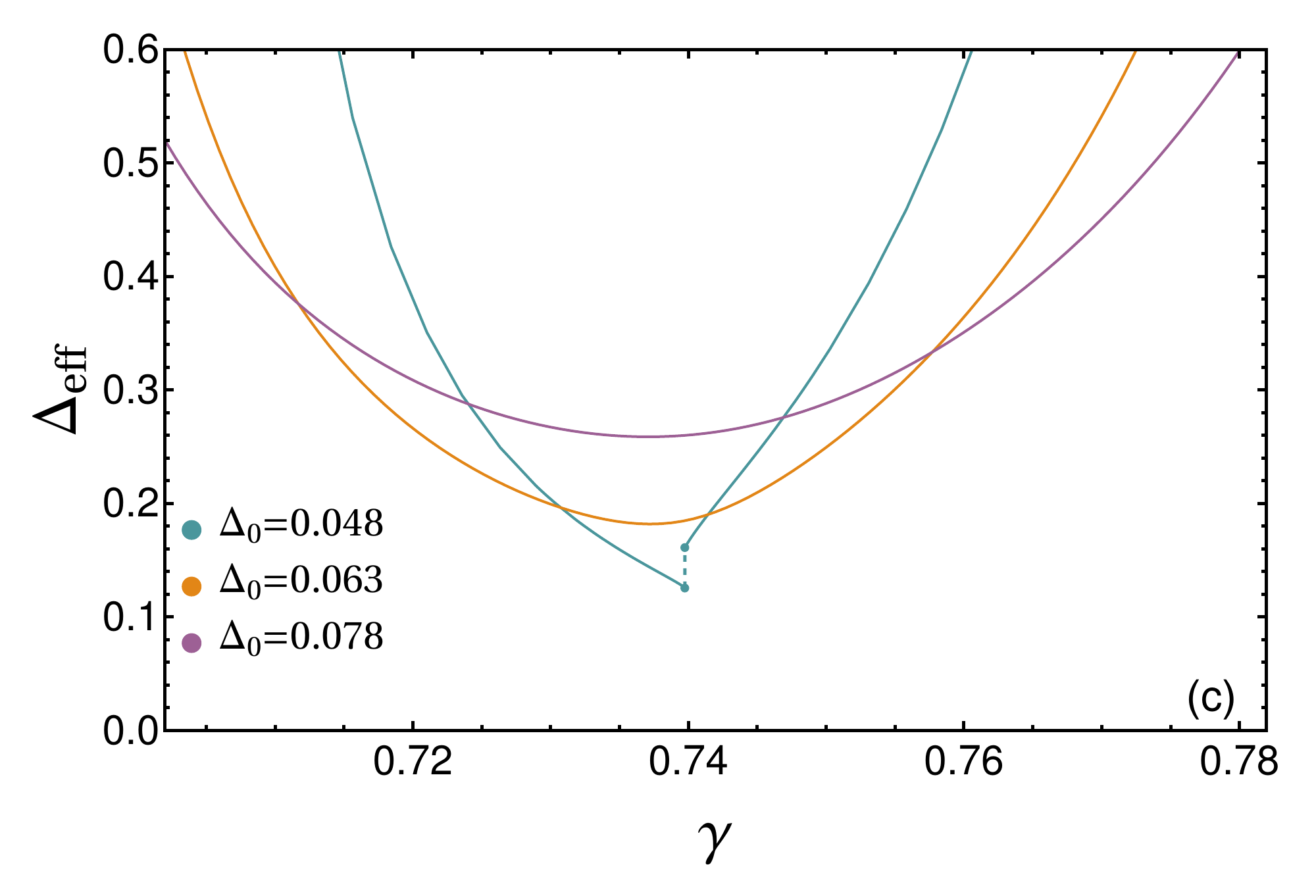}  & \includegraphics[width=.48\textwidth]{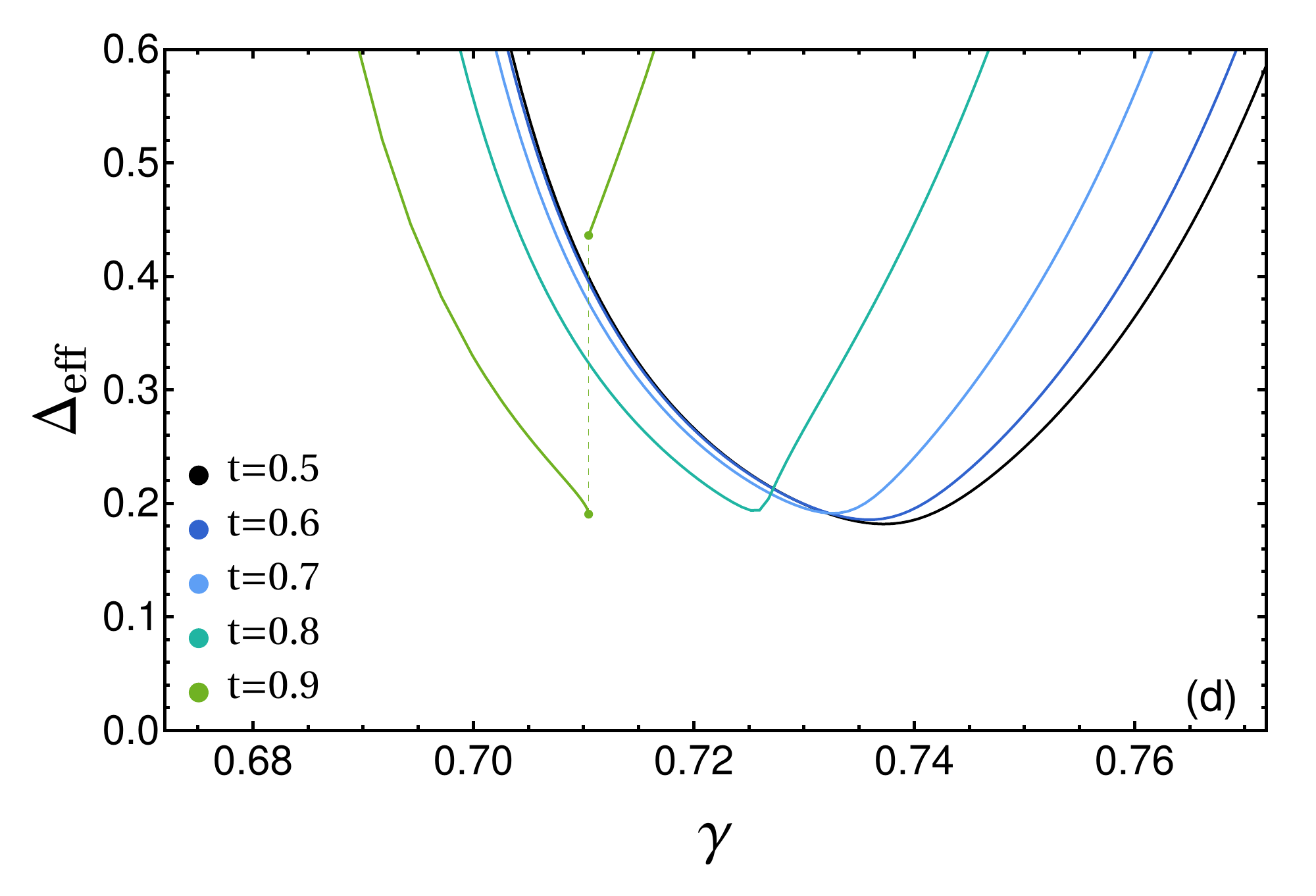} \\

\end{tabular}
\caption{
Effective random-field strength $\Delta_{{\rm eff}}(\gamma)$  for the MF-EPM in the region of the yielding 
transition where the connected susceptibility $\chi_{{\rm con}}(\gamma) \gg 1$. As in previous figures, different distributions of the local random 
jumps $g(\hat x)$ [see Eq.~(\ref{eq_several_gRJ})] and  different distributions of the initial local stresses $P_0(x)$ [see Eq.~(\ref{eq_distrib_initial})] are considered.
(a): $P_0(x)$ is a 2-exponential combination and $g(\hat x)$ is a single exponential; curves are shown for several values of the initial disorder variance $\Delta_0$. (b-d): $P_0(x)$ is a Gaussian and $g(\hat x)$ is either a delta function (no randomness) in (b), a 2-exponential combination at fixed $t=0.5$ and several values of $\Delta_0$ in (c), or a 2-exponential combination at fixed $\Delta_0=0.25$ and several values of $t$ in (d).
In all cases, $2\mu_2=1$ and $<\hat x>=0.92$. When $P_0(x)$ is the combination of 2 exponentials, $\mu_1=0.0222\mu_2$, while in the Gaussian case we chose $\mu_1=0.9\mu_2$ in order to see both ductile and brittle behavior.}
\label{fig_compare_Delta_eff}
\end{figure}

\begin{figure}
\centering
\begin{tabular}{cc}
    \includegraphics[width=.48\textwidth]{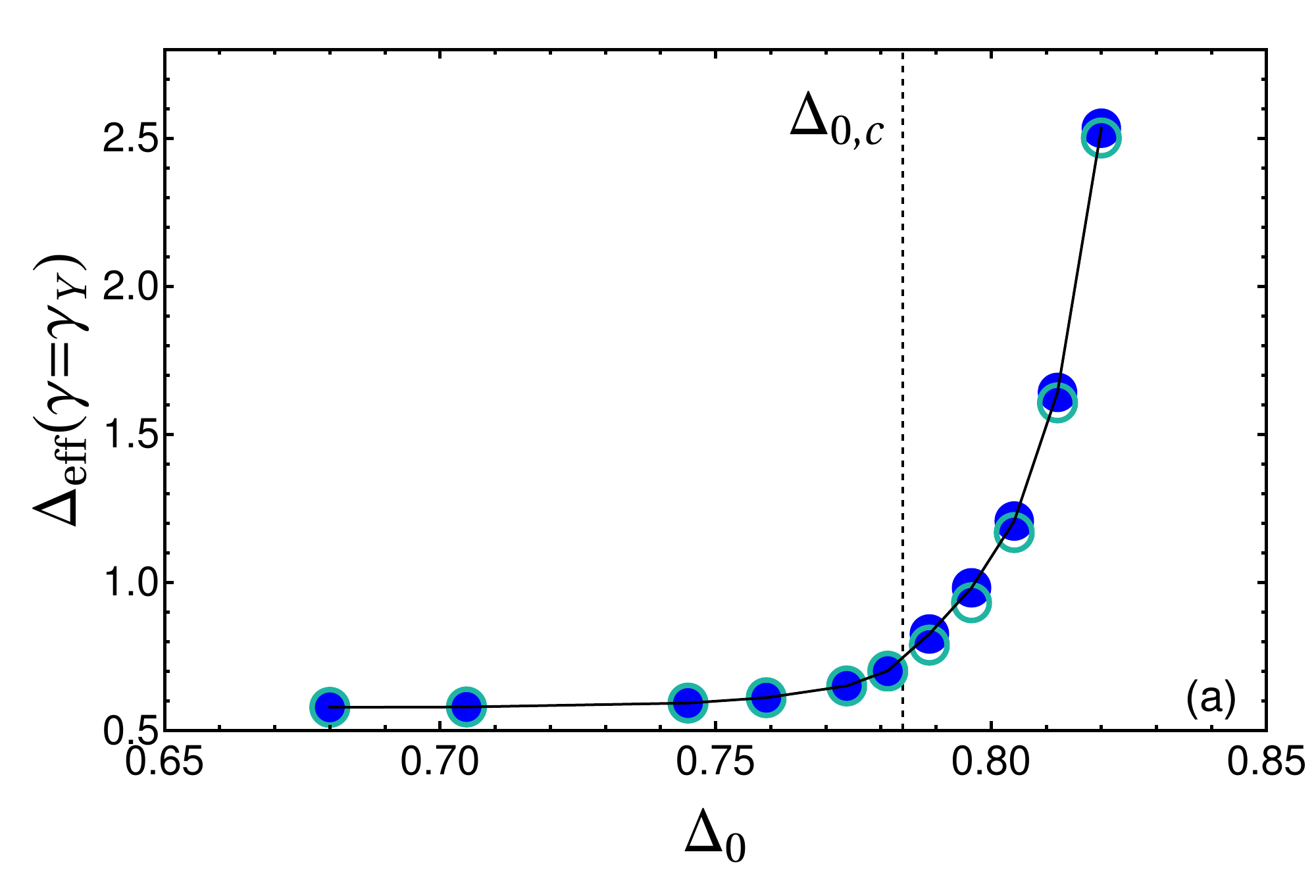}  &     \includegraphics[width=.48\textwidth]{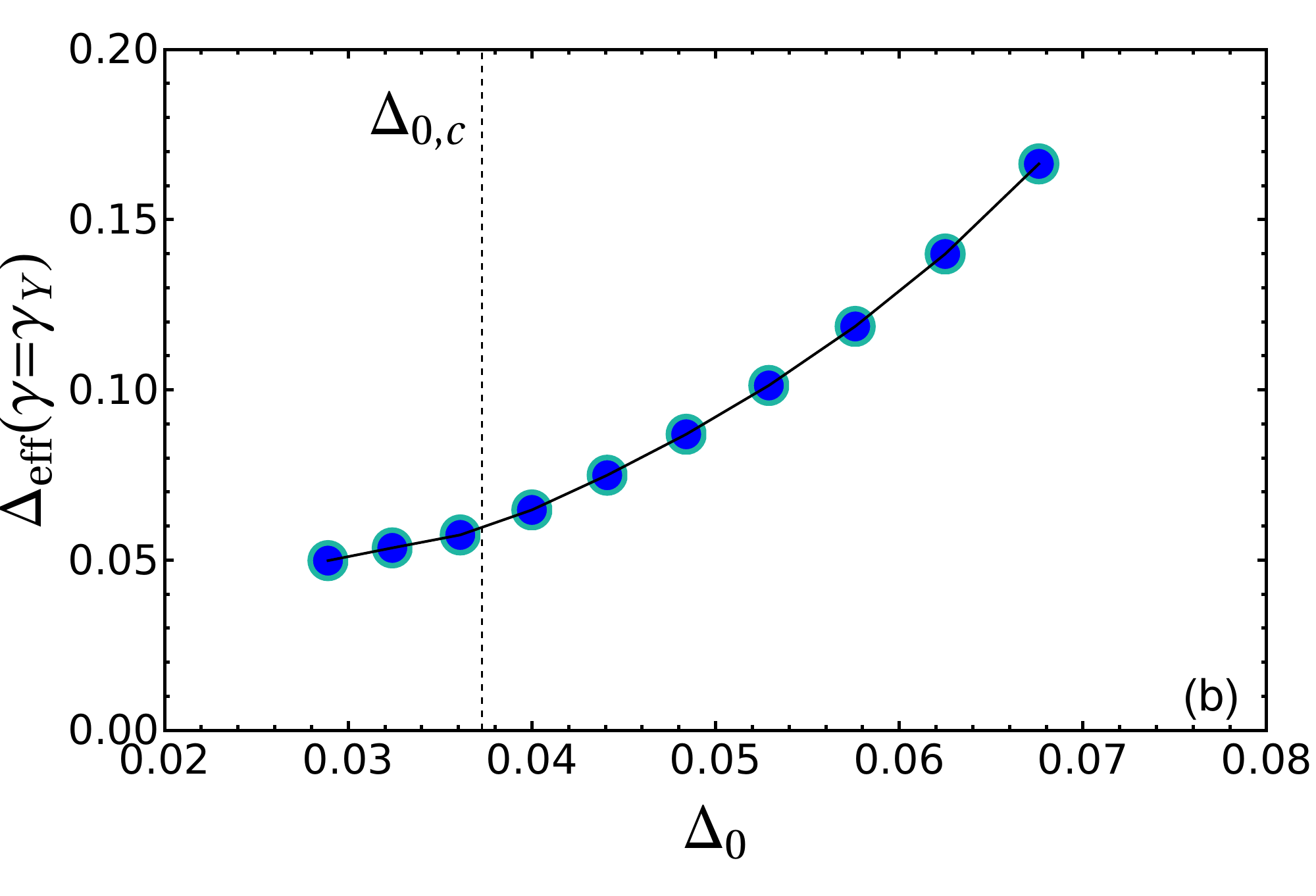} 
    \\
    \includegraphics[width=.48\textwidth]{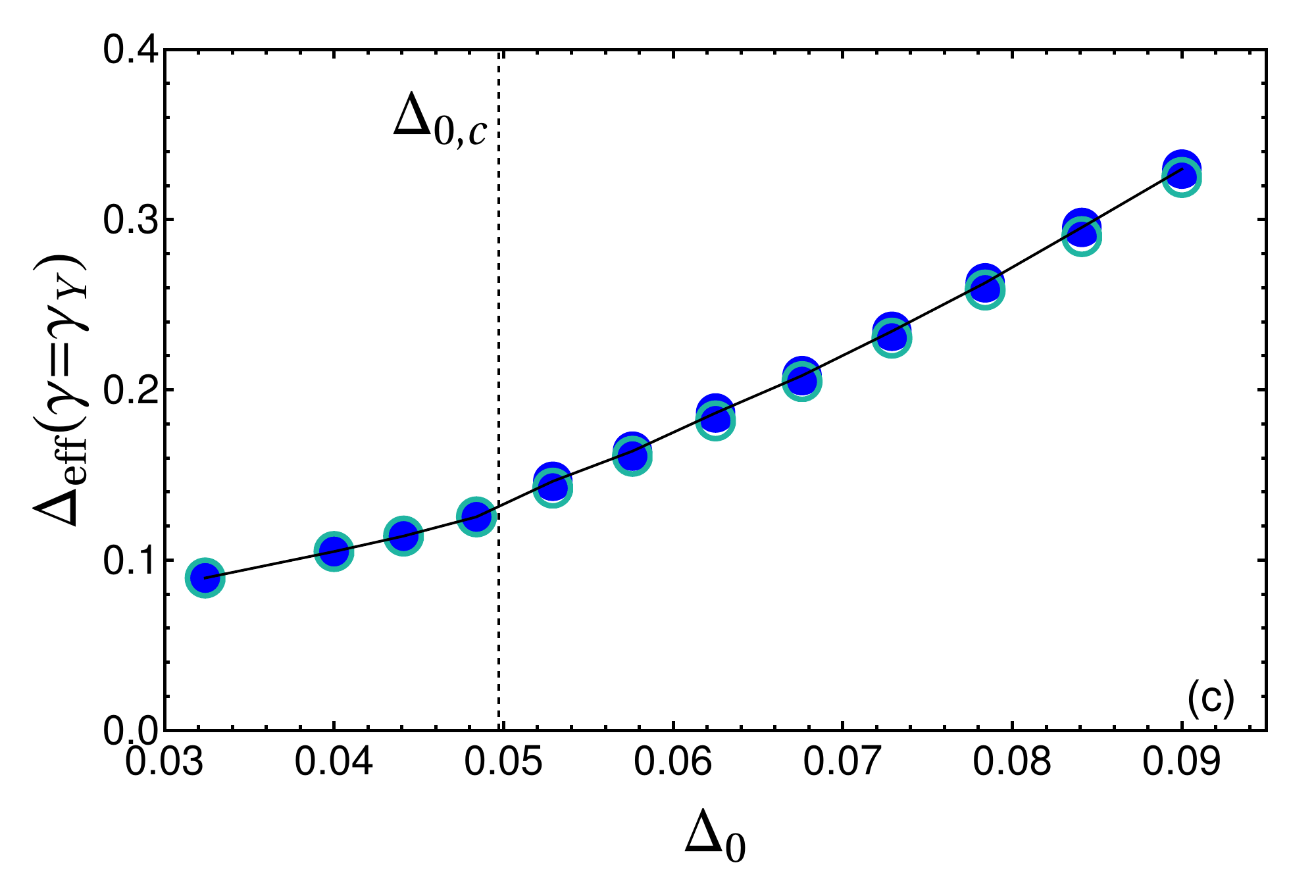}  & \includegraphics[width=.48\textwidth]{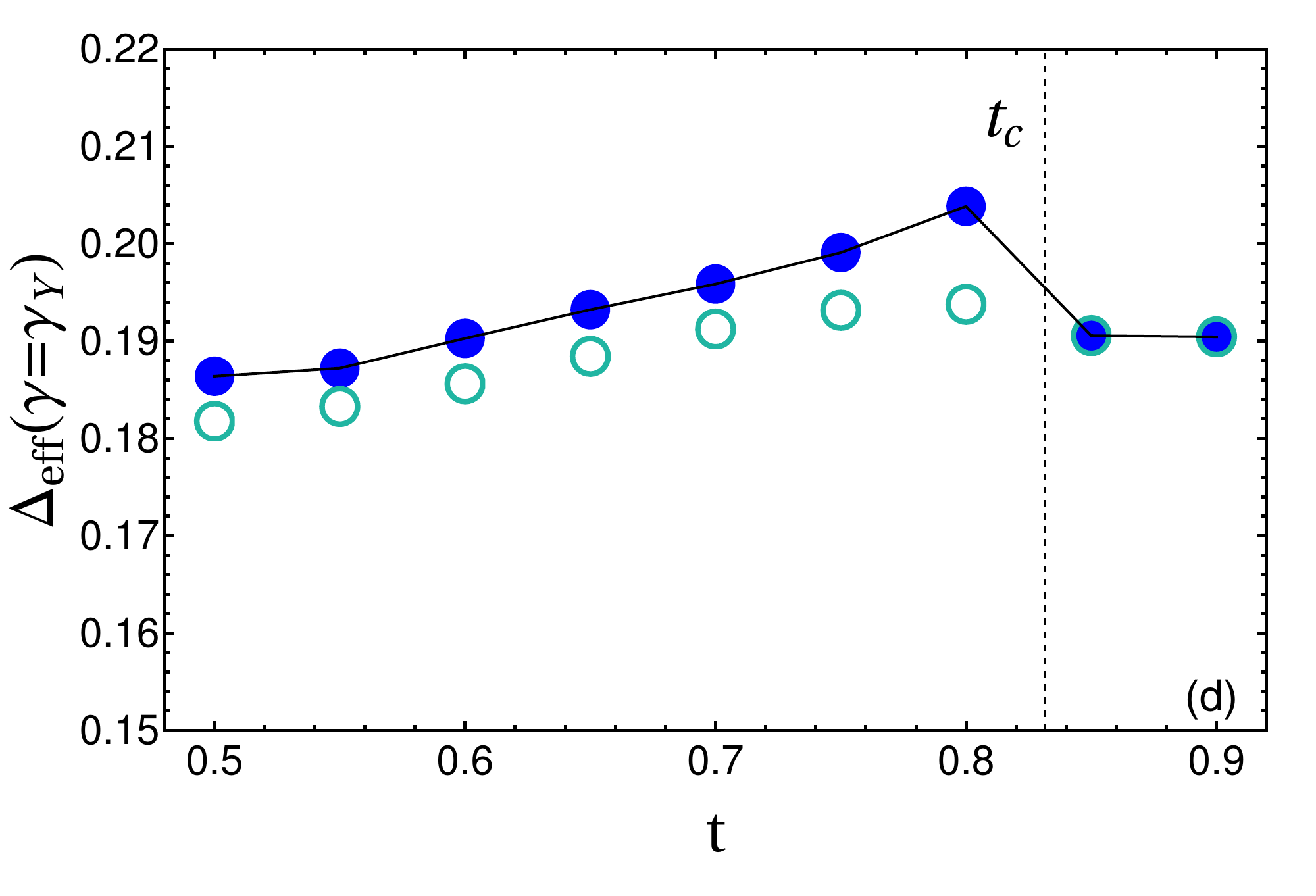} \\
\end{tabular}
\caption{
Variation of the effective random-field variance $\Delta_{{\rm eff}}(\gamma)$ for $\gamma=\gamma_Y$ (filled symbols and full line), at which the connected susceptibility diverges or/and is maximum, and for  $\gamma=\gamma_Y^\Delta$ (open symbols), the value at which $\Delta_{{\rm eff}}(\gamma)$ is minimum. We consider the same distributions of the local random jumps $g(\hat x)$ and of the initial local stresses $P_0(x)$ as in the previous figures. 
(a): $P_0(x)$ is a 2-exponential combination and $g(\hat x)$ is a single exponential; curves are shown for several values of the initial disorder variance $\Delta_0$. (b-d): $P_0(x)$ is a Gaussian and $g(\hat x)$ is either a delta function (no randomness) in (b), a 2-exponential combination at fixed $t=0.5$ and several values of $\Delta_0$ in (c), or a 2-exponential combination at fixed $\Delta_0=0.25$ and several values of $t$ in (d). 
In (b) the evolution is periodic and we only consider the first yielding region. In this case $\gamma_Y^\Delta=\gamma_Y$ over the range shown in the plot and the filled and open symbols exactly coincide. 
In (a-c) the vertical dashed line marks the critical value of the bare disorder.
}
\label{fig_compare_Delta_eff_Y}
\end{figure}

\subsection{Strength of the effective random field}

Having computed the connected and disconnected susceptibilities for the mean-field EPM we can obtain the strength of the effective random field from Eq.~(\ref{eq:strength_effectiveRF}), which here reads 
\begin{equation}
\label{eq:strength_emergentRF}
\Delta_{{\rm eff}}(\gamma)=\frac{\chi_{{\rm dis}}(\gamma)}{\chi_{{\rm con}}(\gamma)^2}\,.
\end{equation}
We have already stressed that for the mean-field EPM, it is the yielding transition which is in the universality class of the (mean-field) RFIM. As a result, one 
expects the above expression to be valid only in the region where yielding takes place, i.e., in a region where the connected susceptibility is positive and is 
large. Actually, Eq.~(\ref{eq:strength_emergentRF}) predicts that $\Delta_{{\rm eff}}$ diverges at a local maximum (overshoot) or minimum of the stress-strain 
curve, which has no physical meaning. The random field is an {\it emergent} property that results from the disorder present in the system (random initial local 
stresses, random local thresholds, random local jumps) but only appears in the region of deformation around yielding.

From Eqs.~(\ref{eq_disc_suscept_3terms}), (\ref{eq_final_gamma_diff}), and (\ref{eq:conn_MF-EPM}) we obtain that when 
$\chi_{{\rm con}}(\gamma), \chi_{{\rm dis}}(\gamma)\gg 1$,
\begin{equation}
    \label{eq_final_strength_emergentRF}
    \begin{aligned}
    & \Delta_{{\rm eff}}(\gamma)\approx  
    \left(\frac{\mu_1}{2\mu_2(\mu_1+\mu_2)[1-<\hat x>P_{y(\gamma)}(0)]} \right)^2 \Big ( <\hat{x}^2>\int_{0}^{y(\gamma)}d y' P_{y'}(0) 
    -<\hat x>^2\left(\int_{0}^{y(\gamma)} dy' P_{y'}(0) \right)^2 + \\& 
   2 <\hat x>\int_{0}^{y(\gamma)} dy' P_{y'}(0) \int_{y'}^{y(\gamma)} dy''\Big[g(0)( <\hat x> - T(y''-y')) - T'(y''-y') +
    \int_0^{y''-y'} d\widehat{y}(<\hat x> - T(\widehat{y}))R'_{y''-y'-\widehat{y}}(0) \Big ] \Big )
    \end{aligned}
\end{equation}
where, as defined before, $T(x)=\int_x^\infty d\hat x \hat x g(\hat x)$, and the functions $P_y(0)$ and $R_y(0)$ have been introduced in Sec.~\ref{sec:MF-EPMaverage}.

We illustrate in Fig.~\ref{fig_compare_Delta_eff} the behavior of $\Delta_{{\rm eff}}(\gamma)$ in the yielding region [region (iii) of Sec.~\ref{subsec_stress-strain}] for the three choices of random-jump distribution and the two choices of initial local-stress distribution, for several values 
of the initial disorder variance $\Delta_0$. We see that the effective variance increases very rapidly for a given value of the initial disorder $\Delta_0$ 
when one moves away from $\gamma_Y$ (which is either the location of the yielding transition when the susceptibilities diverges or the location of the local maximum of the connected susceptibility when the susceptibilities do not diverge).  The region of interest where $\Delta_{{\rm eff}}$ changes by less than a factor of, say, $2$ is very narrow around 
$\gamma_Y$. This results from the emergent character of the random field near yielding, for which the underlying random-field strength that incorporates the whole history of the system depends on the deformation. It is quite different than the behavior of the RFIM shown in Fig.~\ref{fig:eff_dis_RFIM} in which the effective random-field strength 
does not vary much.

We plot in Fig.~\ref{fig_compare_Delta_eff_Y}(a-c) the effective random-field variance as a function of the initial disorder variance $\Delta_0$ for both  $\gamma=\gamma_Y$, the maximum of the connected susceptibility,  and $\gamma=\gamma_Y^\Delta$, the extremum (minimum) of $\Delta_{{\rm eff}}(\gamma)$. Note first that the two locations $\gamma_Y^\Delta$ and $\gamma_y$ are identical when there is a bona fide yielding transition and are otherwise very close, as a result of the proximity of the maxima of the connected and disconnected susceptibilities (insets of Fig.~\ref{fig:gamma_y_gamma_max}): in consequence, there is virtually no difference in the values of the effective random-field strength evaluated at the two locations. We observe that $\Delta_{{\rm eff}}(\gamma_Y\;{\rm or}\;\gamma_Y^\Delta)$ monotonically increases with $\Delta_0$. As physically expected, it also increases as yielding changes from discontinuous to continuous, a discontinuous transition requiring a smaller effective random-field strength, just like in the RFIM. One can also see that $\Delta_{{\rm eff}}(\gamma_Y\;{\rm or}\;\gamma_Y^\Delta)$ is not a unique function of $\Delta_0$ and is sensitive to the various kinds of disorder present in the EPM: although qualitatively similar, the curves in (a), (b) and (c) are quantitatively different. In addition, Fig.~\ref{fig_compare_Delta_eff_Y} (d) illustrates the variation of $\Delta_{{\rm eff}}(\gamma_Y\;{\rm or}\;\gamma_Y^\Delta)$ with the parameter $t$ of the random jump distribution for the same given value of $\Delta_0$. The parameter $t$ has a small but noticeable effect on the effective random-field strength. (At the same time, changing $t$ has a strong influence on the steady state, as can be seen from Fig.~\ref{fig_compare_stress-strain}(d).) Note that the variation with $t$ is nonmonotonic but, in the immediate vicinity of the critical point $t_c$, one again finds that the brittle side of yielding ($t>t_c$) corresponds to smaller values of $\Delta_{{\rm eff}}(\gamma_Y\;{\rm or}\;\gamma_Y^\Delta)$ than the ductile side ($t<t_c$).

\begin{figure}
\centering
\begin{tabular}{cc}
    \includegraphics[width=.48\textwidth]{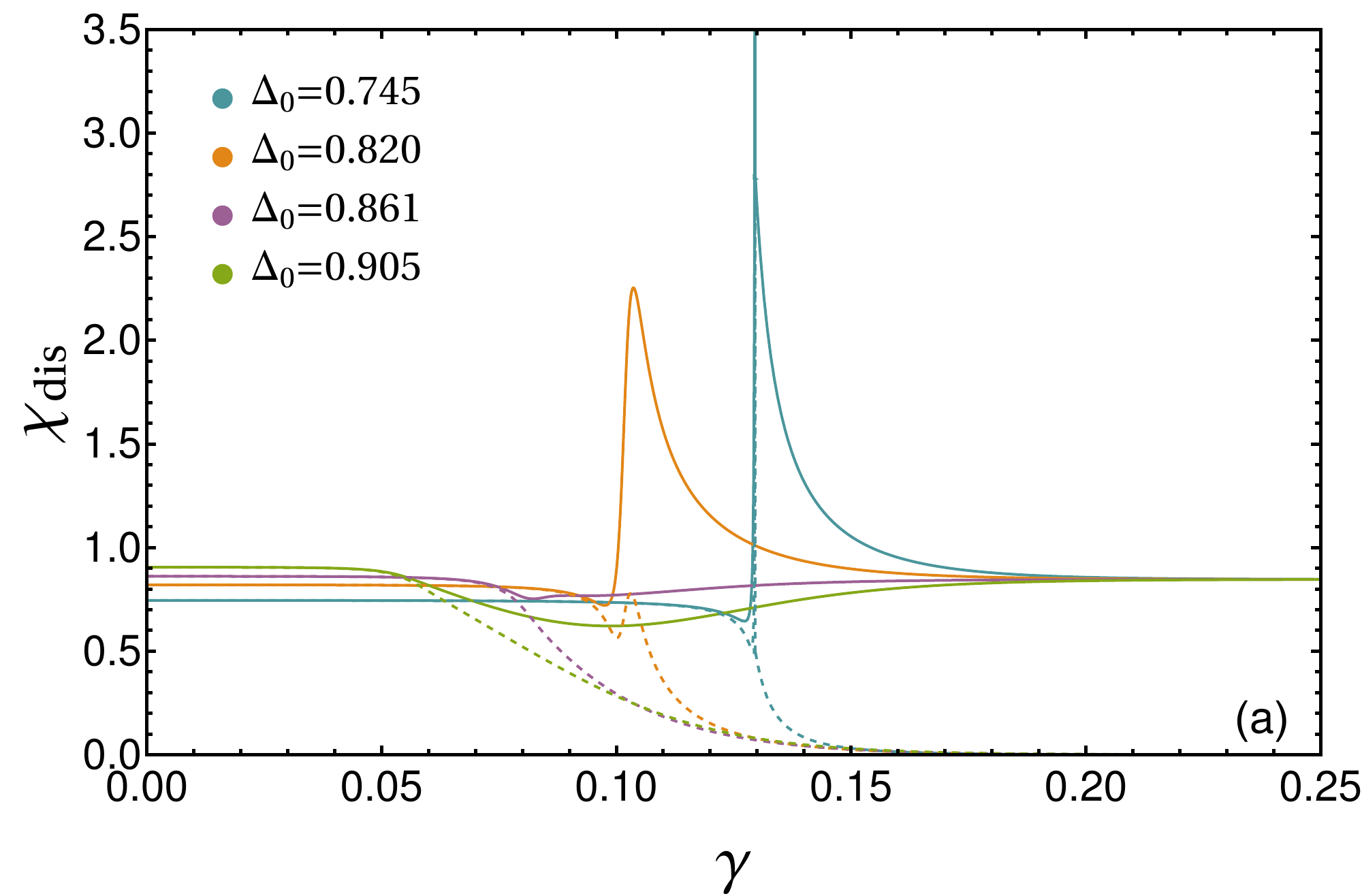}  & \includegraphics[width=.48\textwidth]{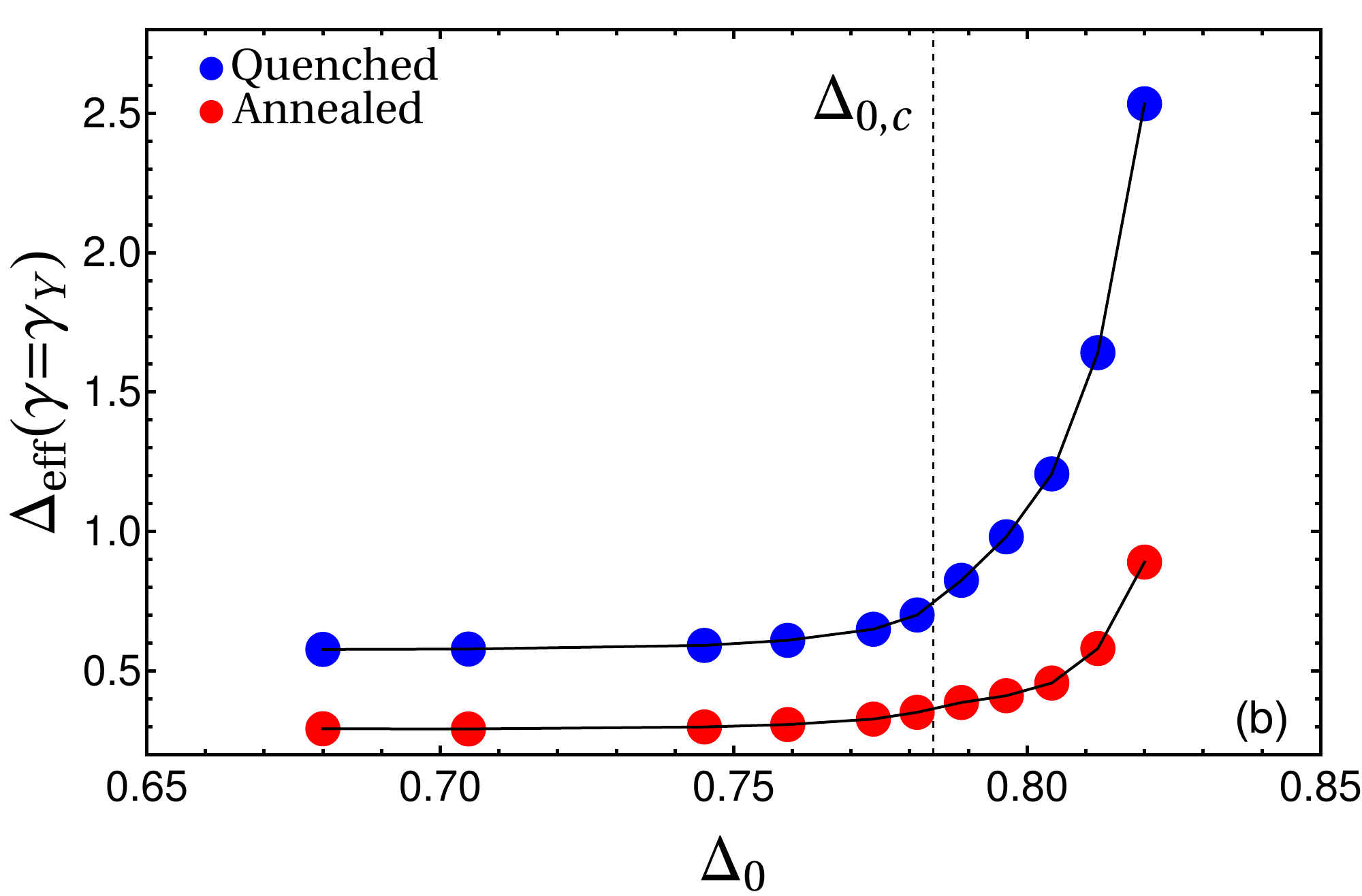}   \\
 \end{tabular}
\caption{Comparison between the quenched and the annealed average over the random jumps: (a) Disconnected susceptibility 
$\chi_{{\rm dis}}(\gamma)$ for several values of the initial disorder variance $\Delta_0$ and (b) variation with $\Delta_0$ of the effective random-field 
variance at yielding $\Delta_{{\rm eff}}(\gamma_Y)$ (note that $\gamma_Y$ is identical for the quenched and the annealed computations and so is the critical value of the bare disorder $\Delta_{0,c}$ which is indicated by the vertical line). The 
distribution $P_0(x)$ is a 2-exponential combination and $g(\hat x)$ is  a single exponential.}
\label{fig_compare_QA}
\end{figure}

To further evince the fact that the emerging random field is not fixed once for all in the initial distribution but also arises from the history of the deformation, e.g., 
through the sequence of random jumps associated with local plastic events, we consider two different ways of computing $\Delta_{{\rm eff}}(\gamma)$: 
one in which the sample-to-sample fluctuations are calculated via a ``quenched'' average over the random jumps (this is the calculation done up to now) 
and one in which we perform an ``annealed'' average over the random jumps, i.e.,
\begin{equation}
\label{eqn:disc_def_annealed}
\chi_{\text{dis}}^{\rm ann}(\gamma) = 
N \overline{\bigg[ \left< m^{\alpha,[\hat{x}]_M}(\gamma)\right>_{[\hat{x}]_M} -  \overline{\left< m^{\alpha,[\hat{x}]_M}(\gamma)\right>_{[\hat{x}]_M}} \bigg]^2}.
\end{equation}
Note that by construction the connected susceptibility is not affected by the change of averaging.
We illustrate the comparison between the quenched and annealed computations over the random jumps for the disconnected susceptibility and for the effective 
random-field variance at yielding, $\Delta_{{\rm eff}}(\gamma_Y)$,  in Fig.~\ref{fig_compare_QA}. One observes that the way the average over the history of 
random jumps is done influences the sample-to-sample quantities and in particular the strength of the effective random field. The latter is slightly smaller with the annealed average.

\section{Conclusion}
\label{sec_conclusion}

We have analytically characterized the sample-to-sample fluctuations in a mean-field elasto-plastic model (EPM) describing the AQS evolution of a 
disordered system under simple shear. We have directly shown that the vicinity of the yielding transition can be described as an effective AQS 
driven (mean-field) RFIM in which the applied magnetic field plays the role of the applied strain. We have shown and emphasized that the 
effective random field that linearly couples to the local stability (i.e., the distance to the local yield stress playing the role of the local order parameter akin 
to the local magnetization in the RFIM) is an emergent property: it is not present at the ``bare" level in the system prior to deformation and rather arises 
at a later stage from a combination of the initial conditions and the history of deformation involving a sequence of local plastic events. We have 
then investigated the influence of the various types of disorder introduced in the EPM to mimic the effect of the local structural inhomogeneity of an
amorphous solid (random initial local stresses, random local stress jumps, ...) on the strength of the effective random field, 
showing that they all contribute. As expected from the analogy with the RFIM, we find that for a decreasing strength of the random field yielding passes 
from a continuous crossover to a discontinuous transition, the two regimes being separated by a critical point.

Although the framework and some general considerations are valid for finite-dimensional systems as well, our study is restricted to mean-field 
models. The advantage is that we are able to obtain an exact analytical solution but one may wonder which of the conclusions reached here hold 
beyond mean-field. The main issue which we have stressed several times is the description of the stress redistribution after a local plastic event 
and therefore the nature of the effective interactions between site variables in the EPM. A proper description should account for the anisotropic, 
quadrupolar-like character of the Eshelby kernel~\cite{barrat_review,eshelby}. This is known to induce marginal stability in the sheared amorphous 
solids~\cite{lin-wyart15} but the degree to which it influences the yielding transition is an unsettled question~\cite{ozawaPNAS,ozawa_2D,ozawa_seeds,fieldingPRL,fielding,lernerMD}. To the least, one expects that the mapping of the yielding 
transition to a random-field Ising model should involve a modified model with Eshelby-like interactions~\cite{saverioMBTZ}.

\appendix

\section{AQS driven mean-field soft-spin RFIM}
\label{app:MF-RFIM}

We consider the mean-field soft-spin RFIM introduced in Sec.~\ref{sec_driven-RFIM} with Hamiltonian
\begin{equation}
\mathcal{H}[\left\{ s_{i} \right\}] = - \frac J{2N} \sum_{i,j\neq i} s_{i} s_{j} + \sum_i V(s_i) - \sum_{i=1}^{N} (h_{i} + H) s_{i}.    
\end{equation}
where $V(s_i)$ is the 2-parabola potential defined in Eq.~(\ref{eqn:potential}). In the AQS evolution the system goes through minima of the Hamiltonian, 
with
\begin{equation}
- \frac J{N} \sum_{j\neq i} s_{j} + V'(s_i) = h_{i} + H.    
\end{equation}
On the ascending branch of the hysteresis curve in which the magnetic field $H$ is adiabatically ramped up when starting from a large negative $H$, 
this directly leads to the solution in Eqs.~(\ref{eqn:spin}) and (\ref{eqn:magnetiz_RFIM}).

The average over the random-field distribution $\rho(h)$, which is equivalent to the average over samples $\alpha$, then yields the average magnetization 
given in Eq.~(\ref{eqn:mean_magnetiz_RFIM}).

More generally, if one wishes to compute the average of quantities that are functions of the sample-dependent magnetization it is convenient to start from 
the identity $\overline{f(m^\alpha)}=\int_{-\infty}^{+ \infty} dm f(m) \delta(m - m^{\alpha})$ and after using Eq.~(\ref{eqn:magnetiz_RFIM}) and standard 
manipulations,
\begin{equation}
\label{eq_f(m)}
\overline{f(m^{\alpha})} \propto \int_{- \infty}^{+ \infty} dm f(m) \int_{-i\infty}^{+i\infty} d\lambda e^{N \mathcal{G} (m,\lambda)},
\end{equation}
with 
\begin{equation}
\mathcal{G} (m,\lambda) = \lambda \left[ m + 1 - \frac{H + J m }{k}  \right] + 
\log\int_{- \infty}^{ -H-Jm + k} dh e^{-\frac{\lambda}{k} h} \rho(h) + \int_{- H - J m + k }^{+ \infty} dh e^{-\frac{\lambda}{k} h - 2 \lambda} \rho(h). 
\end{equation}
In the thermodynamic limit $N\to \infty$, the integrals in Eq.~(\ref{eq_f(m)}) can be treated via Laplace's method by an expansion around the saddle point 
$(m^*,\lambda^*)$ satisfying $\mathcal{G}^{(0,1)} (m^*,\lambda^*)=\mathcal{G}^{(1,0)} (m^*,\lambda^*)=0$, where the superscript indicate partial derivatives 
with respect to the arguments.

In practice, to expand around the saddle-point, we introduce the fluctuations as
\begin{equation}
\begin{aligned}
&m = m^{*} + \frac{\delta m }{\sqrt{N}} \\&    
\lambda= \lambda^{*} +  i\frac{\delta \lambda }{\sqrt{N}}
\end{aligned}
\end{equation}
where
\begin{equation}
\begin{aligned}
&\lambda^*=0, \\&
m^{*}(H)= \frac{k + H}{k - J} - \frac{2k}{k-J}\int_{-\infty}^{-H-Jm^{*}(H) + k } dh \rho(h).
\end{aligned}
\end{equation}
After some algebra, one easily finds that 
\begin{equation}
\begin{aligned}
\label{eq_f(m)_SPfluct}
\overline{f(m^{\alpha})}= f(m^{*}) + \frac{f''(m^{*})}{2N} [-\mathcal{G}_*^{(2,0)}+\frac{(\mathcal{G}_*^{(1,1)})^2}{\mathcal{G}_*^{(0,2)}}]^{-1},
\end{aligned}
\end{equation}
where
\begin{equation}
    \begin{aligned}
    \label{eq_mathcalG_derivatives}
&\mathcal{G}_*^{(2,0)}\equiv \mathcal{G}^{(2,0)}(m^*,\lambda^*) =0 \\&
\mathcal{G}_*^{(1,1)} \equiv \mathcal{G}^{(1,1)}(m^*,\lambda^*) = \left( 1 - \frac{J}{k} \right) - 2 J \rho(H + Jm^{*} -k),\\&
\mathcal{G}_*^{(0,2)}\equiv \mathcal{G}^{(0,2)}(m^*,\lambda^*) 
= \frac{\Delta_B^{2}}{k^{2}} + 4 \int_{-H-Jm^{*} + k}^{+ \infty} dh \rho(h) + \frac 4k\int_{-H-Jm^{*} + k}^{+ \infty} dh h\rho(h) - 4 \left[ \int_{-H-Jm^{*} + k}^{+ \infty} dh \rho(h) \right]^{2},              
\end{aligned}
\end{equation}
and one can check that $\mathcal{G}_*^{(0,2)}>0$.

Consider now the disconnected susceptibility, $\chi_{\rm dis}(H) = N \overline{[m^\alpha(H) - \overline{m^\alpha(H)}]^{2}}$. By applying the above formula in 
Eq.~(\ref{eq_f(m)_SPfluct}), one 
immediately obtains
\begin{equation}
\chi_{\rm disc} = \frac{\mathcal{G}_*^{(0,2)}}{( \mathcal{G}_*^{(1,1)})^2 },
\end{equation}
which after using the explicit expressions in Eq.~(\ref{eq_mathcalG_derivatives}) and $m^*=m+{\rm O}(1/\sqrt N)$ leads to Eq.~(\ref{eqn:disc_MF-RFIM}) 
of the main text.
\\

 \section{Detailed treatment of the mean-field EPM}
 \label{app:solution_MF-EPM}

\subsection{Expression of the disconnected susceptibility}

We derive the expressions of the terms appearing in the disconnected susceptibility as given by Eq.~(\ref{eq_disc_suscept_3terms}).
\\

We first consider the average of the squared fluctuations, $\overline{\left< \delta\widehat y^{\alpha,[\hat{x}]_{M}}(\gamma)^2\right>}_{[\hat{x}]_M}$. As already 
stressed, it is easier to study the evolution as a function of $y$ instead of $\gamma$ [see Eq.~(\ref{eqs_flutuations})]. Then, after integrating 
Eq.~(\ref{eqn:dgamma_disorder}), one has
\begin{equation}
\begin{aligned}
\delta\widehat \gamma^{\alpha,[\hat{x}]_{M}}(y)=
 \frac{-\sqrt N}{2(\mu_1+\mu_2)} \sum_{n=1}^{M} \big [\frac{1}{N} \sum_{i=1}^N \theta(dy-x_{i}^{\alpha,[\hat{x}]_{n}}(ndy)) \hat{x}_i^{n+1} 
 -<\hat x> F_{ndy}(dy) \big ], 
\end{aligned}
\end{equation}
where $y=M dy$ and we recall that $F_y(x)=(1/N)\sum_i \overline{<\theta(x-x_{i}^{\alpha,[\hat{x}]_{M}}(y)) >_{[\hat{x}]_{M}}}$, so that
\begin{equation}
\begin{aligned}
\label{eqn:gam_fluc}
& \frac{4(\mu_1+\mu_2)^2}N\,\overline{\left< \left[  \delta\widehat \gamma^{\alpha,[\hat{x}]_{M}}(y)\right]^2\right>}_{[\hat{x}]_M} =\\& \sum_{n',n''=1}^M 
\Big[\overline{\left< \frac{1}{N^2} \sum_{i,j=1}^N \theta(dy - x_i^{\alpha,[\hat{x}]_{n'}}(n'dy)) \hat{x}_i^{n'+1}  \theta(dy - x_j^{\alpha,[\hat{x}]_{n''}}(n''dy))
 \hat{x}_j^{n''+1}\right>}_{[\hat{x}]_{\tilde n}} - <\hat x>^2 F_{n'dy}(dy)F_{n''dy}(dy)   \Big].
\end{aligned}
\end{equation}
where $\tilde n=\max(n',n'')+1$. 

An important property is that the $x_i$'s evolve independently when considering their evolution with $y$ as the control parameter, so that when $i \neq j$ 
in the sum in the right-hand side of Eq.~(\ref{eqn:gam_fluc}) one can factorize the average of the two $\theta$ functions. After some manipulations this 
leads to
\begin{equation}
\begin{aligned}
\label{eqn:gam_fluc2} 
&4(\mu_1+\mu_2)^2\,\overline{\left< \left[  \delta\widehat \gamma^{\alpha,[\hat{x}]_{M}}(y)\right]^2\right>}_{[\hat{x}]_M} =\\&
\sum_{n',n''=1}^M \Big[\overline{\left<   \theta(dy - x_i^{\alpha,[\hat{x}]_{n'}}(n'dy)) \hat{x}_i^{n'+1}  
\theta(dy - x_i^{\alpha,[\hat{x}]_{n''}}(n''dy)) \hat{x}_i^{n''+1}\right>}_{[\hat{x}]_{\max(n',n'')+1}} - <\hat x>^2 F_{n'dy}(dy)F_{n''dy}(dy)   \Big].
\end{aligned}
\end{equation}
We need to compute the first term for the two different cases $n'=n''$ and $n'<n''$.

When $n'=n''$ one simply has
\begin{equation}
\label{eqn:firstpiece}
\begin{aligned}
        &\sum_{n'=1}^M   \overline{\left<  \theta(dy - x_i^{\alpha,[\hat{x}]_{n'}}(n'dy)) \hat{x}_i^{n'+1}  
        \theta(dy - x_i^{\alpha,[\hat{x}]_{n'}}(n'dy)) \hat{x}_i^{n'+1}\right>}_{[\hat{x}]_{n'+1}} = \\ 
        &= \sum_{n'=1}^M   \overline{\left<    \theta(dy - x_i^{\alpha,[\hat{x}]_{n'}}(n'dy)) \left(\hat{x}_i^{n'+1}\right)^2\right>}_{[\hat{x}]_{n'+1}} 
        =  \int_{0}^{\infty} \hat{x}^2 g(\hat{x}) d\hat{x}\int_{0}^{y} P_{y'}(0) dy'.
\end{aligned}
\end{equation}

The case $n'<n''$ requires some care. First, the factor $\hat{x}_i^{n''+1}$ can be averaged alone, and
\begin{equation}
\begin{aligned}
\label{eq_intermediate_theta}
& \overline{\left<   \theta(dy - x_i^{\alpha,[\hat{x}]_{n'}}(n'dy)) \hat{x}_i^{n'+1}  \theta(dy - x_i^{\alpha,[\hat{x}]_{n''}}(n''dy)) \hat{x}_i^{n''+1}\right>}_{[\hat{x}]_{n''+1}} 
= \\&  
<\hat x> \overline{\left<   \theta(dy - x_i^{\alpha,[\hat{x}]_{n'}}(n'dy)) \hat{x}_i^{n'+1} \left<  \theta(dy - x_i^{\alpha,[\hat{x}]_{n''}}(n''dy))\right>_{[\hat{x}]_{n'+2\to n''}}\right>}_{[\hat{x}]_{n'+1}}.
\end{aligned}
\end{equation}
where $[\hat{x}]_{n'+2\to n''}$ means the sequence of random jumps between $\hat x_{n'+2}$ and $\hat x_{n''}$. From Eq.~(\ref{eq_evolution_theta_y}), 
when using the independence of the random jumps from one plastic event to another, one obtains that
\begin{equation}
\begin{aligned}
&\left< \theta(dy - x_i^{\alpha,[\hat{x}]_{n''}}(n''dy))\right>_{[\hat{x}]_{n'+2\to n''}}=\\&
\left< \theta(2dy - x_i^{\alpha,[\hat{x}]_{n''-1}}((n''-1)dy))\right>_{[\hat{x}]_{n'+2\to n''-1}}-
G(dy)\left< \theta(dy - x_i^{\alpha,[\hat{x}]_{n''}}((n''-1)dy))\right>_{[\hat{x}]_{n'+2\to n''-1}}
\end{aligned}
\end{equation}
where $G(x)=<\theta(\hat x-x)>=\int_x^\infty d\hat x g(\hat x)$, and, by repeating the procedure $(n''-n'-1)$ times,
\begin{equation}
\begin{aligned}
&\left< \theta(dy - x_i^{\alpha,[\hat{x}]_{n''}}(n''dy))\right>_{[\hat{x}]_{n'+2\to n''}}=\\&
\theta((n''-n')dy - x_i^{\alpha,[\hat{x}]_{n'+1}}((n'+1)dy)) - G(dy)\theta((n''-n'-1)dy - x_i^{\alpha,[\hat{x}]_{n'+1}}((n'+1)dy)) \\&
+ \sum_{n'''=1}^{n''-n'-2} S_{n''-n'-2-n'''}(dy) \theta(n''' dy - x_i^{\alpha,[\hat{x}]_{n'+1}}((n'+1)dy)),
\end{aligned}
\end{equation}
with $S_n(x)$ given by the self-consistent equation
\begin{equation}
S_n(x)= [g(x+ndy)-g(ndy)G(x)+\sum_{k=1}^n g((k-1)dy)S_{n-k}(x)]dy,
\end{equation}
or in the continuum limit with $n\to\infty$, $dy\to 0$ with $ndy=y$ and $S_y(x)=S_n(x)/dy$,
\begin{equation}
\label{eq_Sy}
S_y(x)= g(x+y)-g(y)G(x)+\int_{0}^y g(y')S_{y-y'}(x).
\end{equation}
One easily checks that the function $R_y(x)$ introduced in Eq.~(\ref{eqn:R_y}) is the derivative of $S_y(x)$ with respect to $x$.

With the above expressions, by using again Eq.~(\ref{eq_evolution_theta_y}) and then performing the average over $\hat x_{n'+1}$, one has
\begin{equation}
\begin{aligned}
\label{eq_intermediate_theta}
&\left<\hat{x}_i^{n'+1}\theta(dy-x_i^{\alpha,[\hat{x}]_{n''}}(n''dy)) \hat{x}_i^{n''+1}\right>_{[\hat{x}]_{n'+1}} = 
<\hat x>\theta((n''-n'+1)dy - x_i^{\alpha,[\hat{x}]_{n'}}(n'dy))- \\&
T((n''-n')dy)\theta(dy - x_i^{\alpha,[\hat{x}]_{n'}}(n'dy)) - G(dy)\big [<\hat x>\theta((n''-n')dy - x_i^{\alpha,[\hat{x}]_{n'}}(n'dy))-\\&
T((n''-n'-1)dy)\theta(dy - x_i^{\alpha,[\hat{x}]_{n'}}(n'dy)) \big ]
+\sum_{n'''=1}^{n''-n'-2} S_{n''-n'-2-n'''}(dy) \big [<\hat x>\theta((n'''+1)dy - x_i^{\alpha,[\hat{x}]_{n'}}(n'dy))-\\&
T(n'''dy)\theta(dy - x_i^{\alpha,[\hat{x}]_{n'}}(n'dy))\big ] 
\end{aligned}
\end{equation}
with $T(x)=\int_x^\infty d\hat x \hat x g(\hat x)$. The last step is to multiply the above expression by $\theta(dy - x_i^{\alpha,[\hat{x}]_{n'}}(n'dy))$ and to perform 
the average over the sequence of random jumps $[\hat{x}]_{n'}$ and over the samples $\alpha$. After some tedious 
algebra, we find
\begin{equation}
\begin{aligned}
\label{eq_intermediate_theta}
&\overline{\left<\theta(dy - x_i^{\alpha,[\hat{x}]_{n'}}(n'dy)) \hat{x}_i^{n'+1}  \theta(dy - x_i^{\alpha,[\hat{x}]_{n''}}(n''dy)) \hat{x}_i^{n''+1}\right>}_{[\hat{x}]_{n''+1}} 
=[<\hat x>  - T((n''-n')dy)] F_{n'dy}(dy)\\&
- [<\hat x> -T((n''-n'-1)dy)] F_{n'dy}(dy) + F_{n'dy}(dy)\sum_{n'''=1}^{n''-n'-2} S_{n''-n'-2-n'''}(dy)[<\hat x>- T(n'''dy)]. 
\end{aligned}
\end{equation}

Putting everything together and taking the continuum limit we finally obtain
\begin{equation}
    \label{final_gamma_diff}
    \begin{aligned}
&\overline{\left< \left[  \delta\widehat \gamma^{\alpha,[\hat{x}]_{M}}(y)\right]^2\right>}_{[\hat{x}]_M} =  \left(\frac{1}{2(\mu_1+\mu_2)} \right)^2 
\Big ( <\hat{x}^2> \int_{0}^{y} dy'P_{y'}(0)  - <\hat x>^2\left(\int_{0}^{y} dy' P_{y'}(0)\right)^2 + \\&  
2 <\hat x>\int_{0}^{y} dy' P_{y'}(0) \int_{y'}^{y} dy''\left[g(0) [<\hat x> - T(y''-y') ] - T'(y''-y') +\int_0^{y''-y'}d\hat y[<\hat x> - T(\hat{y})]R_{y''-y'-\hat{y}}(0) 
\right ]\Big ) .
    \end{aligned}
\end{equation}
By passing from the variable $y$ to the variable $\gamma$, this then leads to Eq.~(\ref{eq_final_gamma_diff}). One checks that the above expression is 
equal to $0$ when $y=0$ and by using the Laplace transform and expanding one finds that when $y\to \infty$,
\begin{equation}
    \begin{aligned}
&[2(\mu_1+\mu_2)]^2\overline{\left< \left[  \delta\widehat \gamma^{\alpha,[\hat{x}]_{M}}(y)\right]^2\right>}_{[\hat{x}]_M} \to 
 \Delta_0 - \frac{<\hat x^2>^2}{4<\hat x>^2} +\frac{<\hat x^3>}{3<\hat x>}.
    \end{aligned}
\end{equation}
where $\overline{x}=1$ and $\Delta_0=\overline{x^2}-1$.
\\

We now consider the mixed term $\overline{\left< \delta\widehat y^{\alpha,[\hat{x}]_{M}}(\gamma) \delta\widehat m^{\alpha}(0) \right>}_{[\hat{x}]_M}$. Changing 
again from $\gamma$ to $y$, we define
\begin{equation}
\label{eqn:definition_Qy}
Q(y) = \overline{\left< \delta\widehat \gamma^{\alpha,[\hat{x}]_{M}}(y)  \delta\widehat m^{\alpha}(0) \right>}_{[\hat{x}]_M}
\end{equation}
From the equation for $\delta\widehat \gamma^{\alpha,[\hat{x}]_{M}}(y)$ which we have already used above and some straightforward manipulations, we find
\begin{equation}
\begin{aligned}
\label{eq_moreQy}
Q(y) &= -\frac{1}{2(\mu_1+\mu_2)N} \sum_{n=1}^{M} \sum_{i,j=1}^N
\overline{\left<  \theta(dy -x_i^{\alpha,[\hat{x}]_n}(ndy))\hat{x}_i^{n+1}\right>_{[\hat{x}]_M}  [x_j^{\alpha}(0)-1]}\\&
= -\frac{<\hat x>}{2(\mu_1+\mu_2)N} \sum_{i=1}^N
\overline{\sum_{n=1}^{M}\left< \theta(dy -x_i^{\alpha,[\hat{x}]_n}(ndy))\right>_{[\hat{x}]_n}  [x_i^{\alpha}(0)-1]}.
\end{aligned}
\end{equation}
where we have taken advantage of the independence of the site variables when $y$ is the control variable and we have used that by construction 
$\overline{x_j^{\alpha}(0)}=1-\overline{\sigma_j^{\alpha}(0)}=1$.

As for deriving the expression of $\overline{\left< \left[  \delta\widehat \gamma^{\alpha,[\hat{x}]_{M}}(y)\right]^2\right>}_{[\hat{x}]_M}$, we can use the 
equation relating $\theta(x -x_i^{\alpha,[\hat{x}]_n}(ndy))$ at step $n$ to the same function at earlier steps. After taking the continuum limit, one easily 
finds that
\begin{equation}
\label{eq_evolution_theta_av}
\left< \theta(x -x_i^{\alpha,[\hat{x}]_n}(y))\right>_{[\hat{x}]_n}=\theta(x+y -x_i^{\alpha}(0))-G(x)\theta(y -x_i^{\alpha}(0))+
\int_0^y dy' S_{y-y'}(x)\theta(y' -x_i^{\alpha}(0)),
\end{equation}
with $S_y(x)$ defined in Eq.~(\ref{eq_Sy}). Inserting Eq.~(\ref{eq_evolution_theta_av}) in Eq.~(\ref{eq_moreQy}) and performing the average 
over the samples, i.e., over the $x_i^\alpha(0)$ leads to
\begin{equation}
\begin{aligned}
\label{eq_finalQy}
&Q(y)
= \\&
-\frac{<\hat x>}{2(\mu_1+\mu_2)}\int_0^ydy'\Big [(y'-1)P_0(y')+g(0)\int_0^{y'}dy''(y''-1)P_0(y'')+\int_0^{y'}dy''R_{y'-y''}(0)\int_0^{y''}d\hat y(\hat y-1)P_0(\hat y)\Big ],
\end{aligned}
\end{equation}
from which one finally obtains Eq.~(\ref{eqn_mixed_EPM}). One finds that the above expression is $0$ when $y=0$ and it goes to 
\begin{equation}
\begin{aligned}
Q(y) \to \frac{1}{2(\mu_1+\mu_2)}\Delta_0
\end{aligned}
\end{equation}
when $y\to \infty$.
\\

\subsection{Solving different models}
\label{app:diff_mod}

\subsubsection{One- and two-exponential distributions of random jumps}

The results are most easily obtained by using a Laplace transform on the variable $y$. We consider here the 1- and 2-exponential distributions of 
random jumps defined in Eq.~(\ref{eq_several_gRJ}). Their Laplace transforms read
\begin{equation}
\begin{aligned}
\label{eq_laplace_RJ}
&\hat g^{1{\rm exp}}(s)=\frac 1{1+<\hat x>s},\\&
\hat g^{2{\rm exp}}(s)=\frac 1{[1+t s][1+(<\hat x>-t)s]}.
\end{aligned}
\end{equation}

One easily finds from Eqs.~(\ref{eqn:generalP}, \ref{eqn:R_y}) that
\begin{equation}
\begin{aligned}
\label{eq_sol_laplace}
&\hat R_s(x)=-g(x)+s e^{sx}\frac{[\hat g(s)-\int_0^xdy e^{-sy}g(y)]}{1-\hat g(s)},\\&
\hat P_s(x)=e^{sx}[\hat P_0(s)-\int_0^x dy e^{-sy}\hat P_0(y)]+\hat P_0(s)e^{sx}\frac{[\hat g(s)-\int_0^xdy e^{-sy}g(y)]}{1-\hat g(s)},
\end{aligned}
\end{equation}
where a ``hat" indicates a Laplace transform with respect to $y$. It is easy to check that for the single-exponential distribution, $\hat R_s(x)=0$. 
Furthermore, since from Eq.~(\ref{eq_sol_laplace})
$\hat P_s(0)=\hat P_0(s)/[1-\hat g(s)]$, one has
\begin{equation}
\label{eqn:P2exp_solution}
P^{2{\rm exp}}_y(0) = P_0(y) + \frac{1}{<\hat x>}\int_0^{y}dy' P_0(y')[1 -   e^{-\frac{<\hat x>}{t(<\hat x>-t)}(y-y')}]
\end{equation}
which can be solved once an initial distribution $P_0(x)$ has been chosen. (The single-exponential result is simply obtained by dropping the last 
term in the bracket inside the integral.) The numerical solution for all the averaged quantities (average stress, connected susceptibility) is then 
easily obtained.

The contributions to the sample-to-sample fluctuations can be cast in the form
\begin{equation}
    \begin{aligned}
&[2(\mu_1+\mu_2)]^2\overline{\left< \left[  \delta\widehat \gamma^{\alpha,[\hat{x}]_{M}}(y)\right]^2\right>}_{[\hat{x}]_M} = \\&
\mathcal L_y^{-1}\Big \{<\hat{x}^2> \frac{\hat P_0(s)}{s[1-\hat g(s)]}
 - <\hat x>^2\int_0^\infty dy e^{-sy}\left(\int_{0}^{y} dy' P_{y'}(0)\right)^2 -2 <\hat x>\frac{\hat P_0(s)}{s[1-\hat g(s)]^2}g'(s)\Big \},
    \end{aligned}
\end{equation}
and
\begin{equation}
\begin{aligned}
(\mu_1+\mu_2)\overline{\left< \delta\widehat \gamma^{\alpha,[\hat{x}]_{M}}(y)  \delta\widehat m^{\alpha}(0) \right>}_{[\hat{x}]_M}
= \frac{<\hat x>}{2}\mathcal L_y^{-1}\Big \{\frac{[\hat P'_0(s)+\hat P_0(s)]}{s[1-\hat g(s)]}\Big \}.
\end{aligned}
\end{equation}
where $\mathcal L_y^{-1}$ denotes the inverse Laplace transform. 

Again, since $P_y(0)$ is expressed in terms of $P_0(y)$, one can compute from the above expressions the disconnected susceptibility for the 
2-exponential and single-exponential random-jump distributions once an initial distribution $P_0(x)$ is specified.
\\

\subsubsection{Fixed-jump model}

For the fixed-jump model [see Eq.~(\ref{eq_several_gRJ})] the only randomness is in the initial condition. Eq.~(\ref{eqn:microscopic_dy_evolution_EPM_RJ}) 
simplifies to
\begin{equation}
\begin{aligned}
\label{eqn:microscopic_dy_evolution_EPM}
dy^{\alpha} &= 2 \mu_{2} d\gamma + \frac{\mu_{2}<\hat x>}{N(\mu_{1}+\mu_{2}) } \sum_{i=1}^{N} \theta( dy^{\alpha }-x_{i}^{\alpha}(\gamma) )\\
    &=  2 \mu_{2} d\gamma + x_c F_y^{\alpha}(dy^{\alpha}),
\end{aligned}
\end{equation}
and similar simplifications occur for the other equations.
\\

Consider first the averaged quantities. We work with $y$ as the control parameter and we start with the equation for the (averaged) cumulative probability, 
$F_y(x)=\int_0^x dx' P_y(x')$,
\begin{equation}
    \label{eqn:q_evolution}
    \partial_{y} F_y(x) = \partial_{x} F_y(x) - P_y(0) \theta(<\hat x> -x).
\end{equation}
In this subsection, to alleviate the notation, we replace $<\hat x>$ by $\hat x$ (there is anyhow no randomness in the jumps). To solve the above equation 
we have to consider the two cases $x>\hat x$ and $x<\hat x$ separately.
\\

\begin{itemize}
    \item  $x > \hat x$
\end{itemize}

Eq.~(\ref{eqn:q_evolution}) then reduces to
\begin{equation}
    \label{eqn:q_ev_sopraxbar}
    \partial_{y} F_y(x) =\partial_{x} F_y(x),
\end{equation}
which implies that $F_y(x)$ is a function of only $x+y$. Using the condition at $y=0$ gives 
\begin{equation}
    F_y(x)=F_0(x+y),
\end{equation}
Note that the condition that $F_y(x)\to 1$ when $y\to \infty$ is properly satisfied. The condition comes from the fact that at large strain (or $y$), 
all the sites have undergone at least one plastic event, so that their stability cannot be larger than $\hat x$ (whereas this could be possible at small $y$ if 
$P_0(x)$ extends over all values of $x$). As a result, $F_y(x)$ being the cumulative distribution, it must be equal to $1$.
\\

\begin{itemize}
    \item  $x < \hat{x}$
\end{itemize}
Then,
\begin{equation}
\left ( \partial_{y} - \partial_{x} \right) F_y(x) = -P_y(0),
\end{equation}
so that $F_y(x)$ is a function of $y$ and $x+y$. It can be written as  $F_y(x) = P(x+y) - H(y)$, with
\begin{equation}
    H(y) = \int_0^{y} dy' P_{y'}(0)
\end{equation}
and $P(x+y)$ yet to be determined. The condition that $F_y(0) = 0$ then imposes that $F_y(x)=H(x+y)-H(y)$. One expects that the function $H(y)$ is 
piecewise continuous and one easily checks that it takes the form
\begin{equation}
        H(y) = H_{0}(y) + \sum_{n = 1}^{\infty} H_{0}(y-n\hat{x}) \theta(y-n\hat{x}),
\end{equation}
with $H_0(y)$ defined between $0$ and $\hat x$.

From the above results, one finds the solution for all values of $x$ as
\begin{equation}
    \label{eqn:sol_q_EPM}
    F_y(x) =  F_{0}(y+x) + \theta(\hat{x}-x) \left[ - F_{0}(y) + 
    \sum_{n = 1}^{\infty} \left( F_{0}(y+x-n\hat{x}) \theta(y+x-n\hat{x}) -F_{0}(y-n\hat{x}) \theta(y-n\hat{x}) \right) \right],
\end{equation}
and by deriving with respect to $x$,
\begin{equation}
    \label{eqn:sol_p_EPM}
    P_{y}(x) = P_{0}(y+x) + \theta(\hat{x}-x)\sum_{n=1}^{\infty} \theta(y+x - n\hat{x})P_{0}(y+x-n\hat{x}).
\end{equation}
This solution is a periodic function of $y$ for large values of $y$ since the contribution from the first term vanishes when $y \xrightarrow{}\infty$. The 
periodicity of this function (of period $\hat x$) is clearly unphysical and disappears as soon as one introduces randomness in the jumps (or in the thresholds). 

The expressions of the average stress and of the connected susceptibility versus the strain $\gamma$ are then derived by using the general formulas given in the main text. Note that by combining Eq.~(\ref{eqn:sol_p_EPM}) and Eqs.~(\ref{eqn:gamma_fullaverage},\ref{eq_sigma_y_general}) one finds
\begin{equation}
\begin{aligned}
&\gamma(y+\hat x)-\gamma(y)=\frac 1{2\mu_2}[\hat x-x_c\int_0^{y+\hat x}dx' P_0(x')], \\&
\sigma(y+\hat x)-\sigma(y)=\int_{y+\hat x}^{\infty}dx' P_0(x')[\hat x+2(1+y-x')],
\end{aligned}
\end{equation}
so that when $y\gg 1$, $\gamma(y+\hat x)-\gamma(y)\to \hat x \frac{\mu_1}{2\mu_2(\mu_1+\mu_2)}$ and $\sigma(y+\hat x)-\sigma(y)\to 0$.
\\

We next study the disconnected susceptibility,
\begin{equation}
\label{eqn:disc_epm_nojumps}
\chi_{\rm dis}(\gamma)= \Delta_0 + \left( \frac{\mu_1}{\mu_2}\right)^{2} \overline{\delta\widehat y^{\alpha}(\gamma)^{2}} + 
2 \left( \frac{\mu_1}{\mu_2}\right)\overline{\delta\widehat y^{\alpha}(\gamma) \delta \widehat m^{\alpha}(0) },
\end{equation}
by considering as before the fluctuations with $y$ as the control parameter. 

We first calculate the squared-fluctuation term. One has
\begin{equation}
\label{eq_noJ_squared1}
    \overline{\delta\widehat \gamma^{\alpha}(y)^{2}} = \left( \frac{x_{c}}{2 \mu_{2}} \right)^{2} N  \sum_{n'=0}^{M} \sum_{n''=0}^{M} 
    \Big[\overline{ F^{\alpha}_{n'dy}(dy) F^{\alpha}_{n''dy}(dy)} - F_{n'dy}(dy) F_{n''dy}(dy)\Big],
\end{equation}
where $Mdy=y$. The function $F_{y',y''}(x',x'')=\overline{ F^{\alpha}_{y'}(x') F^{\alpha}_{y''}(x'')}$ satisfies evolution equations that generalize 
Eq.~(\ref{eqn:q_evolution}):
\begin{equation}
\label{eqn:finding_H}
\begin{aligned}
&\partial_{y'} F_{y',y''}(x',x'') = \partial_{x'} F_{y',y''}(x',x'') -  \partial_{x'} F_{y',y''}(x,x')\bigg|_{x'=0} \theta(\hat{x}-x') \\& 
\partial_{y''} F_{y',y''}(x',x'')= \partial_{x''} F_{y',y''}(x',x'') - \partial_{x''} F_{y',y''}(x,x')\bigg|_{x''=0} \theta(\hat{x}-x'').
    \end{aligned}
\end{equation}
Furthermore, one can introduce the function $H_{y',y''}(x',x'')$ through
\begin{equation}
\label{eq_noJ_defH}
F_{y',y''}(x',x'')= \left(1 - \frac{1}{N}\right) F_{y'}(x') F_{y''}(x'') + \frac{1}{N} H_{y',y''}(x',x''), 
\end{equation}
where we have used the independence of the evolution with $y$ of the variables on different sites.

The above equations for $F_{y',y''}(x',x'')$ translate into equations for $H_{y',y''}(x',x'')$. They can be solved by using the procedure followed above 
for $F_y(x)$ and considering separately the cases $x',x''<\hat x$, $x'<\hat x<\hat x''$ or $x''<\hat x<\hat x'$, and $\hat x<x',x''$. One can then show that
\begin{equation}
    H_{y',y''}(x',x'') =  U(y'+x', y''+x'') - U(y'+x',x'') -U(x',y''+x'') + U(y',y''),
\end{equation}
with 
\begin{equation}
    \begin{aligned}
        U(z',z'') =& F_{0}(\min(z',z'')) + \sum_{n=1}^{\infty} F_{0}(\min(z' -n\hat{x},z''))\theta \Big( z'-n\hat{x} \Big) + \sum_{n=1}^{\infty} F_{0}(z',\min(z'' -n\hat{x}))\theta \Big( z''-n\hat{x} \Big)  +\\
    &+\sum_{n'=1}^{\infty} \sum_{n''=1}^{\infty} F_{0}(\min(z' -n'\hat{x},z''-n'' \hat{x}))\theta \Big( z'-n'\hat{x} \Big) \theta \Big( z''-n''\hat{x} \Big).
    \end{aligned}
\end{equation}

From Eqs.~(\ref{eq_noJ_squared1}, \ref{eq_noJ_defH}), one can write the averaged squared fluctuations as
\begin{equation}
   \overline{\delta\widehat \gamma^{\alpha}(y)^{2}} = \left( \frac{x_{c}}{2 \mu_{2}} \right)^{2}   \sum_{n'=0}^{M} \sum_{n''=0}^{M} 
   \Bigg[H_{n'dy, n''dy}(x',x'') - F_{n'dy}(dy) F_{n''dy}(dy)\Bigg],
\end{equation}
which after expanding in $dy$ and taking the continuum limit gives
\begin{equation}
\label{eqn:final_Gamma}
    \overline{\delta\widehat \gamma^{\alpha}(y)^{2}} =  \left( \frac{x_{c}}{2 \mu_{2}} \right)^{2} [ U(y,y) - H(y)^{2}],
\end{equation}
where $H(y)$ has been defined before.

We now compute the cross term $\overline{\delta \widehat \gamma^\alpha(y)\delta \widehat m^\alpha(0)}$ which can be expressed as 
\begin{equation}
\overline{\delta \widehat \gamma^\alpha(y)\delta \widehat m^\alpha(0)}=\frac{x_{c}\sqrt N}{2 \mu_{2}} \sum_{n=1}^{M}   
\overline{ \Big[F_{n \, dy} (dy) - F_{n \, dy}^{\alpha}(dy)  \Big] [m^{\alpha}(0)-1]},
\end{equation}
with
\begin{equation}
m^{\alpha}(0) =  \int_{0}^{\infty} dx' x' \frac{\partial}{\partial x'} F_{0}^{\alpha}(x').
\end{equation}
After some lengthy but straightforward manipulations, we find that
\begin{equation}
\overline{\delta \widehat \gamma^\alpha(y)\delta \widehat m^\alpha(0)}= \frac{x_{c}}{2 \mu_{2}} \sum_{n=0}^{\infty} 
 \int_{0}^{y-n\hat{x}}dy' P_{0}(y') ( 1-y' ). 
\end{equation}

Using the formulas to switch from fluctuations at constant $y$ to fluctuations at constant $\gamma$ and collecting all the terms, we finally obtain 
an expression for the disconnected susceptibility.
\\

\subsection{Annealed average over the random jumps}
\label{app:RJ_averages}

We go back to the generic case where the local stress jumps are random and we consider an annealed calculation in which one first averages 
all quantities over the random jumps, i.e., 
\begin{equation}
    A^{\alpha,[\hat{x}]_{M}}_y(x) \rightarrow A^{\alpha}_y(x) = \left< A^{\alpha,[\hat{x}]_{M}}_y(x) \right>_{[\hat{x}]_M}.
\end{equation}
Clearly, this procedure does not change the result for the average stress nor for the connected susceptibility. However, this modifies the 
sample-to-sample fluctuations and the disconnected susceptibility, which is now written
\begin{equation}
\begin{aligned}
\chi_{\text{dis}}^{\rm (ann)}(\gamma) =
\Delta_0 +  \left(\frac{\mu_1}{\mu_2} \right)^2 \overline{\left<  \delta \widehat y^\alpha(\gamma)\right>^2_{[\hat{x}]_M}} + 
2 \left (\frac{\mu_1}{\mu_2}\right ) \overline{\left<  \delta \widehat y^\alpha(\gamma)\right>^2_{[\hat{x}]_M}\delta \widehat m^{\alpha}(0)},
\end{aligned}
\end{equation}
One can see that the only term that is different from the quenched-average calculation is the squared-fluctuation term (the second one in the right-hand 
side). We thus compute this term by changing again from $\gamma$ to $y$ for the control parameter.

We introduce the random-jump averaged fluctuation
\begin{equation}
 \delta \widehat \gamma^{\alpha}(y)=\left< \delta \widehat \gamma^{\alpha,[\hat{x}]_M}(y)\right>_{[\hat{x}]_M} = 
 \frac{x_c}{2 \mu_2}  \sum_{n=1}^M \left(  F_{ndy}^{\alpha}(dy) -  F_{ndy}(dy)   \right),
\end{equation}
with
\begin{equation}
    F_{y}^{\alpha}(x) = \left< F_{y}^{\alpha,[\hat{x}]_M}(x)\right>_{[\hat{x}]_M} = F_0^{\alpha}(x+y) -G(x)F_0^{\alpha}(y) +
    \int_0^y dy' F_0^{\alpha}(y')S_{y-y'}(x).
\end{equation}

When computing the sample-averaged squared fluctuations we will have to consider the average of products of functions $ F_{y}^{\alpha}(x)$. We 
therefore derive the general formula
\begin{equation}
\begin{aligned}
&\overline{F^{\alpha}_{y'}(x)F^{\alpha}_{y''}(x)} =\left(1 - \frac{1}{N}\right)F_{y'}(x)F_{y''}(x) + \\&
+\frac{1}{N} \Big[ \frac{1}{2} \left ( F_0(\min(y',y'')+x) + G^{2}(x)F_0(\min(y',y'')) + \int_0^{y'} \int_0^{y''} d\widehat{y}'d\widehat{y}'' 
F_0(\min(\widehat{y}',\widehat{y}'')) S_{y'-\widehat{y}'}(x) S_{y''-\widehat{y}''}(x) \right ) +  \\&
- G(x) F_0(\min(x+y',y'')) +\int_0^{y''}d\widehat{y}[F_0(\min(y'+x,\widehat{y})) -G(x)F_0(\min(y',\widehat{y}))]S_{y''-\widehat{y}}(x)  + y' \leftrightarrow y'' \Big],
\end{aligned}
\end{equation}
where $y' \leftrightarrow y''$ denotes the term obtained by exchanging $y'$ and $y''$.

To simplify the calculation, we restrict ourselves to the single-exponential random-jump distribution. Then,
\begin{equation}
    F_{y}^{\alpha}(x) =  F_0^{\alpha}(x+y) -e^{-\frac{x}{<\hat x>}}F_0^{\alpha}(y).
\end{equation}
and 
\begin{equation}
\begin{aligned}
&\overline{F^{\alpha}_{y'}(x)F^{\alpha}_{y''}(x)} =\left(1 - \frac{1}{N}\right)F_{y'}(x)F_{y''}(x) + \\ 
        &+\frac{1}{N}\left[ F_0(x+\min(y',y'')) + e^{-\frac{2x}{<\hat x>}}F_0(\min(y',y'')) - e^{-\frac{x}{<\hat x>}} (F_0(\min(y',x+y'')) + F_0(\min(y'',x+y'))\right].
\end{aligned}
\end{equation}
We can now compute the fluctuations as 
\begin{equation}
\begin{aligned}
\overline{ \delta \widehat \gamma^{\alpha}(y)^2}& = N\left(\frac{<\hat x>}{2 (\mu_2+\mu_1)} \right)^2 \sum_{n',n''=1}^M \left(  \overline{F_{n'dy}^{\alpha}(dy)
 F_{n''dy}^{\alpha}(dy)} -  F_{n'dy}(dy)F_{n''dy}(dy)   \right)  = \\&= \left(\frac{<\hat x>}{2 (\mu_2+\mu_1)} \right)^2 \sum_{n',n''=1}^M\left[ F_0(dy+dy\min(n',n'')) 
 + e^{-\frac{2dy}{<\hat x>}}F_0(dy\min(n',n''))+\right. \\&\left.- e^{-\frac{dy}{<\hat x>}} (F_0(dy\min(n',1+n'')) + 
 F_0(dy\min(n'',1+n')) -F_{n'dy}(dy)F_{n''dy}(dy)\right].
\end{aligned}
\end{equation}
Some care is needed to consider separately the cases $n'=n''$ and $n'<n''$ (or $n''<n'$), and we finally arrive at the sought-for expression,
\begin{equation}
    \label{final_gamma_annealed}
    \begin{aligned}
\overline{\delta \widehat y^\alpha(\gamma)^2} =
\left(\frac{<\hat x>}{2(\mu_2+\mu_1)} \right)^2 \Big[F_0(y)  +  \frac{2}{<\hat x>}\int_{0}^{y} dy'P_{y'}(0)(y - y' ) 
-  \left(\int_{0}^{y}dy' P_{y'}(0) \right)^2\Big].
    \end{aligned}
\end{equation}
From this and previous results we can compute the annealed disconnected susceptibility as a function of $\gamma$.
\\

\section{Direct mapping between correlation functions in the mean-field EPM and a mean-field RFIM}
 \label{app_direct-mapping}

In this appendix we push further the idea of a direct mapping between correlation functions in the mean-field EPM and a mean-field RFIM which was discussed in Sec.~\ref{subsec_contrasting}. As already explained, this can only be done in a restricted range of the driving parameter, in the vicinity of the 
yielding transition. In particular, the range must be small enough that (i) the connected susceptibility of the mean-field EPM is positive and large 
[region (iii) in the main text, i.e., between the overshoot and the steady state] and that, moreover, (ii) the sites in the EPM yield only once. We 
define this range as $[\gamma_0, \gamma_{\rm m}]$, which translates in an interval $[H_0, H_{\rm m}]$ for the RFIM. At $\gamma_0$ the 
state of the EPM is given by the exact evolution starting from $\gamma=0$. However, we have more freedom for the RFIM. The only requirement 
is that the system be in a metastable state at $H_0$, and we can relax the condition used in the main text that this state is the one obtained by 
ramping up the external field from $-\infty$ and by following the ascending (lower) branch of the hysteresis loop. In the following we first express the 
evolution of the mean-field EPM and mean-field RFIM in terms of the initial condition at $\gamma_0$ or $H_0$ only.

\subsection{Mean-field EPM starting from $\gamma_0>0$}

For simplicity, as in Sec.~\ref{subsec_contrasting}, we consider a model in which randomness is only in the initial condition, with 
$g(\hat x) = \delta(\hat x-<\hat{x}>)$. This is not expected to alter the qualitative behavior of the system in the close vicinity of the yielding transition 
(of course, as shown in Sec.~\ref{subsec_stress-strain}, it does for the steady state). We again switch the control parameter from $\gamma$ to $y$ 
and then study the evolution between $y_0$ and $y_{\rm m}$, with $y_0=\overline{y^\alpha(\gamma_0)}$ and 
$y_{\rm m}=\overline{y^\alpha(\gamma_{\rm m})}$. 
Due to the assumption of a single jump per site, the local stability evolves according to 
\begin{equation}
\begin{aligned}
\label{eqn:dynamic_evolution_equation_EPM}
   x_i^{\alpha}(y) &=
    x_i^{\alpha}(y_0) - (y-y_0), & {\rm when}\; (y-y_0)< x_i^{\alpha}(y_0),\\&
    = x_i^{\alpha}(y_0) - (y-y_0) + <\hat{x}>, & {\rm when}\;  (y-y_0)> x_i^{\alpha}(y_0),
    \end{aligned} 
\end{equation}
which is the equivalent of Eq.~(\ref{eq_contrast_EPM}) in main text, with the local stability as a function of $y$ instead of $\gamma$. One immediately 
derives for the volume-averaged quantity [compare with Eq.~(\ref{eq_def_volume-averaged})]
\begin{equation}
\label{eqn_volume-averaged}
  \widetilde{m}^\alpha(y) = \widetilde{m}^\alpha(y_0) - (y-y_0) + <\hat{x}> \int_{0}^{y-y_0} dx P^\alpha_{y_0}(x),
\end{equation}
where $P^\alpha_{y_0}(x)=(1/N)\sum_i\delta(x-x_i^\alpha(y_0))$, and, after averaging over the samples,
\begin{equation}
\label{eqn:x_evolution}
  \widetilde{m}(y) = \overline{x}(y_0) - (y-y_0) + <\hat{x}> \int_{0}^{y-y_0} dx P_{y_0}(x),
\end{equation}
To go back from $y$ to the original control parameter $\gamma$, we use
\begin{equation}
\label{eqn:gamma-gamma0}
\begin{aligned}
2 \mu_2 [\gamma^{\alpha}(y) -\gamma^\alpha(y_0)] &= y - y_0 
-\frac{x_c}{N} \sum_i \int_{y_0}^y dy' \delta(-x_i^{\alpha}(y')) \\&
=y - y_0 -\frac{x_c}{N} \sum_i \int_{y_0}^y dy' \delta(-x_i^{\alpha}(y_0)+y'-y_0)
\end{aligned}
\end{equation}
where the second equation is obtained by taking into account the fact that only one jump per site is allowed. This is true provided $y-y_0$ is small 
enough (in particular, one should have $y-y_0 \ll <\hat{x}>$). After averaging, one has
\begin{equation}
2\mu_2 [\gamma(y) - \gamma_0 ]= (y-y_0) - x_c \int_{0}^{y-y_0}dy' P_{y_0}(y'),
\end{equation}
which can be inverted to give
\begin{equation}
\Delta y(\gamma) = y(\gamma)-y_0 = 2 \mu_2 (\gamma -\gamma_0)+ x_c \int_0^{\Delta y(\gamma)} dy' P_{y_0}(y'),
\end{equation}
and by derivation,
\begin{equation}
\frac{\partial \Delta y(\gamma)}{\partial \gamma } = \frac{2 \mu_2}{1-x_c P_{y_0}( \Delta y(\gamma))}\,,
\end{equation}
where one should keep in mind that $\Delta y$ also depends on $\gamma_0$.
 
The connected susceptibility $ \chi_{\rm con}(\gamma) = m'(\gamma)$ can then be expressed as
\begin{equation}
\begin{aligned}
\label{eqn:epm_conn}
    \chi_{\rm con}(\gamma) = \frac{\partial \widetilde{m}(\Delta  y)}{\partial \Delta y } \frac{\partial \Delta y(\gamma)}{\partial \gamma}    
    =  \frac{2 \mu_2 \big(-1 +  <\hat{x}> P_{y_0}( \Delta y(\gamma))\big)}{1-x_c P_{y_0}(\Delta y(\gamma))}.
\end{aligned}
\end{equation}
One can notice by comparing with the exact expression in Eq.~(\ref{eq:conn_MF-EPM}) that the single-jump approximation leads to the substitution 
of $P_{y_0+\Delta y}(0)$ with $P_{y_0}(\Delta y)$. This makes sense, since if a site is unstable at $y$, with $x_i(y)=0$, it should be the first time since 
$y_0$, meaning that it evolved only elastically with, as a consequence, $x_i(y_0)=y-y_0$. The condition that the connected susceptibility is positive in the 
relevant interval of $\gamma$ or $y$ implies that
\begin{equation}
    P_{y_0}(\Delta y) > \frac{1}{<\hat{x}>},
\end{equation}
and puts a bound on the possible values of $y_0$ and $y_{\rm m}$.

We now proceed with the computation of the sample-to-sample fluctuations. In what follows we set $2\mu_2=1$ to simplify the notation. Following the same 
procedure as in Sec.~\ref{sec_suscept_EPM}, we introduce the reduced fluctuations $\delta \widehat{m}^{\alpha}(\gamma)$, 
$\delta \widehat{\widetilde{m}}^{\alpha}(y)$, $\delta \widehat{y}^{\alpha}(\gamma)$, and $\delta \widehat{\gamma}^{\alpha}(y)$. One still has
\begin{equation}
\label{eqn:fluctuations_of_m_at_H_RFIM}
    \delta \widehat{m}^{\alpha}(\gamma) = \delta \widehat{\widetilde{m}}^{\alpha}(y(\gamma)) + \widetilde{m}'(y(\gamma) )\delta \widehat{y}^{\alpha}(\gamma),
\end{equation}
and therefore,
\begin{equation}
    \delta \widehat{m}^{\alpha}(\gamma) -  \delta \widehat{m}^{\alpha}(\gamma_0) = \delta \widehat{\widetilde{m}}^{\alpha}(y(\gamma)) - 
    \delta \widehat{\widetilde{m}}^{\alpha}(y_0) + \widetilde{m}'(y(\gamma) )\delta \widehat{y}^{\alpha}(\gamma) -
    \widetilde{m}'(y_0 )\delta \widehat{y}^{\alpha}(\gamma_0).
\end{equation}
By combining Eq.~(\ref{eqn:gamma-gamma0}) and (\ref{eqn_volume-averaged}), we derive
\begin{equation}
\gamma^{\alpha}(y) -\gamma^{\alpha}(y_0) = \left( 1-\frac{x_c }{<\hat{x}>}\right) \Delta y  -\frac{x_c}{<\hat{x}>}[\widetilde{m}^{\alpha}(y)-\widetilde{m}^{\alpha}(y_0)],
\end{equation}
which can be inverted to give
\begin{equation}
    y^{\alpha}(\gamma)-  y_0 = \frac{<\hat{x}>}{<\hat{x}>-x_c}(\gamma - \gamma_0) + \frac{x_c}{<\hat{x}>-x_c}[m^{\alpha}(\gamma)-m^{\alpha}(\gamma_0)].
\end{equation}
The fluctuations of $y$ can then be rewritten as 
\begin{equation}
\delta \widehat{y}^{\alpha}(\gamma) = \delta \widehat{y}^{\alpha}(\gamma_0) + \frac{x_c}{<\hat{x}>-x_c}\left[ \delta \widehat{m}^{\alpha}(\gamma) - 
\delta \widehat{m}^{\alpha}(\gamma_0) \right],
\end{equation}
from which we get 
\begin{equation}
\delta \widehat{m}^{\alpha}(\gamma) =  \delta \widehat{m}^{\alpha}(\gamma_0)  + \frac{\delta \widehat{\widetilde{m}}^{\alpha}(y(\gamma)) - 
\delta \widehat{\widetilde{m}}^{\alpha}(y_0) + (\widetilde{m}'(y(\gamma))-\widetilde{m}'(y_0))  \delta \widehat{y}^{\alpha}(\gamma_0)}{1 - \frac{x_c}{<\hat{x}>-x_c}\widetilde{m}'(y(\gamma))}.
\end{equation}

The disconnected susceptibility $\chi_{\rm dis}(\gamma) = \overline{\delta \widehat{m}^{\alpha}(\gamma)^2}$ can then be expressed as
\begin{equation}
\begin{aligned}
\label{eq_disc_sum}
\chi_{\rm dis}(\gamma) =& \chi_{\rm dis}(\gamma_0) + 2 \chi_{\rm con}(\gamma)\frac{ \overline{\delta \widehat{m}^{\alpha}(\gamma_0) 
[\delta \widehat{\widetilde{m}}^{\alpha}(y(\gamma)) -\delta \widehat{\widetilde{m}}^{\alpha}(y_0) + (\widetilde{m}'(y(\gamma))- \widetilde{m}'(y_0)) 
\delta \widehat{y}^{\alpha}(\gamma_0)]} }{\frac{<\hat{x}>}{<\hat{x}> - x_c}\widetilde{m}'(y(\gamma)) } \\&
+\chi_{\rm con}(\gamma)^2  \frac{\overline{\big [\delta \widehat{\widetilde{m}}^{\alpha}(y(\gamma)) -\delta \widehat{\widetilde{m}}^{\alpha}(y_0) + 
(\widetilde{m}'(y(\gamma))- \widetilde{m}'(y_0)) \delta \widehat{y}^{\alpha}(\gamma_0)\big ]^2} }{\left(\frac{<\hat{x}>}{<\hat{x}> - x_c}
\widetilde{m}'(y(\gamma)) \right)^2}, 
\end{aligned}
\end{equation}
where we have introduced the connected susceptibility given by Eq.~(\ref{eqn:epm_conn}). As already stressed, we are interested in a range 
$[\gamma_0,\gamma_{\rm m}]$ such that $\chi_{\rm con}(\gamma) \gg 1$. Then the variance of the effective random field 
$\Delta_{\rm eff}(\gamma)=\chi_{\rm dis}(\gamma)/\chi_{\rm con}(\gamma)^2$ is obtained as
\begin{equation}
\label{eq_variance_eff}
\Delta_{\rm eff}(\gamma) \approx \frac{\overline{\big(\delta \widehat{\widetilde{m}}^{\alpha}(y(\gamma)) -\delta \widehat{\widetilde{m}}^{\alpha}(y_0) 
+ (\widetilde{m}'(y(\gamma))- \widetilde{m}'(y_0)) \delta \widehat{y}^{\alpha}(\gamma_0)\big)^2} }{\left[\frac{<\hat{x}>}{<\hat{x}> - x_c}\widetilde{m}'(y(\gamma)) \right]^2},
\end{equation}
where all quantities can be calculated as done in the main text and in Appendix~\ref{app:solution_MF-EPM}.

\subsection{Mean-field RFIM starting from $H_0$}

We consider a starting point $H_0$ which is finite and between the two coercive fields, so that the sites can be either on the upper or lower branch of 
their hysteresis loop, since they both represent local minima of the Hamiltonian. The sites that have already flipped (upper branch) can only contribute elastically 
to the magnetization curve (remember however that the RFIM has a ``negative elasticity'' compared to the EPM), so that the plastic-like behavior associated 
with changes of local minimum depends only on the fraction of sites that start on the lower branch. (Notice that, as mentioned above,  such an initial configuration 
is not reachable by starting from a magnetic field that goes to $-\infty$.)

We start by introducing a variable $y$, analogous to that used in the EPM,
\begin{equation}
\label{eqn:control_relation_RFIM}
y^{\alpha}(H)= H + J m^{\alpha}(H) = H + \frac{J }{N k}\sum_i \phi_i^{\alpha}(H).
\end{equation}
where we have defined for convenience $\phi_i^{\alpha}=ks_i^{\alpha}$.
Inverting the above relation gives
\begin{equation}
    H^{\alpha}(y) = y - \frac{J }{N k}\sum_i \phi_i^{\alpha}(y).
\end{equation}

To describe the initial preparation of the system at $H=H_0$ or $y=y_0$ we consider that all the site variables are independently distributed according to 
the distribution $P_{y_0}(\phi)$, which plays the same role as $P_{y_0}(x)$ in the EPM. The constraint that the site with local magnetization $\phi$ 
correspond to a local minimum of the Hamiltonian imposes the following form:
\begin{equation}
\begin{aligned}
\label{eqn:p_y0}
P_{y_0}(\phi)=&  \theta(\phi-2k) \big[\rho_-(\phi-k-y_0) + \rho_+(\phi-k-y_0)\big] 
+\theta(-\phi-2k) \big[\rho_-(\phi+k-y_0) + \rho_+(\phi+k-y_0)\big] \\&
+ \theta(\phi) \theta(2k-\phi) \rho_+(\phi-k-y_0) + \theta(-\phi) \theta(2k+\phi) \rho_-(\phi+k-y_0),
\end{aligned}
\end{equation}
where we have defined the probability distributions $\rho_-(h)$ and $\rho_+(h)$ for having a random field $h$ and being in the lower and upper branch, 
respectively (which, again, means that the local magnetization has or has not flipped from the left minimum of the $2$-parabola potential to the right 
one: see also Fig.~\ref{fig_EPM-RFIM_sketch}). The situation considered in the main text corresponds to taking $y_0 \to -\infty$ and $\rho_+(h)=0$. The 
two distributions $\rho_{-}$ and $\rho_+$ satisfy
\begin{equation}
\begin{aligned}
\label{eq_constraints_distrib}
&\int_{-\infty}^{\infty}dh \rho_-(h)  = a, \; \int_{-\infty}^{\infty}dh \rho_+(h) = 1-a,  \\&
\int_{-\infty}^{\infty}dh\, h\rho_-(h)= \int_{-\infty}^{\infty} dh\,h \rho_+(h) = 0, \\&
\rho_+(h<-k-y_0) = 0, \; \rho_-(h>k-y_0) = 0,
\end{aligned}
\end{equation}
where the last two conditions come from the fact that a large negative (respectively, a large positive) local random field forces the local magnetization to 
be in the lower (respectively, the upper branch). Note that $a \in [0,1]$ is a free parameter.

In what follows we will consider the case $y_0>0$, which we anticipate to be the relevant one. Taking Eq.~(\ref{eq_constraints_distrib}) into account, 
the expression of $P_{y_0}(\phi)$ then simplifies to 
\begin{equation}
P_{y_0}(\phi) =  \theta(\phi-2k) \big[\rho_-(\phi-k-y_0) + \rho_+(\phi-k-y_0)\big]
 + \theta(\phi) \theta(2k-\phi) \rho_+(\phi-k-y_0) + \theta(-\phi) \rho_-(\phi+k-y_0).
\end{equation}

The evolution of the local magnetization $\phi_i^\alpha$ as a function of $y$ reads
\begin{equation}
\begin{aligned}
\phi^{\alpha}_i(y) &= \phi^{\alpha}_i(y_0) + y-y_0 , \;\;{\rm when}\;\;  \phi^{\alpha}_i(y_0) >2k\; {\rm or}\; \phi^{\alpha}_i(y_0) < -(y-y_0) \\&
     =\phi^{\alpha}_i(y_0) + y-y_0 + 2k, \;\; {\rm when}\; \;0>\phi^{\alpha}_i(y_0)> -(y-y_0),  
\end{aligned}\end{equation}
if $i$ belongs to the sites on the lower branch between $y_0$ and $y$, and 
\begin{equation}
    \phi^{\alpha}_i(y) = 
    \phi^{\alpha}_i(y_0) + y-y_0
\end{equation}
otherwise. As in the EPM, if one considers $y$ as our control parameter, each site evolves independent from the others. After averaging, one finds
\begin{equation}
\begin{aligned}
\phi(y)=k \widetilde m(y) &= \overline{\phi^{\alpha}_1}(y_0)+y-y_0 + 2k \int_{-(y-y_0)}^{0} {\rm d}x P_{y_0}(x)  \\&
= \overline{\phi^{\alpha}_1}(y_0)+y-y_0 + 2k \int_{-(y-y_0)}^{0} {\rm d}x \rho_-(x+k-y_0).
\end{aligned}
\end{equation}
From this expression and the average of Eq.~(\ref{eqn:control_relation_RFIM}), $y(H)= H+(J/k)\phi(y(H))$ one can compute the connected 
susceptibility
\begin{equation}
\chi_{\rm con}(H)=\frac 1k\phi'(y)y'(H)=\frac{\phi'(y(H))}{k-J\phi'(y(H))}.
\end{equation}

We now consider the disconnected susceptibility $\chi_{\rm dis}(H)=\overline{[\delta\widehat m^\alpha(H)]^2}$. We again use the same procedure of 
relating the fluctuations at constant $H$ to the fluctuations at constant $y$. If the bare disorder (random-field distribution) is independent of the applied 
magnetic field $H$, the sample-to-sample fluctuations at a given $H$ can be formally expressed in a way that do not explicitly depend on the starting point $H_0$, and one has
\begin{equation}
\label{eq_def_variance_RFIM}
\Delta_{\rm eff}(H)= \frac{\overline{[\delta \widehat{\widetilde{m}}^{\alpha}(y(H))]^2}}{\widetilde{m}'(y(H))^2}.
\end{equation}
Actually, the disconnected susceptibility $\chi_{\rm dis}(H)$ can be cast as the sum of three terms as in Eq.~(\ref{eq_disc_sum}) for the EPM, but in the present case 
$\delta \hat{y}^{\alpha}(H_0)$ is simply given by $J \delta \widehat{\widetilde{m}}^{\alpha}(y_0)$ and the sum of three terms simplifies to the above equation. Note however that $\delta \widehat{\widetilde{m}}^{\alpha}(y(H))$ and $\widetilde{m}'(y(H))$ do depend on the choice of initial condition in $H_0$, as we show below.

The sample-to-sample fluctuations at fixed $y$ can be calculated by using the evolution equations given above and after some lengthy but straightforward 
manipulations we arrive at
\begin{equation}
\begin{aligned}
&\overline{[\delta \widehat{\widetilde{m}}^{\alpha}(y(H))]^2}=\left (\frac 1k\right )^2 [\overline{\phi^{\alpha}_i(y)^2} - \overline{\phi^{\alpha}_i(y)}^2] \\&
= 
1 - (1-2a)^2  + \frac 1{k^2}\int_{-\infty}^{\infty}dh h^2 [\rho_-(h)+\rho_+(h)]+ \frac 4k \int_{k-y(H)}^{\infty}dh h\rho_-(h) 
- 4 \int_{k-y(H)}^{\infty}dh \rho_-(h)[2a-1+\int_{k-y(H)}^{\infty}dh \rho_-(h)]
\end{aligned}
\end{equation}
which is valid for $H\geq H_0$. By combining this expression with Eq.~(\ref{eq_def_variance_RFIM}) one obtains the disconnected susceptibility. Note that when 
$y_0\to -\infty$, $\rho_+(h)=0$, $\rho_-(h)=\rho(h)$ (as a result, $a=1$), and one exactly recovers the expression in Eq.~(\ref{eqn:disc_MF-RFIM}), 
after using $y(H)=H+J m(H)$ and $\Delta_B=\int_{-\infty}^{\infty}dh h^2 \rho(h)$.

\subsection{Mapping between the two models}

We first consider the mapping of the mean-field EPM to a mean-field RFIM at the average level, i.e., for the stress-strain curve and the connected susceptibility. It 
is convenient to look at the expressions parametrized by the control parameter $y$ (which is then taken as the identical in both models). For the EPM, one has
\begin{equation}
\begin{aligned}
\label{eqn:analogy_average_epm}
&m'(y)= -1 + <\hat{x}> P^{\rm EPM}_{y_0}(y-y_0)\\&
\frac{\partial \gamma }{\partial y}=\left (\frac{1-\frac{x_c}{<\hat x>}}{2 \mu_2}\right )[1-\frac{x_c}{<\hat{x}>-x_c}m'(y)].
\end{aligned}
\end{equation}
and for the RFIM,
\begin{equation}
\begin{aligned}
&m'(y)= \frac{1}{k} + 2 P^{\rm RFIM}_{y_0}(-(y-y_0)) \\&
H'(y)=1-J m'(y).
\end{aligned}
\end{equation}
To map the EPM on the RFIM we then define an interaction and a magnetic field in the RFIM as
\begin{equation}
\begin{aligned}
&J=\frac{x_c}{<\hat{x}>-x_c} \\&
H=\frac{2\mu_2}{1-\frac{x_c}{<\hat x>}}\gamma,
\end{aligned}
\end{equation}
and we choose the initial distribution such that
\begin{equation}
\label{eqn:analogy_main}
2P^{\rm RFIM}_{y_0}(-(y-y_0)) =2\rho_-(-y+k)= <\hat x>  P^{\rm EPM}_{y_0}(y-y_0) - \big( 1+\frac{1}{k}\big).
\end{equation}
The choice for $\rho_+(h)$ is still free and will not affect the evolution of the average magnetization, but the above equation puts a constraint on the maximum value of 
$y$: the right-hand side should indeed always be larger than 0, since on the left there is a probability. This restricts intervals $[y_0,y_{\rm m}]$ over which the analogy 
can hold.

To pursue the analogy one would like to describe the sample-to-sample fluctuations and tune the mean-field RFIM such that  
$\chi_{\rm{dis}}^{\rm{RFIM}}(y) \approx \chi_{\rm{dis}}^{\rm{EPM}}(y)$, in the limit where $\chi_{\rm{con}}\gg1$. Since $\rho_-(h)$ is already essentially fixed by the mapping at the level of the averaged quantities, we have some freedom with $\rho_+(h)$. This turns out not to be sufficient to match the two disconnected susceptibilities. A possible solution to this issue could be to consider an effective RFIM in which the bare disorder evolves with the applied magnetic field, i.e., with a variance $\Delta_B(H)$. This seems reasonable but we have not tried to implement this scenario. Note again that this mapping at the level of correlation functions or susceptibilities does not imply a direct mapping between individual dynamical trajectories.

\begin{figure}
\centering
    \includegraphics[width=.48\textwidth]{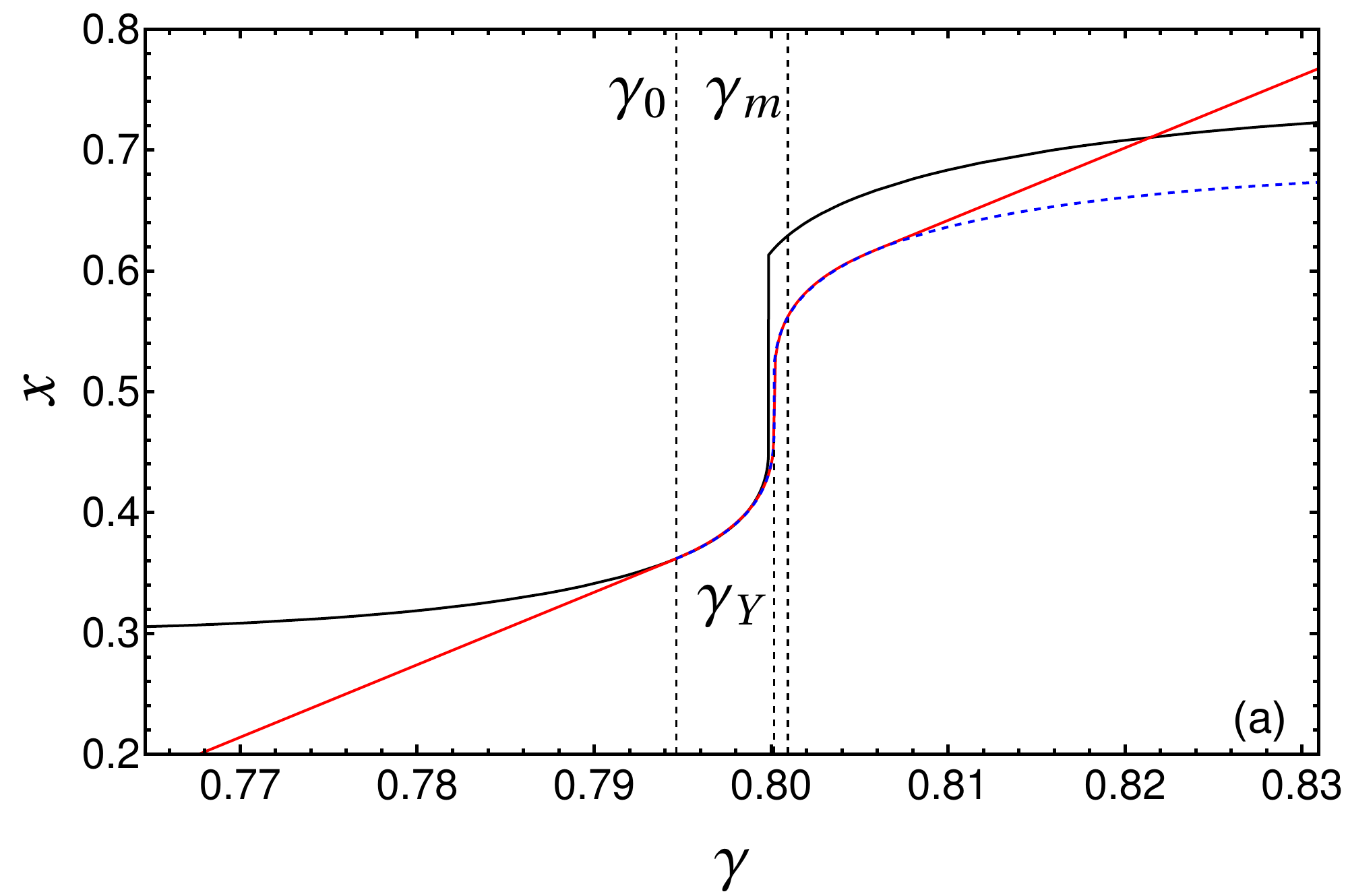}  
    \includegraphics[width=.48\textwidth]{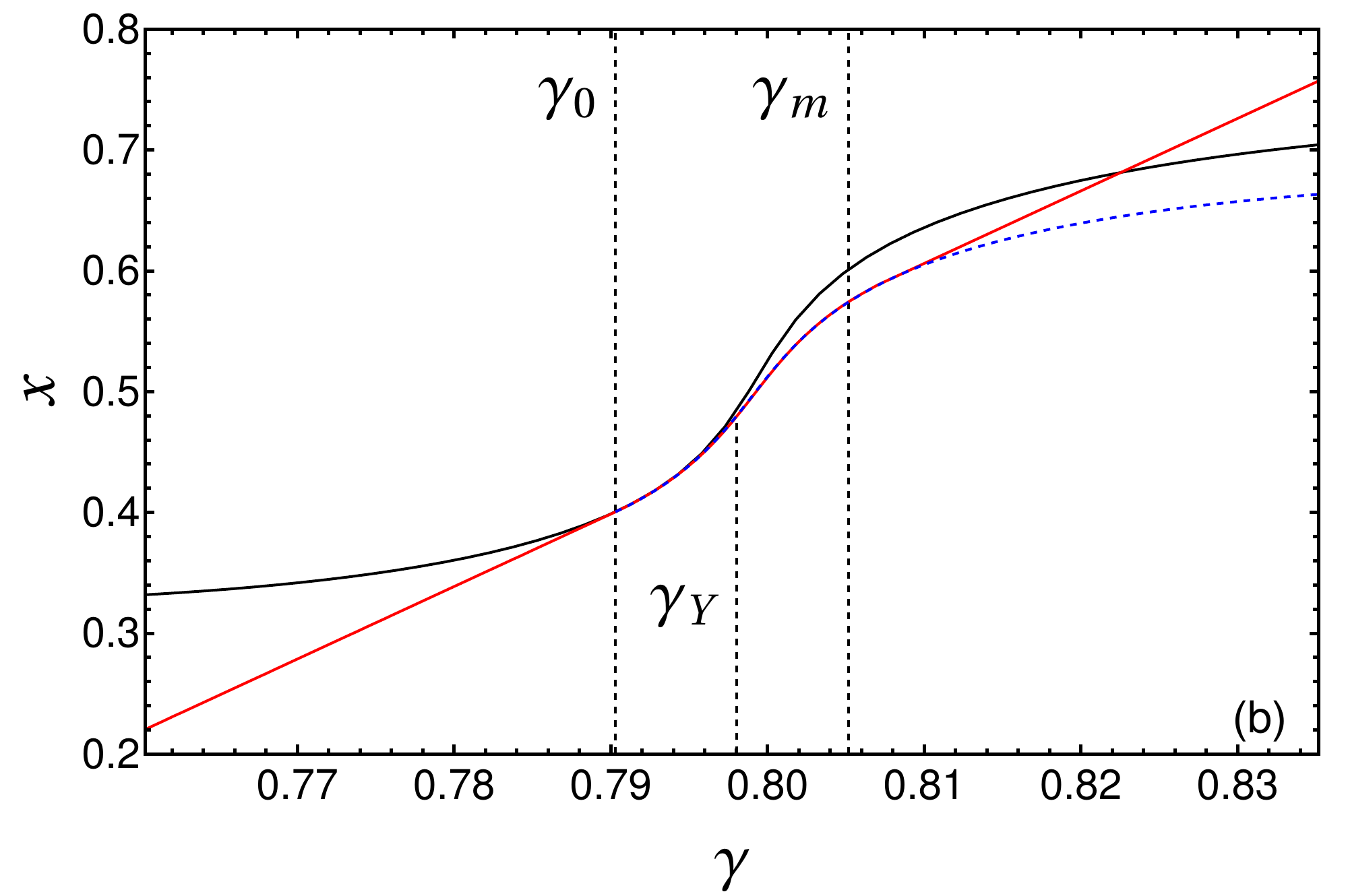}
    \\
    \includegraphics[width=.48\textwidth]{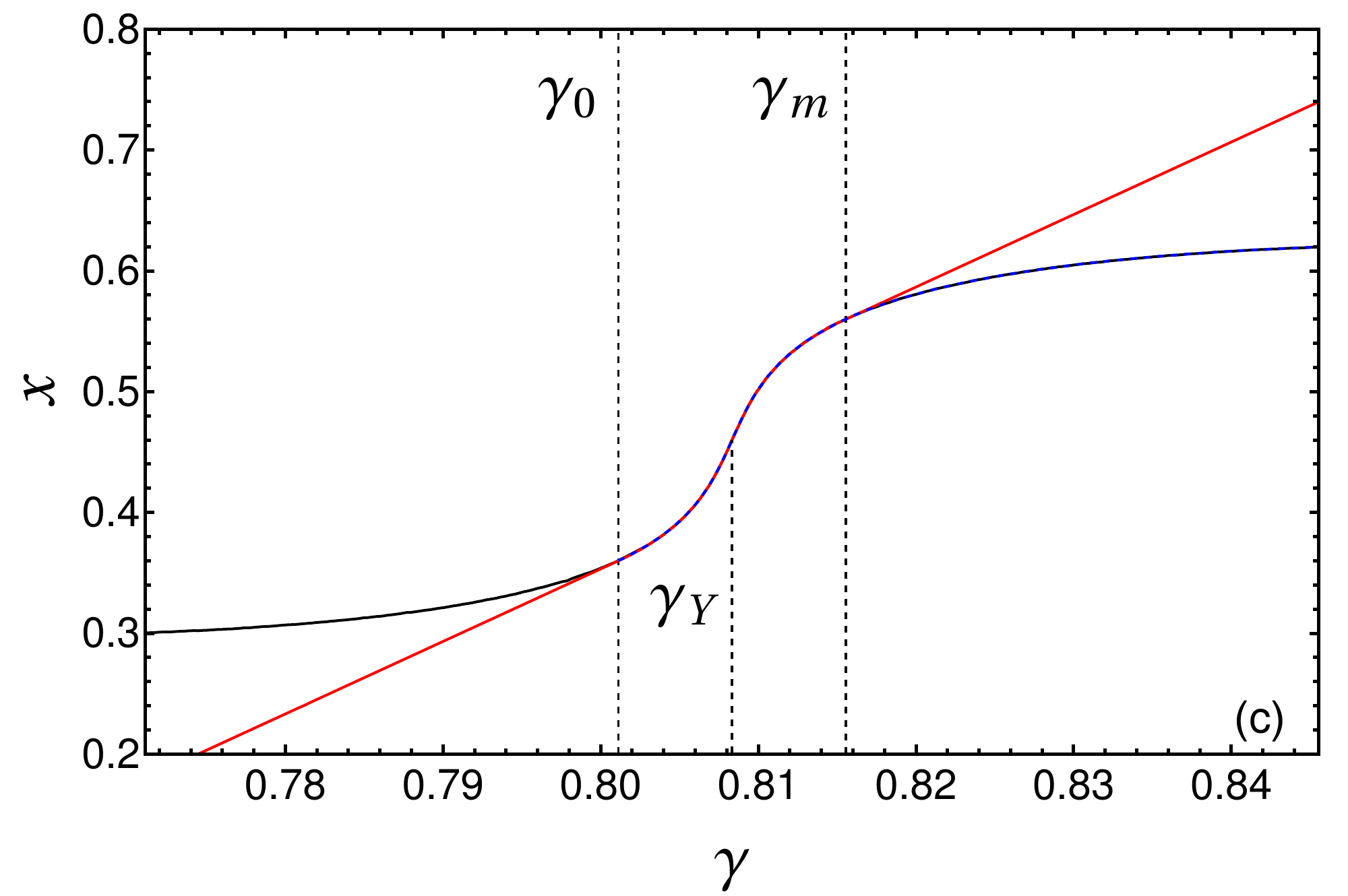}
\caption{Average stability $x(\gamma)=1-\sigma(\gamma)$ for the mean-field EPM. We focus on the elastic branch and the yielding region. Top row: Model with a 2-exponential  distribution of random jumps, in the weak-disorder (brittle) case (a) and in the continuous case with a ductile yielding (b). Here, $2\mu_2=1$, $\mu_1=0.7$ and $<\hat{x}>=0.92$. Bottom row (c): Model with randomness only in the initial conditions (no random jumps). In all cases the exact solution is displayed as a full black curve and the solution obtained under the assumption of a single jump per site as a dashed blue curve [they coincide in (c)]. The red curve represents the equivalent RFIM which exactly coincides with the single-jump EPM curve between the two vertical lines at $\gamma_0$ and $\gamma_{\rm m}$. We also indicate the location $\gamma_Y$ of the (discontinuous) yielding  transition in (a) or the maximum of the connected susceptibility in (b) and (c).}
\label{fig:mapping}
\end{figure}

\subsection{Results and test of the hypothesis of a single plastic event per site}

We illustrate the mapping of the mean-field EPM with the assumption of a single jump per site in the vicinity of the yielding transition to the mean-field 
RFIM at the average level in Fig.~\ref{fig:mapping}. We display in panel (a) a weak-disorder case with a brittle (discontinuous) yielding and and in panel (b) a value of the disorder close to the critical point. In both cases we also show the exact solution for the EPM. One can see that the assumption of a single jump per site is virtually exact from the initial condition at $\gamma_0$ up to yielding 
and then starts to deteriorates, as expected. The direct mapping to the RFIM is found to hold in a narrow region around yielding. The ``negative elasticity'' of 
the RFIM prevents a broader application.

We also note that the single-jump hypothesis is correct for a wide range of values of $y$ in the case without random jumps (see Fig.~\ref{fig:mapping}c, where the full black curve and the dashed blue one superimpose). A direct calculation shows that it is exact for $y\leq \hat x$.
\\

\newpage
\begin{acknowledgements}
We thank Giulio Biroli and Alberto Rosso for fruitful discussions.
\end{acknowledgements}


\begin{thebibliography}{85}
\bibitem{footnote_depinning}
A solution to observe an overshoot in the depinning problem is to consider a disorder that breaks the statistical translational invariance, for instance by modifying the distribution of the position of the wells so that it varies  as the interface progresses: Middleton's theorem then no longer applies.
\end{thebibliography}

\end{document}